\documentclass[12pt]{article}
\usepackage[utf8]{inputenc}
\usepackage[T1]{fontenc}
\usepackage{textcomp}
\usepackage{gensymb}
\usepackage{amsmath}
\usepackage[margin=1in]{geometry}
\usepackage{graphicx}
\usepackage{amssymb}
\usepackage{authblk}
\usepackage{blindtext}
\allowdisplaybreaks
\usepackage{graphics}
\usepackage{float}
\usepackage{setspace}
\usepackage{amsthm}
\theoremstyle{plain}

\usepackage{algorithm}
\usepackage{algpseudocode}

\newcommand{\bbeta}{ \mbox{\boldmath $ \beta $} }

\newcommand{\btheta}{ \mbox{\boldmath $ \theta $} }

\newcommand{\bSigma}{ \mbox{\boldmath $\Sigma$} }

\newcommand{\bLam}{ \mbox{\boldmath $\Lambda$} }

\newcommand{\bgamma}{ \mbox{\boldmath $\gamma$} }

\newcommand{\bone}{\textbf{1}}

\newcommand{\bZ}{\textbf{Z}}
\newcommand{\bz}{\textbf{z}}

\newcommand{\bk}{\textbf{k}}

\newcommand{\bh}{\textbf{h}}

\newcommand{\bI}{\textbf{I}}

\newcommand{\bM}{\textbf{M}}

\newcommand{\bs}{\textbf{s}}
\newcommand{\bS}{\textbf{S}}

\newcommand{\bV}{\textbf{V}}

\newcommand{\bR}{\textbf{R}}

\usepackage{subcaption}
\usepackage{amssymb}
\usepackage{hhline}
\RequirePackage[OT1]{fontenc}
\RequirePackage{amsthm,amsmath}
\RequirePackage{natbib}
\RequirePackage[colorlinks,citecolor=blue,urlcolor=blue]{hyperref}

\begin{document}

\title{Improving Piecewise Linear Snow Density Models through Hierarchical Spatial and Orthogonal Functional Smoothing}
\author[1]{Philip A. White \thanks{Corresponding Author: pwhite@stat.byu.edu}}
\author[2]{Durban G. Keeler \thanks{durban.keeler@gmail.com}}
\author[1]{Daniel Sheanshang\thanks{ danielsheanshang@gmail.com }}
\author[2]{Summer Rupper \thanks{ summer.rupper@geog.utah.edu}}

\affil[1]{Department of Statistics, Brigham Young University, USA}
\affil[2]{Department of Geography, University of Utah, USA}

\maketitle

\begin{abstract}

Snow density estimates as a function of depth are used for understanding climate processes, evaluating water accumulation trends in polar regions, and estimating glacier mass balances. The common and interpretable physically-derived differential equation models for snow density are piecewise linear as a function of depth (on a transformed scale); thus, they can fail to capture important data features. Moreover, the differential equation parameters show strong spatial autocorrelation. To address these issues, we allow the parameters of the physical model, including random change points over depth, to vary spatially. We also develop a framework for functionally smoothing the physically-motivated model. To preserve inference on the interpretable physical model, we project the smoothing function into the physical model's spatially varying null space. The proposed spatially and functionally smoothed snow density model better fits the data while preserving inference on physical parameters. Using this model, we find significant spatial variation in the parameters that govern snow densification.

\end{abstract}



\section{Introduction}\label{sec:intro}

Antarctica is the world's largest freshwater reservoir and, due to climatic changes, is linked to significant contributions to sea level rise \citep[see][]{thomas2004accelerated,hock2009mountain,deconto2016contribution}. To quantify and predict sea-level changes, polar scientists use snow density measurements or estimates to track Antarctic water accumulation \citep[see e.g.,][]{medley_constraining_2014, koenig_annual_2016, lewis_recent_2019, keeler_probabilistic_2020} and total mass balance trends \citep[see e.g.,][]{li_modeling_2011, csatho_laser_2014, shepherd_trends_2019, smith_pervasive_2020}. Understanding such mass changes are primary research targets within the scientific community, with critical implications for current and future sea-level rise \citep{ipcc_climate_2013}. Thus, snow density measurements and estimates are essential. However, due to the remoteness and extreme environments of Antarctica, snow density measurements are costly to obtain, making improved utilization of existing data of paramount concern \citep{eisen_ground-based_2008}. Therefore, studying and estimating snow density is an important research target in itself \citep{herron_firn_1980,helsen_elevation_2008, arthern_situ_2010,horhold_densification_2011, verjans2020bayesian,keenan2021physics}.

In this analysis, our primary goal is estimating how the snow density response varies over West Antarctica using physically-motivated thermal models. In most studies, the snow density response to climate, and in particular temperature, is generally assumed to be constant over space \citep{herron_firn_1980,  horhold_densification_2011, verjans2020bayesian}.
However, understanding spatial variation in the snow density response would allow scientists to identify regions that are most likely to be affected by rising temperatures and associated shifts in climate.
In addition to understanding the relationship between snow density and climate, we aim to improve simple and interpretable thermal models through functional smoothing (over the depth domain), while preserving inference on the physical model.

Most methods for estimating snow density build upon differential equation models for snow densification that account for thermal compaction \citep{herron_firn_1980, horhold_densification_2011,verjans2020bayesian}. Empirically, densification patterns change at various \textit{critical densities} or change points due to the interplay of particle rearrangement and plasticity change \citep{gow1975, herron_firn_1980, maeno1983, martinerie1992, salamatin2009}. To account for these snow density patterns, densification models often use three change points that separate four densification stages \citep[e.g.,][]{horhold_densification_2011}, where each stage is governed by a two-parameter temperature-dependent reaction rate. Together, the differential equation solution is piecewise linear (on a transformed scale) with four stages. Importantly, several studies hypothesize that the last three densification stages have similar thermal properties \citep[see][for discussion]{horhold_densification_2011}, and we investigate this claim for these data.
Snow density models of this type are generally extensions of the \cite{herron_firn_1980} model (which we call the HL model). This physical model has interpretable parameters, but, because the HL model is piecewise linear, it can fail to capture important snow density patterns (See Section \ref{sec:data}). Therefore, preserving inference on the physical model while expanding functional flexibility is our primary target.

The parameters of the HL model are assumed to be constant over spatial domains \citep{herron_firn_1980, horhold_densification_2011}; however, the physics related to surface snow density, thermal reaction rates, and critical densities depend on snow type, surface conditions, and many climatic factors which change spatially \citep{alley1982,johnson1998,freitag2004}. Therefore, surface snow densities, reaction rates, and critical densities are, in fact, not constant over space, something discussed by \cite{horhold_densification_2011} and in Section \ref{sec:data}.
Despite these findings, many current models rely on simple global fitting constraints averaged across all sites, often spanning both poles \citep{herron_firn_1980,helsen_elevation_2008, arthern_situ_2010, verjans2020bayesian}.
Capturing the spatial variation in these parameters (change points, reaction rate parameters, and surface snow density) would therefore improve the accuracy and general applicability of snow density models to more widespread regions. 


Our modeling approach focuses on addressing the HL model's shortcomings (piecewise linearity, global reaction rates, and fixed change points), while preserving inference on the HL model's parameters.
To allow the reaction rates and densification change points to vary over space, we adopt the spatially varying coefficient model \citep{gelfand2003} on transformed HL model parameters, including a surface density parameter, four two-parameter reaction rates, and three random change points. We call this model the spatially varying snow density model (SVSD). Estimating mechanistic or differential equation parameters statistically is well studied in spatial and spatiotemporal statistics (\cite{wikle2010general}; \cite{lindgren2011explicit}; and \cite{cressie2015statistics}, Chapter 6), and our approach follows that of \cite{wikle2010general} closely. From an applied perspective, spatial smoothing is particularly important for the HL model (and other thermal models) because site-specific temperature-dependent rate parameters are only identifiable through hierarchical spatial smoothing.

To account for deviations from piecewise linearity, a weakness of the physical model, we add a flexible, spatially-correlated functional component to the SVSD model. As is well discussed in the spatial confounding literature \citep[see, e.g.,][]{griffith2003spatial, reich2006effects, tiefelsdorf2007semiparametric, hodges2010, paciorek2010importance, hughes2013dimension, murakami2015random}, flexible random effects can interfere with inference on other interpretable model terms. However, in our application, we do not address spatial confounding; instead, we address the challenge of preserving inference on a functional model (the SVSD model) whose span varies spatially. Thus, we are addressing \emph{functional confounding} at each site because, due to spatially varying parameters, the model's span or column space is site-specific. We resolve functional confounding for the spatially varying model by orthogonalizing the smooth functional component with respect to the SVSD model at every site. This approach ensures that the flexible functional smoothing lies in the orthogonal column (null) space of the SVSD model, providing increased flexibility while preserving inference on the parameters of the SVSD model. 

We continue by discussing the snow densification processes and our snow density dataset in Section \ref{sec:data}. In this section, we also fit the \cite{herron_firn_1980} model without adaptations to motivate the need for spatial and functional smoothing for these data \citep[see, e.g.,][for examples of more recent use]{horhold_densification_2011, verjans2020bayesian}. In Section \ref{sec:arrhenius}, we present our modeling framework based on the physics and data attributes discussed in Section \ref{sec:data} that incorporates spatial and functional smoothing to an otherwise piecewise linear global model. In Section \ref{sec:models}, we detail our model comparison approach (Section \ref{sec:mod_comp}), our final model (Section \ref{sec:final_mod}), as well as details on our prior distributions, model fitting, and prediction (Section \ref{sec:model_fit}). Based on our final model, we present results in Section \ref{sec:res}. The data and code that support the findings of this study are openly available at \url{https://github.com/philawhite/Piecewise_linear_smoothing}.

\section{Snow Densification Processes and Data}\label{sec:data}

\subsection{The Physics of Snow Densification}\label{sec:snow}

Snow densification is a thermal reaction with a temperature-dependent reaction rate called an \emph{Arrhenius} constant, $$k= A e^{-\frac{E}{RT}},$$
where $T$ is the temperature (in Kelvin), $R \approx 8.314 \ \text{J} \ \text{K}^{-1} \ \text{mol}^{-1}$ is the ideal gas constant, $E > 0$ is the energy of activation, and $A > 0$ is a pre-exponentiation factor. The energy of activation $E$ determines the sensitivity of the thermal reaction rate to temperature. Smaller $E$ corresponds to higher temperature sensitivity in the reaction rate, while larger $E$ is connected with lower temperature sensitivity.

A snow core extends from the surface of the ice sheet to some depth (2 to 140 meters in this dataset) beneath the surface, and density measurements are then taken along the full depth profile of the core. For a single snow core with many snow density estimates, the Arrhenius constant $k$ is statistically identifiable, but parameters $A$ and $E$ that define $k$ are not. Therefore, many cores with different temperatures are needed to identify $A$ and $E$, a process that could be carried out regionally to account for geographic differences. However, such regional core groupings are arbitrary. Snow densification models also rely on multiyear surface mass balance (SMB), the average rate of snow accumulation, in m w.e./yr (meters water equivalent per year) because increased snowfall increases snow compaction rates. 

Densification rates change at various depth-dependent \textit{critical densities} due to the interplay of particle rearrangement and plasticity change \citep{gow1975, herron_firn_1980, maeno1983, martinerie1992, salamatin2009}. Densification models often use three critical densities \citep{horhold_densification_2011}, giving a piecewise linear model with four stages, each with a unique temperature-dependent slope corresponding to the densification rate over depth. In this manuscript, we focus on extensions of the interpretable HL model \citep{herron_firn_1980}, a highly cited and commonly used snow density model \citep[see][as examples of recent use]{horhold_densification_2011,verjans2020bayesian}. Like many other models, the HL model can be written as a linear combination of piecewise linear basis functions on a transformed scale. From these data, we wish to estimate densification rates over depth, infer change points in density where densification rates change, and estimate temperature-dependent physical subcomponents of the densification rate (discussed in Section \ref{sec:arrhenius}).

The HL model is defined in stages, where the behavior of the model changes at critical densities where various physical forces alter the pattern of densification \citep[see][for review and discussion]{horhold_densification_2011}. The HL model defines rates of densification using Arrhenius constants for two stages: for densities less than $ 0.55$ g/cm$^3$ and between $ 0.55$ g/cm$^3$ and $0.80$ g/cm$^3$. The model is
\begin{equation}\label{eq:herron}
\begin{aligned}
 &\rho_I k_1 x + \log\left( \frac{y(0)}{ \rho_I - y(0)}  \right),   &\text{for } x \in [0,\kappa_1),  \\
  &\frac{\rho_I k_2 (x - \kappa_1) }{\sqrt{SMB}}+ \log\left( \frac{y(\kappa_1)}{ \rho_I - y(\kappa_1)}  \right),    & \text{for } x \in [\kappa_1,\kappa_2), 
\end{aligned}
\end{equation}
where $k_1 = 11 \exp\left\lbrace -10160/RT\right\rbrace$, $k_2 = 575 \exp\left\lbrace -21400/RT\right\rbrace$, and $T$ is the temperature 10-m below the surface. Depths $\kappa_1$ and $\kappa_2$, corresponding to the densities $0.55$ and $0.80$ g/cm$^3$, can be extracted from \eqref{eq:herron} as
\begin{equation}\label{eq:change_depths}
\begin{aligned}
\kappa_1 &= \frac{1}{ \rho_I k_1} \left[ \log\left( \frac{0.55}{\rho_I - 0.55} \right) - \log\left( \frac{y(0)}{\rho_I - y(0)} \right) \right],  \\
\kappa_2 &= \kappa_1 + \frac{\sqrt{SMB}}{ \rho_I k_2} \left[ \log\left( \frac{0.80}{\rho_I - 0.80} \right) - \log\left( \frac{0.55}{\rho_I - 0.55} \right) \right].
\end{aligned}
\end{equation}

When comparing the HL model to field data, \cite{horhold_densification_2011} find good agreement to data at some sites and poor agreement at other sites. Fitting site-specific HL models, we observe similar issues for our data. Moreover, we find spatial patterns in the surface densities and the Arrhenius constants, as we discuss in Section \ref{sec:ourdata}. For these reasons, we propose spatially varying extensions of \eqref{eq:herron} in Section \ref{sec:arrhenius}. 

More recent models for deeper snow cores often assume that there are three critical densities at $ 0.55$, $0.73$, and between $0.82-0.84$ g/cm$^3$  \citep[see][for review and discussion]{horhold_densification_2011}, which we call $\rho_{1}$, $\rho_{2}$, and $\rho_{3}$. For each critical density, $\rho_{1}$, $\rho_{2}$, and $\rho_{3}$, there is a corresponding critical depth $\kappa_{1}$, $\kappa_{2}$, and $\kappa_{3}$. Using high-resolution density measurements, \cite{horhold_densification_2011} find that the first critical density $\rho_{1}$ ranged between 0.45 and 0.60 g/cm$^3$ depending on the site, rather than always occurring at 0.55 g/cm$^3$. Moreover, \cite{horhold_densification_2011} argue that the density transitions at critical densities $\rho_{2}$ and $\rho_{3}$ are only ``weakly apparent.''

Rather than a two-stage model, we use a four-stage model that incorporates more recent extensions of the HL model with three random critical densities $\rho_{1}$, $\rho_{2}$, and $\rho_{3}$. As discussed, because the transitions at $\rho_{2}$ and $\rho_{3}$ are weak, we assume that the hierarchical mean parameters for the second, third, and fourth stages are equal, but we do not constrain the site-specific parameters to be equal. We connect the physical and statistical models by writing the four-stage version of \eqref{eq:herron} as a generalized piecewise linear multiple regression model. Let $\alpha$ be an intercept, $\bz(x)$ be depth-specific covariates with corresponding Arrhenius constants $\bk$, and $\bone(\cdot)$ be the indicator function. The model can be written as 
\begin{equation}\label{eq:linear_mod}
\begin{aligned}
&\log\left( \frac{y(x)}{ \rho_I - y(x)} \right) = \alpha + \bz(x)^T \bk,\\
&z_1(x) = \rho_I\min(x,\kappa_1), \\
&z_2(x) = \frac{\rho_I\min(x - \kappa_1,\kappa_2- \kappa_1)}{\sqrt{SMB}} \bone\left( x > \kappa_1 \right),
\end{aligned}
\qquad
\begin{aligned}
&\alpha = \log\left( \frac{y(0)}{ \rho_I - y(0)}  \right) \\
&z_3(x) = \frac{\rho_I \min(x - \kappa_2,\kappa_3- \kappa_2)}{\sqrt{SMB}} \bone\left( x > \kappa_2 \right), \\
&z_4(x) = \frac{\rho_I (x - \kappa_3)}{\sqrt{SMB}} \bone\left( x > \kappa_3 \right).
\end{aligned}
\end{equation} 

Although frequently assumed to be constant over spatial domains \citep{herron_firn_1980, horhold_densification_2011,verjans2020bayesian}, the physics related to thermal reaction rates and critical densities depend on snow type and conditions which change spatially \citep{alley1982,johnson1998,freitag2004}. Therefore, Arrhenius constants and critical densities are not constant over space, something discussed by \cite{horhold_densification_2011} and in Section \ref{sec:ourdata}.
Despite these findings, many current models rely on simple global fitting constraints averaged across all sites, often spanning both poles \citep{herron_firn_1980,helsen_elevation_2008, arthern_situ_2010, verjans2020bayesian}.
Capturing the spatial variation in these parameters would therefore improve the accuracy and general applicability of snow density models to more widespread regions.
In Section \ref{sec:final_mod}, we present the hierarchical spatial model corresponding to \eqref{eq:linear_mod}.

\subsection{Data}\label{sec:ourdata}

This dataset consists of $N = 14,844$ density measurements from 57 snow/ice cores at $n_s = 56$ unique locations. These data come from four different field campaigns: the East Antarctic Plateau \citep{albert_extreme_2004}, the Siple Dome project \citep{lamorey_waiscores}, the Satellite Era Accumulation Traverse \citep{burgener_observed_2013}, and the US portion of the International Trans-Antarctic Scientific Expedition \citep{mayewski_international_2005}. We refer to these field campaigns as EAP, SDM, SEAT, and US-ITASE, respectively. 

We denote density measurements as $y(\bs_{i},x) > 0$, where we index cores by $i$, the core's location as $\bs_{i}$, and depth with $x$. We let $\mathcal{S}$ denote the collection of the 56 unique core sites (there is one site with two cores; i.e., $\bs_i = \bs_{i'}$). Because we use our model to estimate parameters where we do not have data, we let $\bs$ represent a generic spatial location. To specify campaign-specific parameters, we define $m_i$ to be an indicator for which field campaign acquired the $i$th core. 

The precise methods and techniques of density measurement are generally similar but differ somewhat across campaigns. Density measurements involve measuring the mass and volume of sections of a core. Density measurements are averaged over some length of the snow core; therefore, density variability relies not only on the mass and volume measurements themselves but also on the length of the core section used. We let $dx_i$ be the length of the core used to procure a density measurement for core $i$ (i.e., the maximum depth of that core divided by the number of density measurements). Due to increased averaging, we expect smoother snow density curves with longer $dx_i$. The core lengths for the EAP and SEAT campaigns are internally consistent in that, if $m_i = m_{i'}$, $dx_i = dx_{i'}$. The SDM and US-ITASE expeditions do not use the same core lengths $dx_i$ to measure density for each core, an issue we address through a heteroscedastic error model.
Most density measurements were taken in the field, which, unsurprisingly, poses several challenges. Many of the cores display anomalous density measurements that are likely erroneous, motivating the use of error distributions that down-weight these anomalies. Altogether, these data attributes motivate using an error model with heavy tails, random effects to account for research group differences, and weighting that accounts for differences in the core length used to measure snow density.

\begin{figure}[H]

        \vspace{-8mm}

\begin{center}
\includegraphics[width=0.38\textwidth]{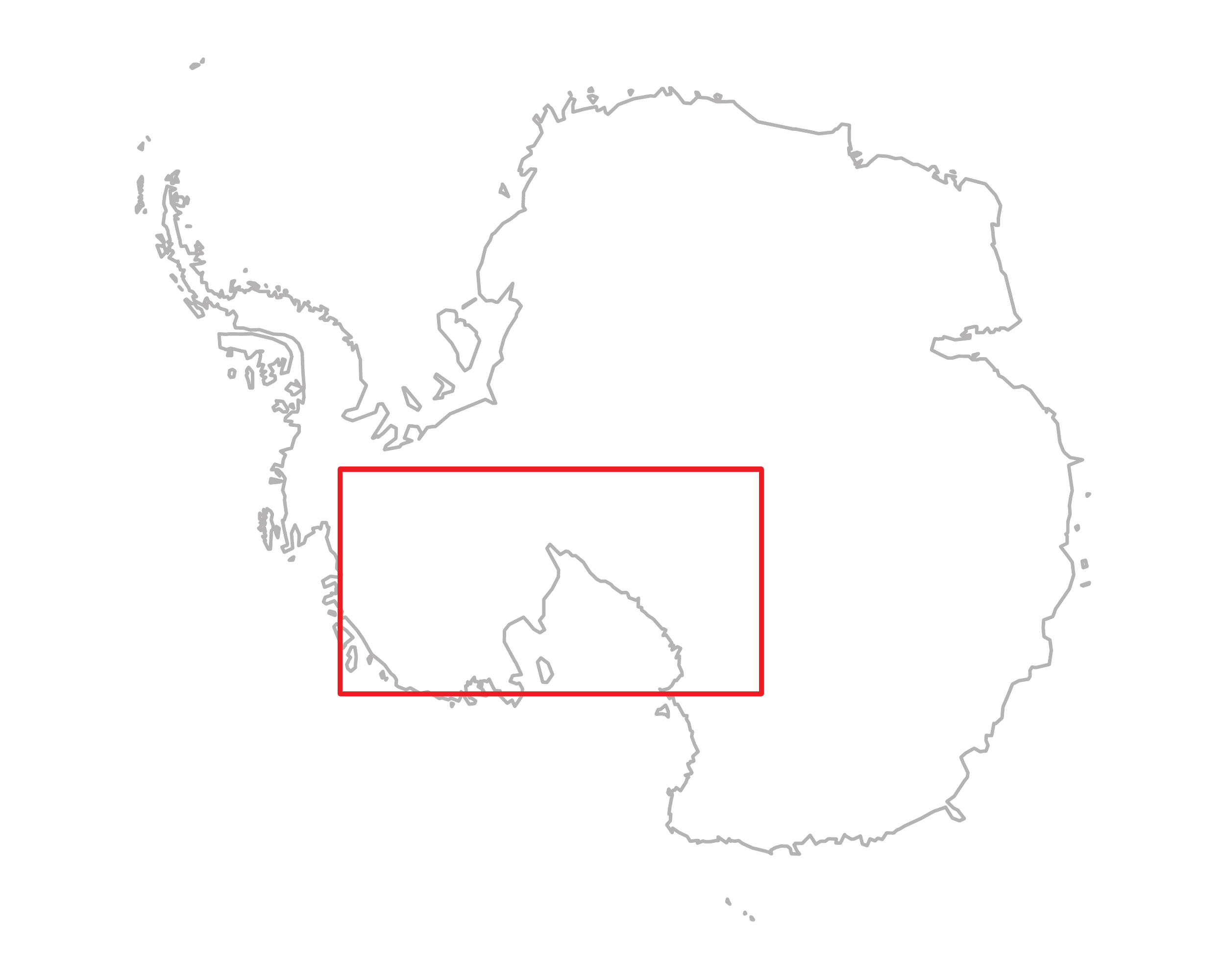}
\includegraphics[width=0.38\textwidth]{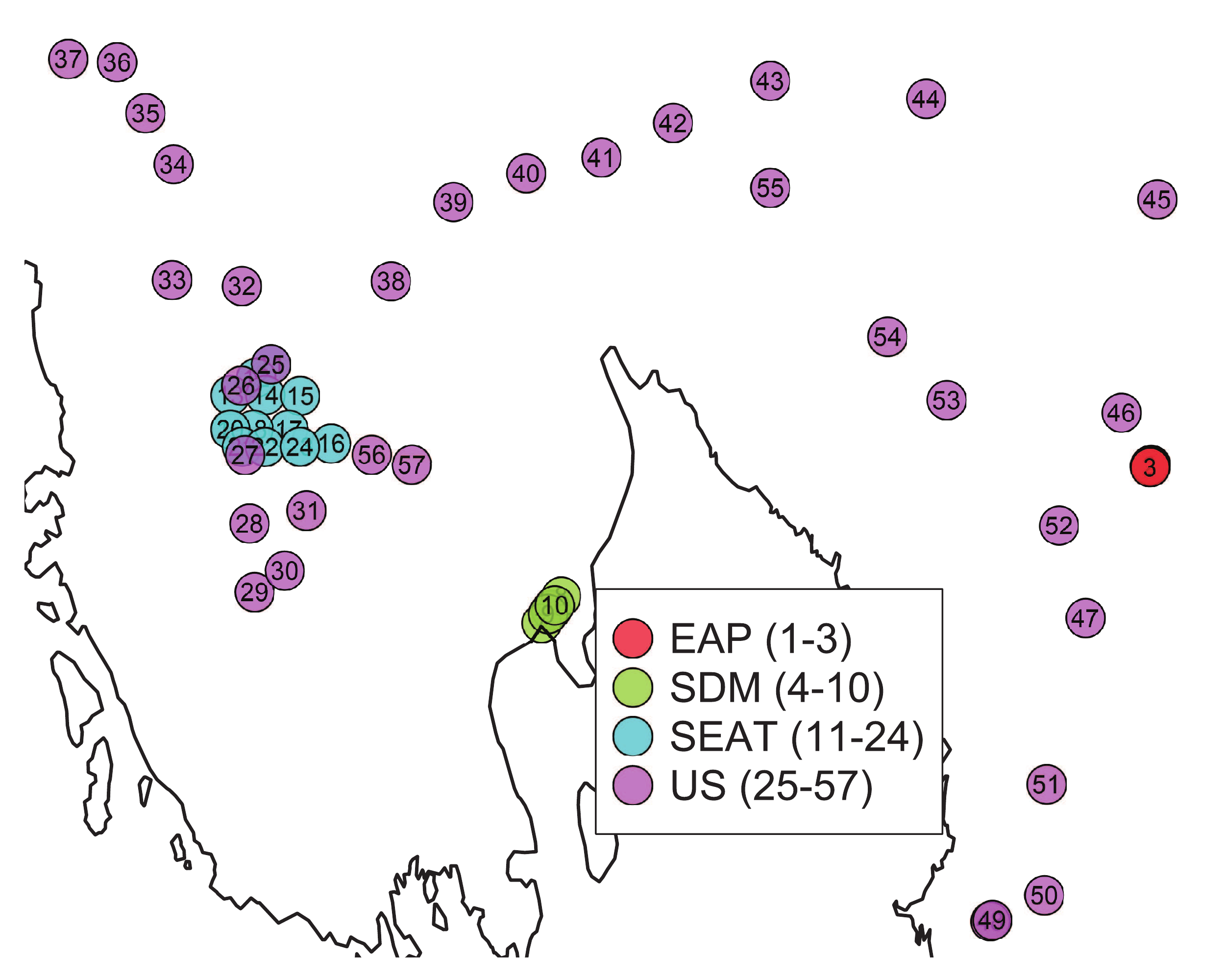}
\includegraphics[width=0.45\textwidth]{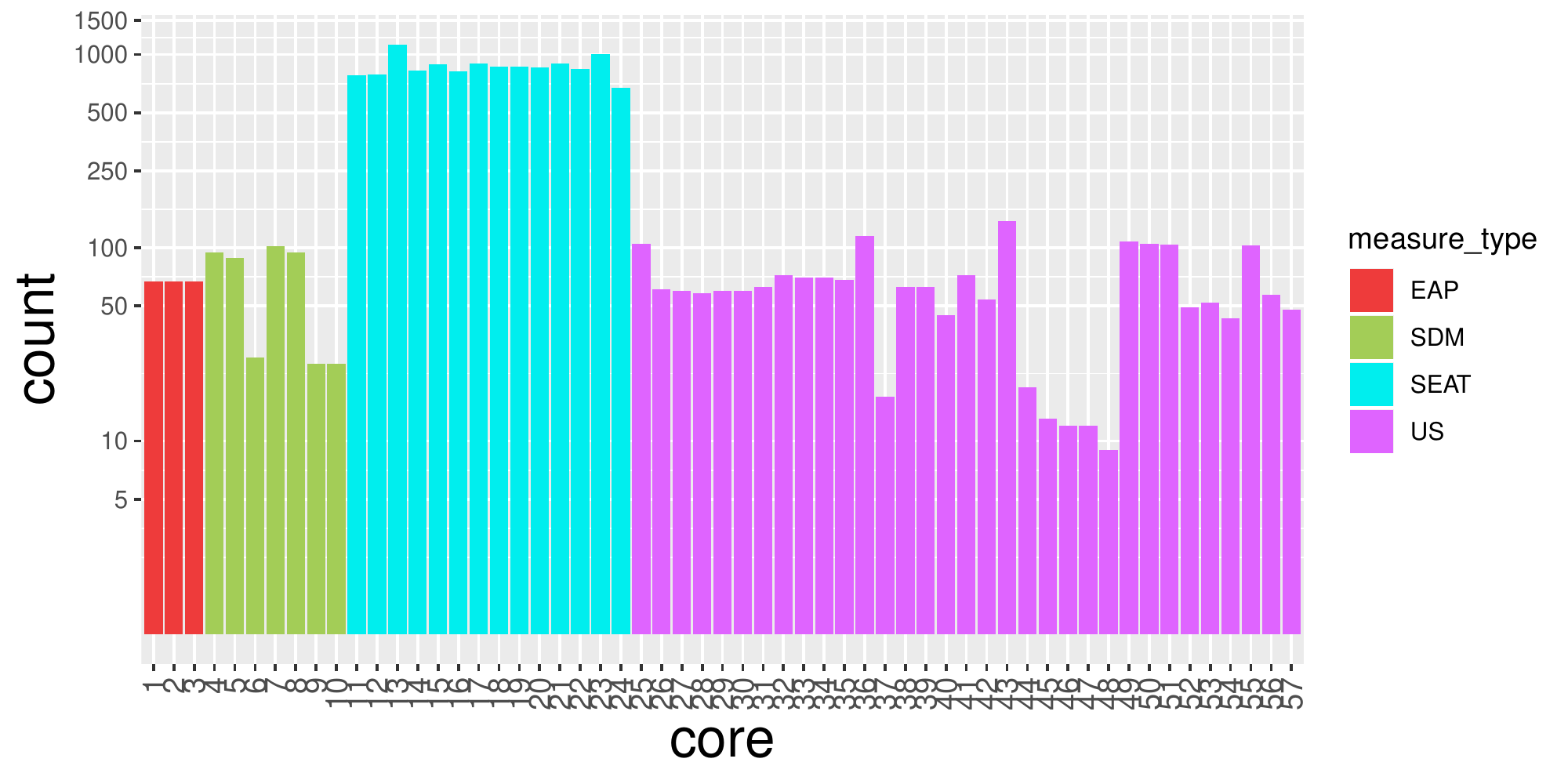}
\includegraphics[width=0.45\textwidth]{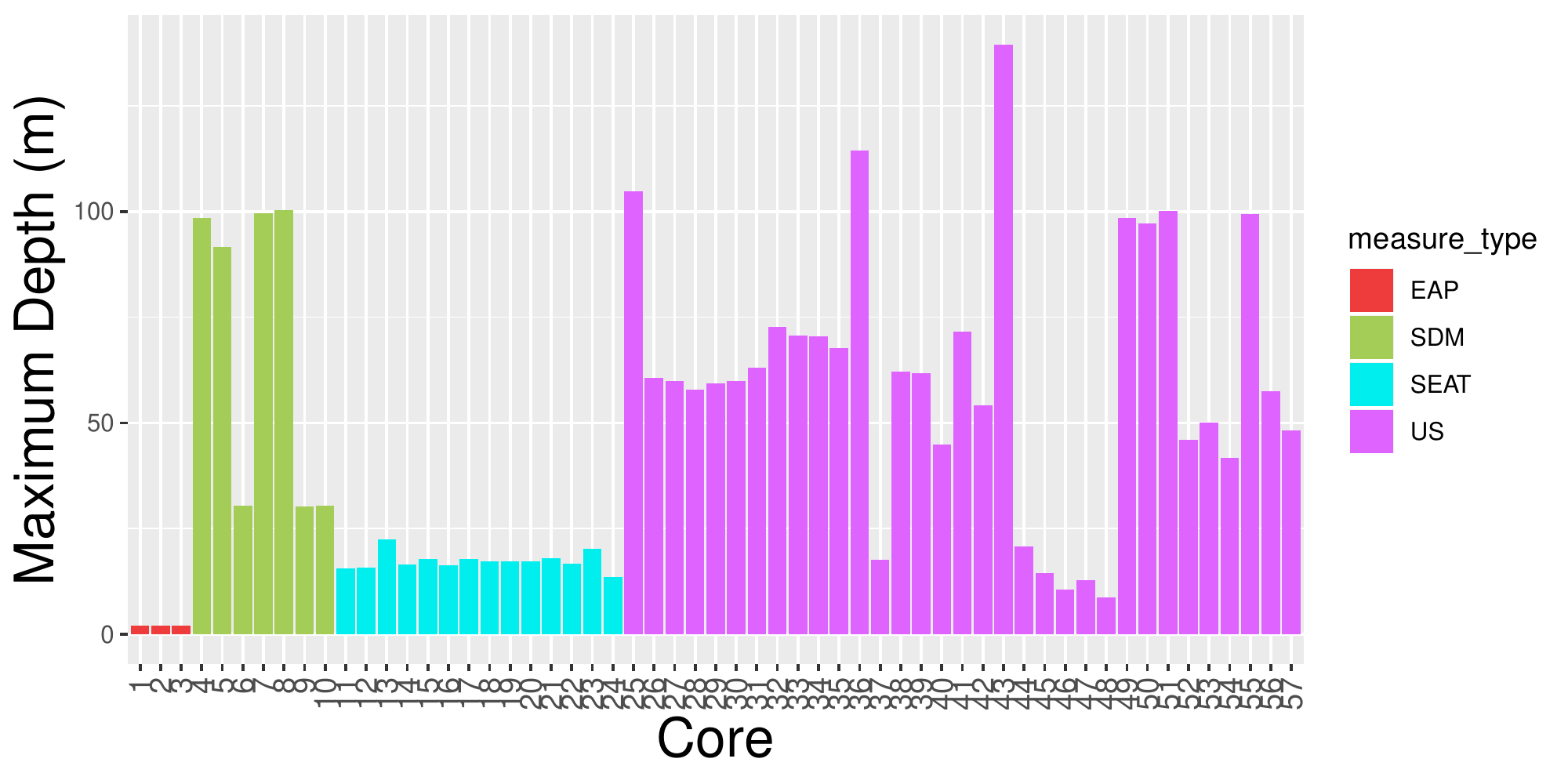}
\includegraphics[width=0.24\textwidth]{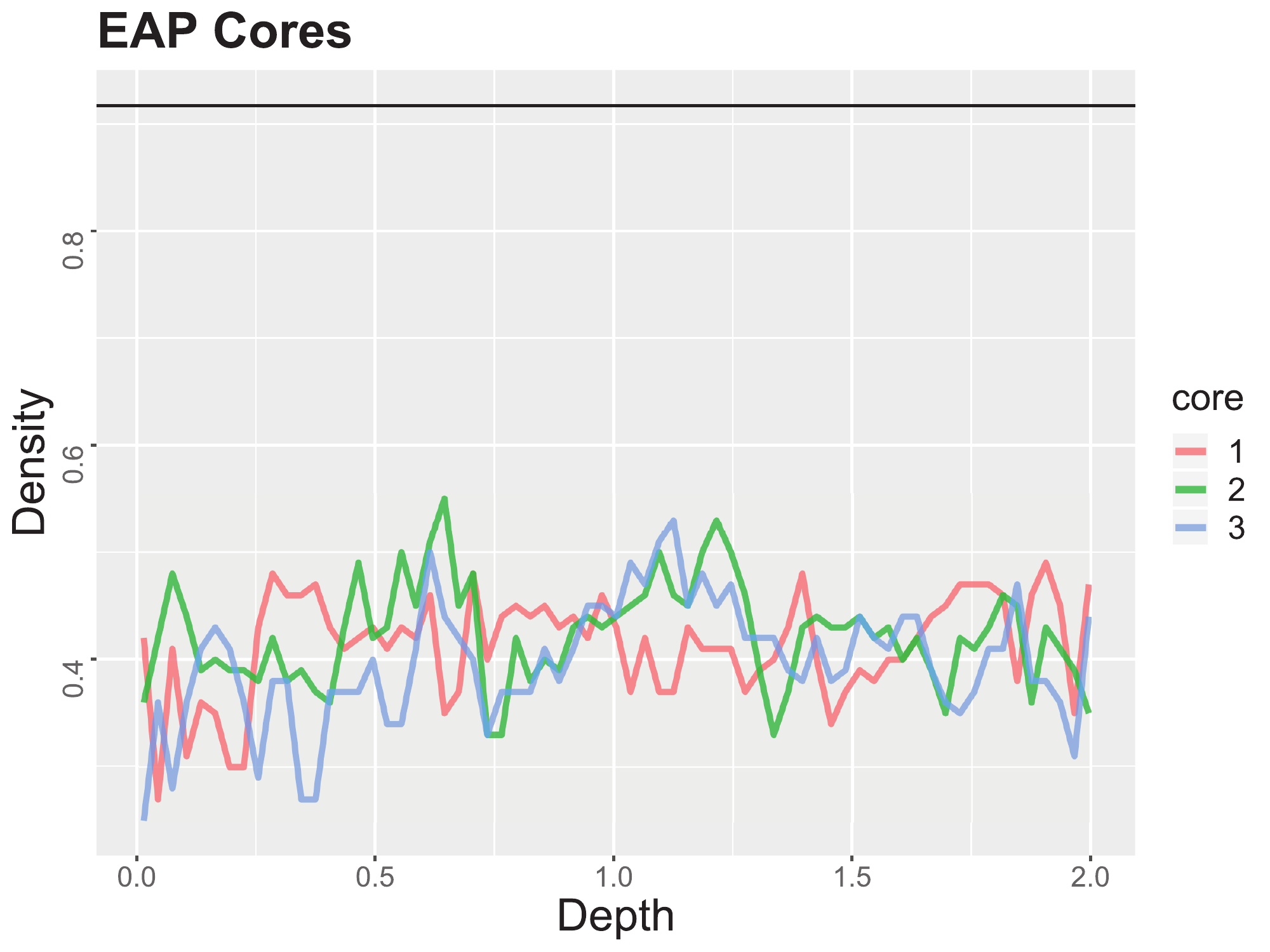}
\includegraphics[width=0.24\textwidth]{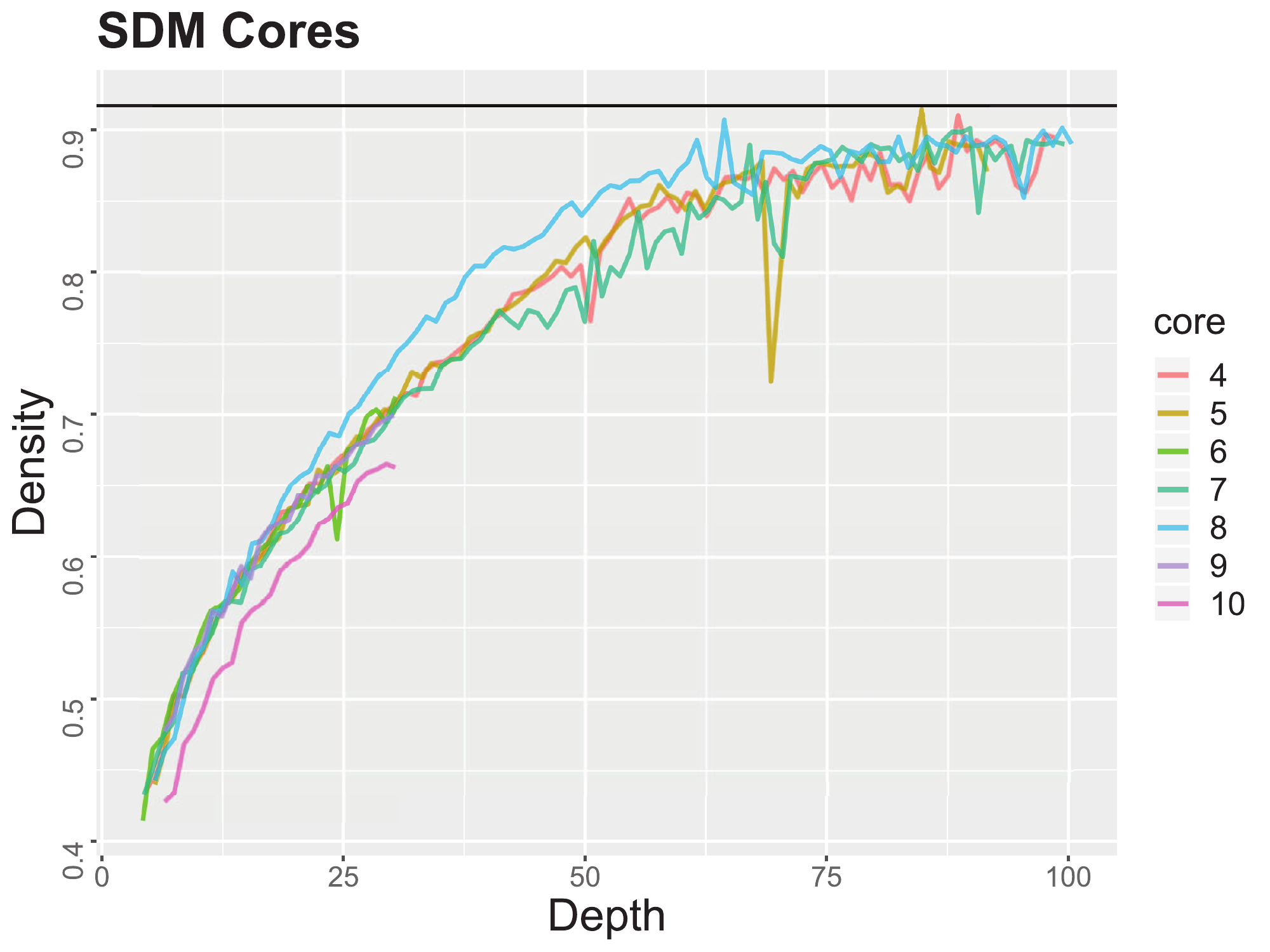}
\includegraphics[width=0.24\textwidth]{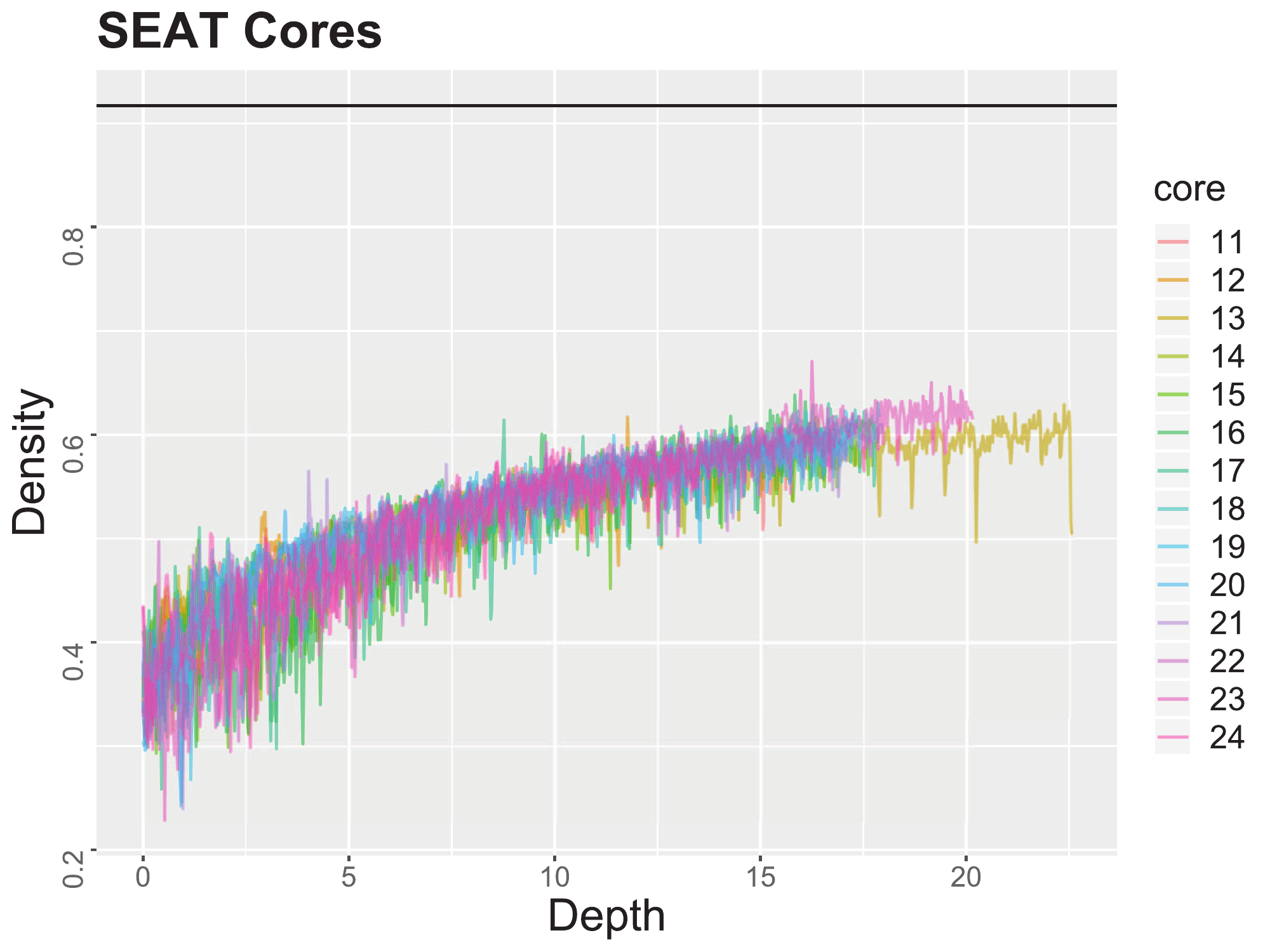}
\includegraphics[width=0.24\textwidth]{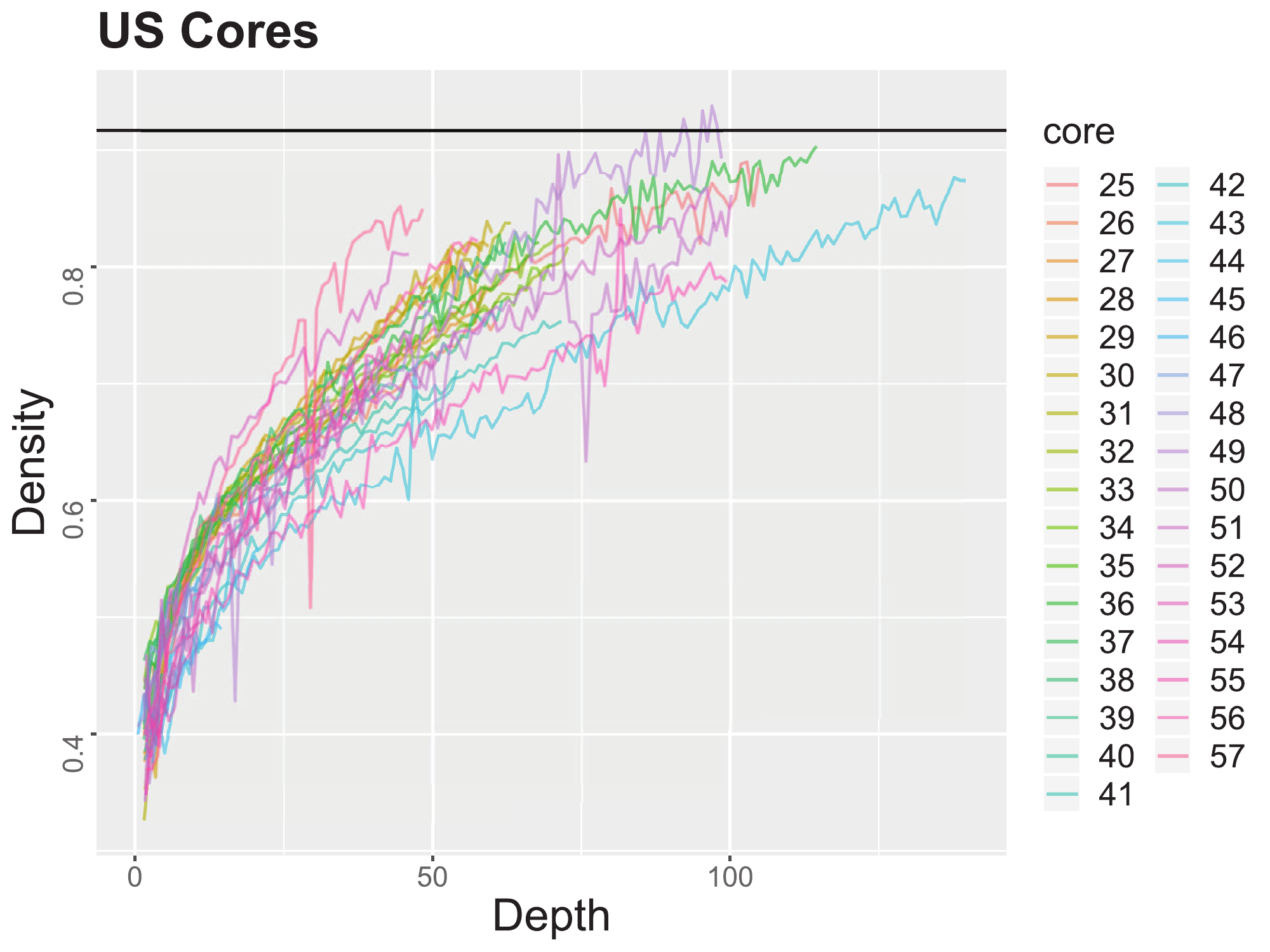}
\end{center}
        \vspace{-5mm}
\caption{(Top-Left) Region of Antarctica where snow/ice cores are located. (Top-Right) Location of core sites with colors indicating measurement type. (Middle-Left) The number of density measurements at each core, and (Middle-Right) Maximum depth in m obtained by each core. (Bottom) Density measurements over depth by core, grouped by expedition. Note that the depth scales of these measurement types differ greatly. The horizontal line indicates the density of solid ice.}\label{fig:locs}

\end{figure}

In Figure \ref{fig:locs}, we plot the locations of the cores, the number of density measurements from each core, maximum depth of the core, and the observed density measurements as a function of depth for each site. Although snow density cannot exceed $\rho_I$, some measurements violate this limit, suggesting that our model for $y(\bs_{i},x)$ must permit values greater than $\rho_I$, while a reasonable model for the mean function should constrain it between $(0,\rho_I]$.

\begin{figure}[H]

\begin{center}

\includegraphics[width=0.24\textwidth]{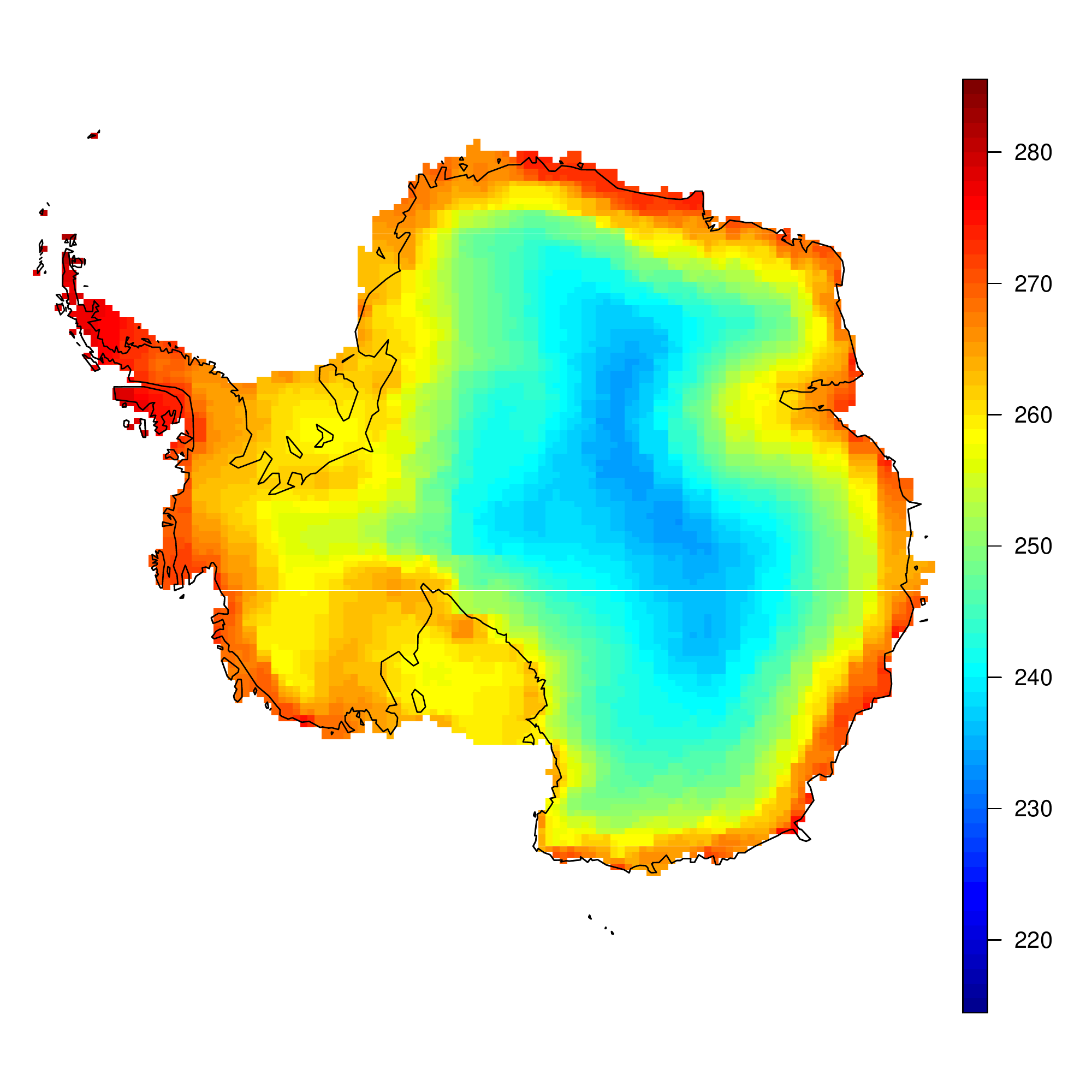}
\includegraphics[width=0.24\textwidth]{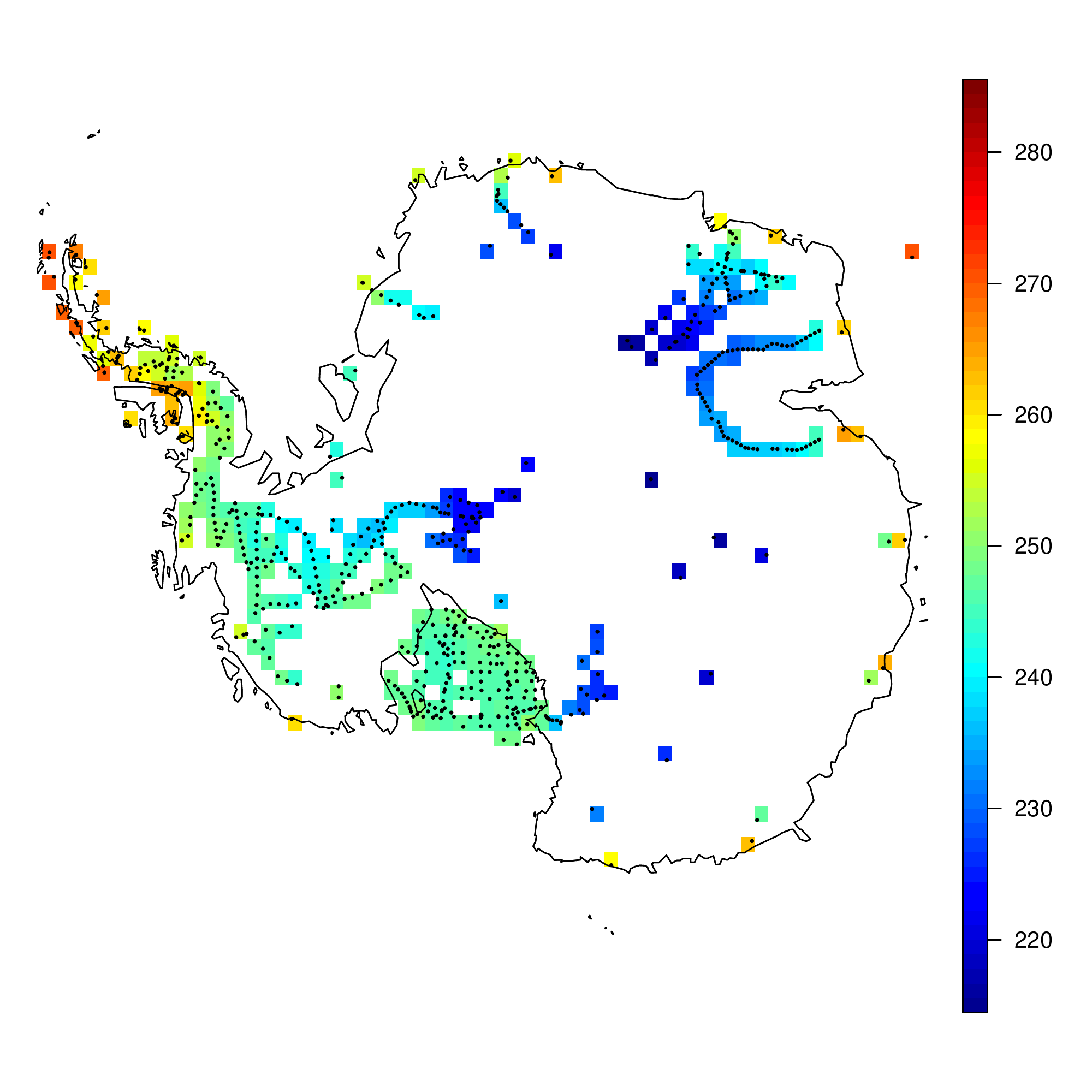}
\includegraphics[width=0.24\textwidth]{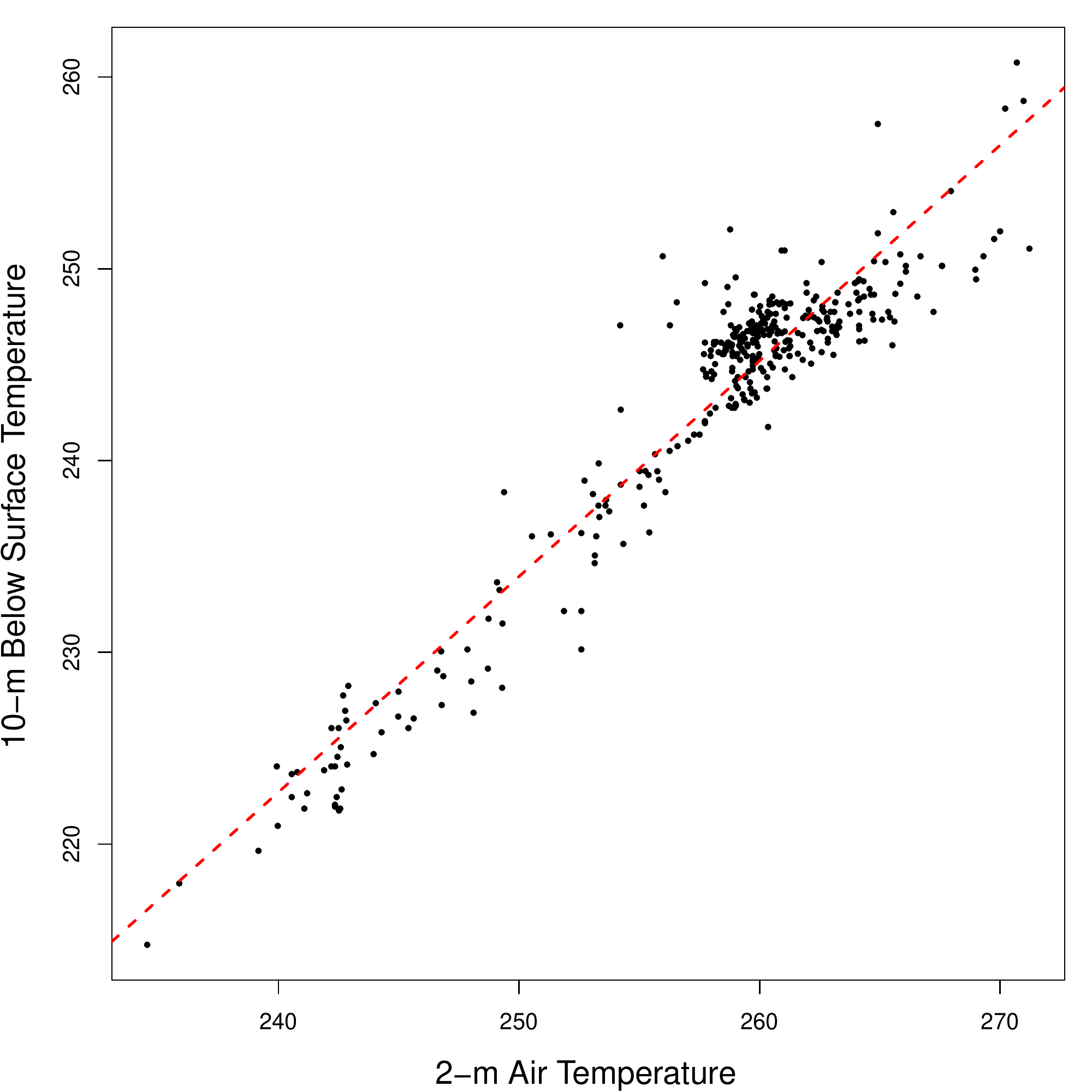}
\includegraphics[width=0.24\textwidth,height = 0.255\textwidth]{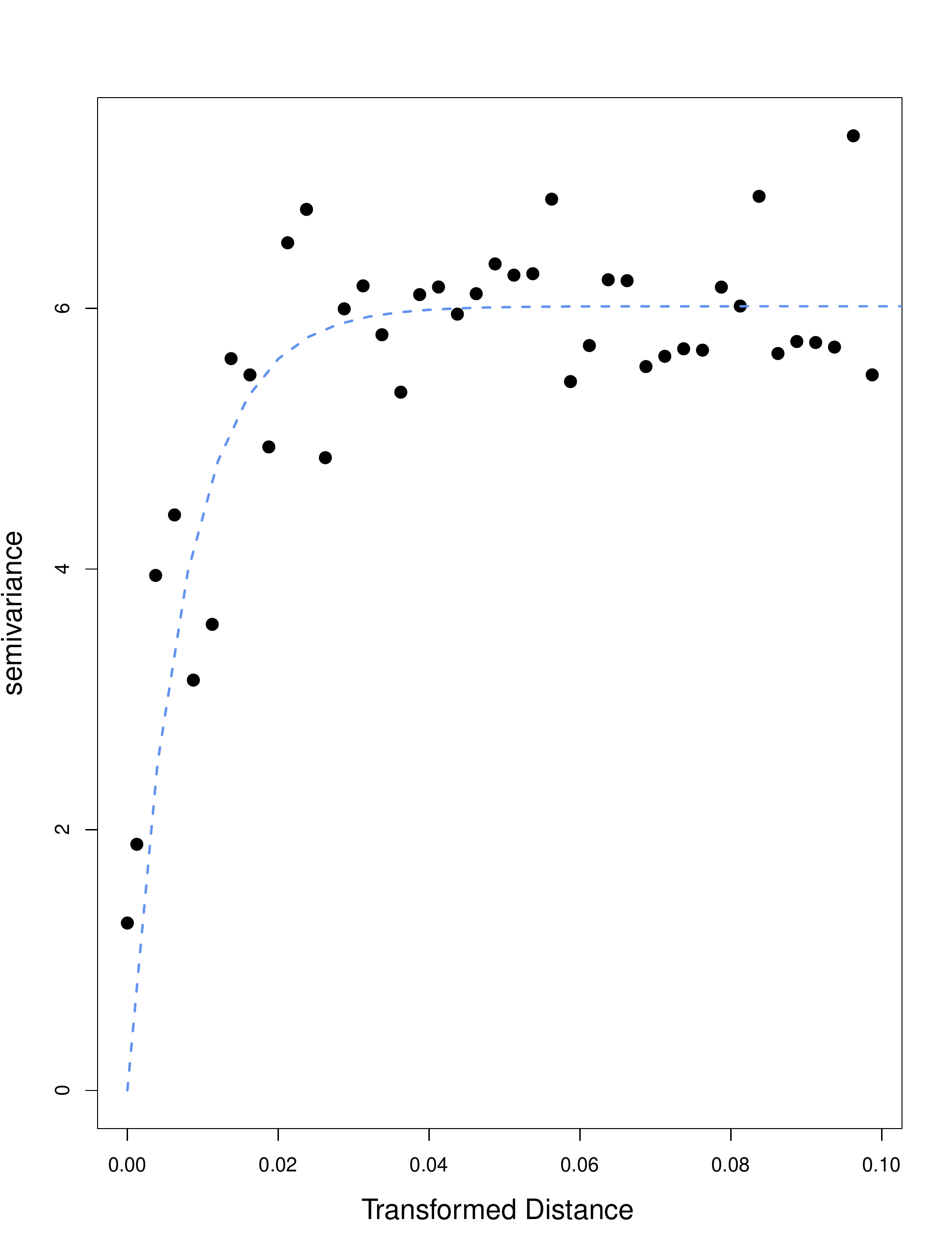}
\includegraphics[width=0.4\textwidth]{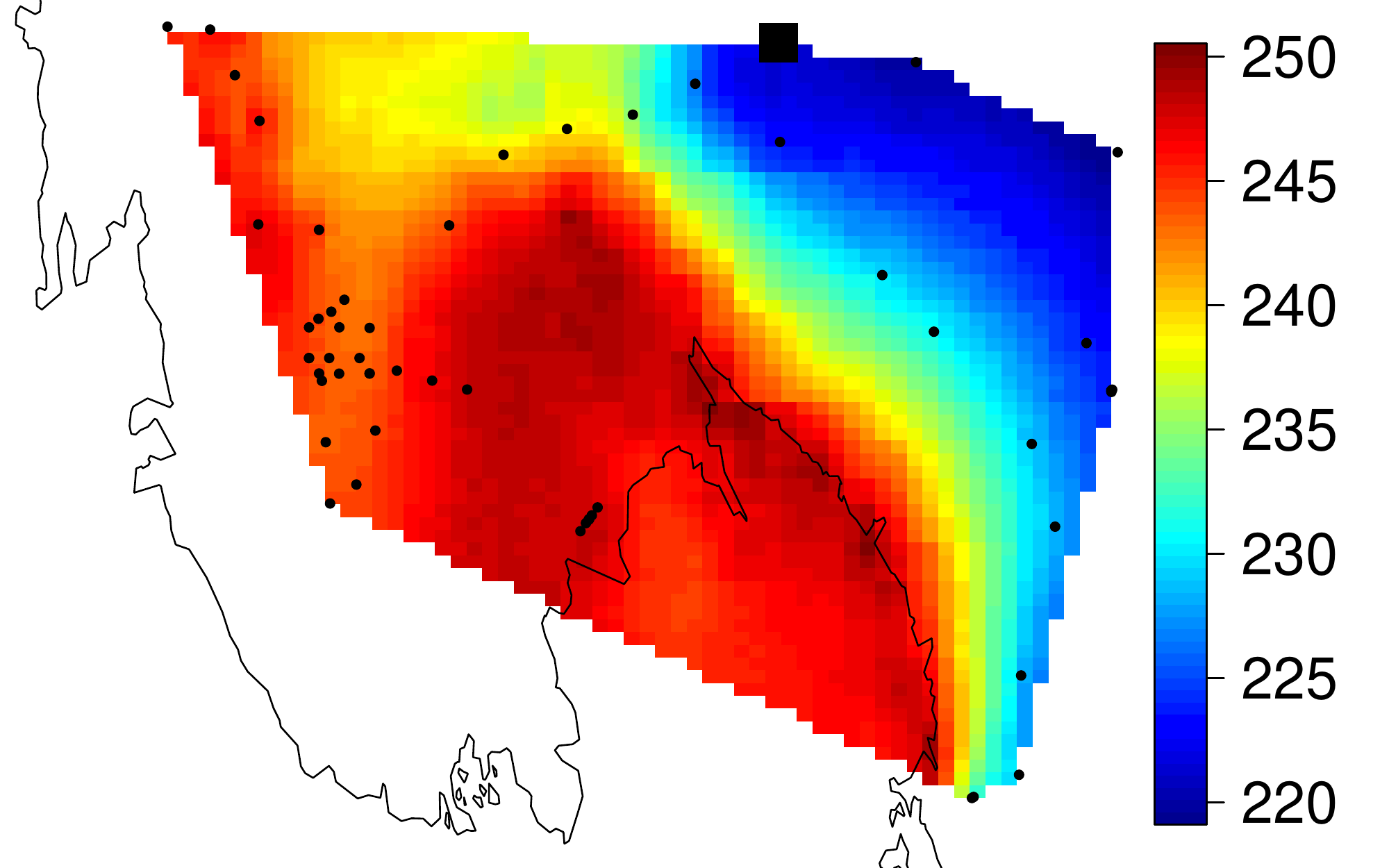}
\includegraphics[width=0.4\textwidth]{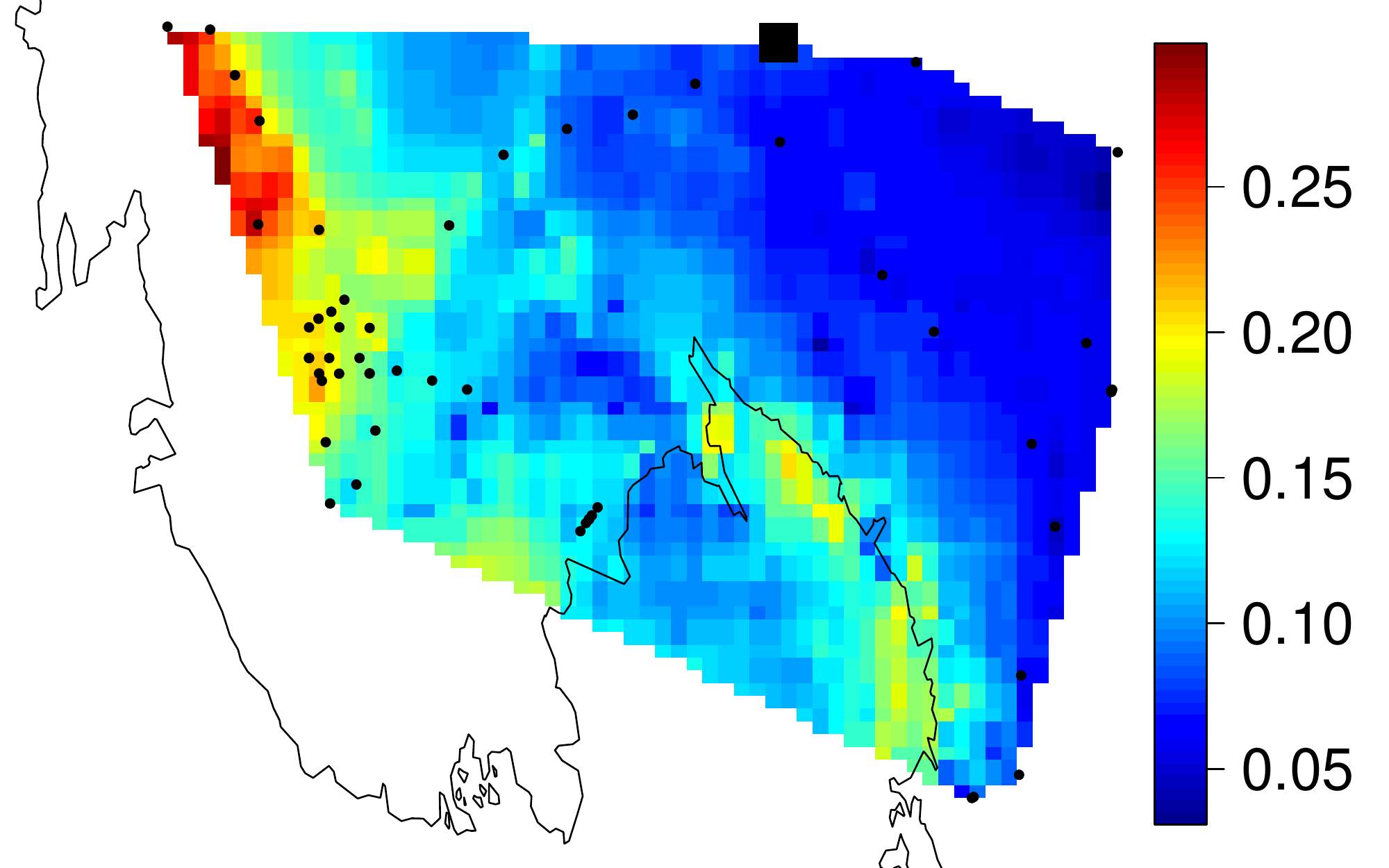}

\end{center}

        \vspace{-3mm}

\caption{(Top-Left) Smoothed ERA-Interim 2-m air temperature averaged from 1979-2014. (Top-Center-Left) Temperature data 10-m below the surface. (Top-Center-Right) Relationship between nearest 2-m air temperature to observed temperature data 10-m below the surface. (Top-Right) Semivariogram for the residuals for temperature 10-m below the surface with a fit to an exponential semivariogram. The estimated (Bottom-Left) Temperature 10 m below the surface and (Bottom-Right) SMB, where the square shows the south pole. }\label{fig:temps}

\end{figure}

As discussed, the HL model uses surface mass balance (SMB$(\bs)$) and the average temperature 10-m below the surface ($T(\bs)$). We have access to a dataset of temperature 10-m below the surface \citep{bohlander2001}; however, these data do not extend to all core sites. To estimate temperature 10-m below the surface, we use ERA-Interim 2-m air temperature data averaged from 1979 to 2014 \citep{molteni1996}, which we call $T_{(2)}(\bs)$ to distinguish it from $T(\bs)$. The ERA-Interim 2-m air temperature data are spatially complete but very coarse; therefore, we use a smoothed temperature surface \citep{white2019}. In Figure \ref{fig:temps}, we plot the 2-m air temperature, observed temperature 10-m below the surface, and a scatterplot of these temperatures with a least-square regression line.

The average 2-m air temperature is higher than the temperature below the surface, and the relationship between the two temperatures appears to be linear with some heteroscedasticity. The simple linear regression model has an $R^2 = 0.923$. Although there is relatively little variability in temperature 10-m below not explained by 2-m air temperature, there is also spatial autocorrelation in the residuals for temperature 10-m below the surface given 2-m air temperature. Thus, we use a simple spatially varying intercept model to estimate $T(\bs)$ at the 56 unique core locations and use its posterior mean for the HL model and our extensions. We use the posterior mean SMB from \cite{white2019} who use a spatial generalized linear model to interpolate $\text{SMB}(\bs)$. The posterior means of $T(\bs)$ and $\text{SMB}(\bs)$ are plotted in Figure \ref{fig:temps}. While we could model density, temperature, and SMB jointly, joint modeling poses significant computational challenges.

\begin{figure}[H]
\begin{center}
\includegraphics[width=0.24\textwidth]{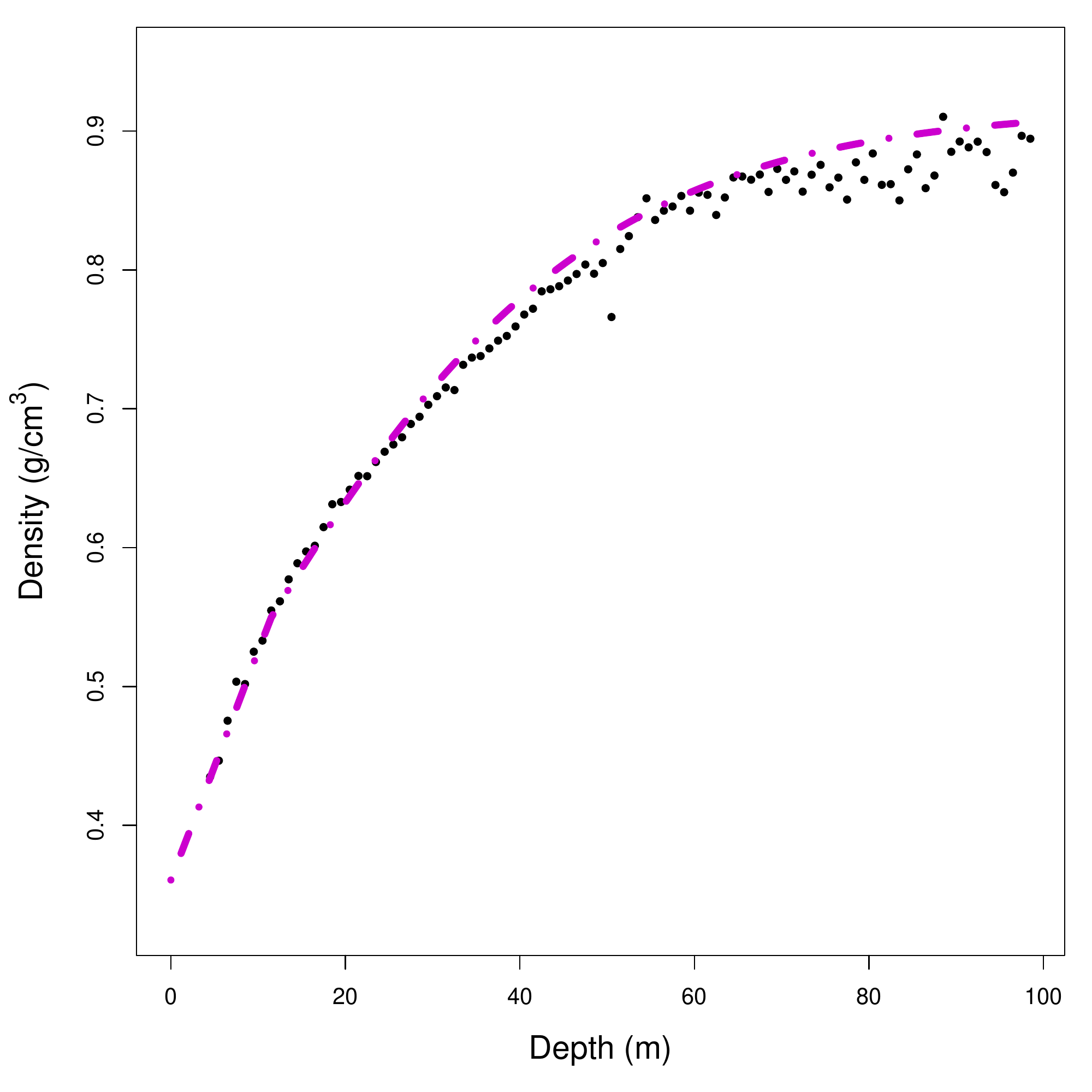}
\includegraphics[width=0.24\textwidth]{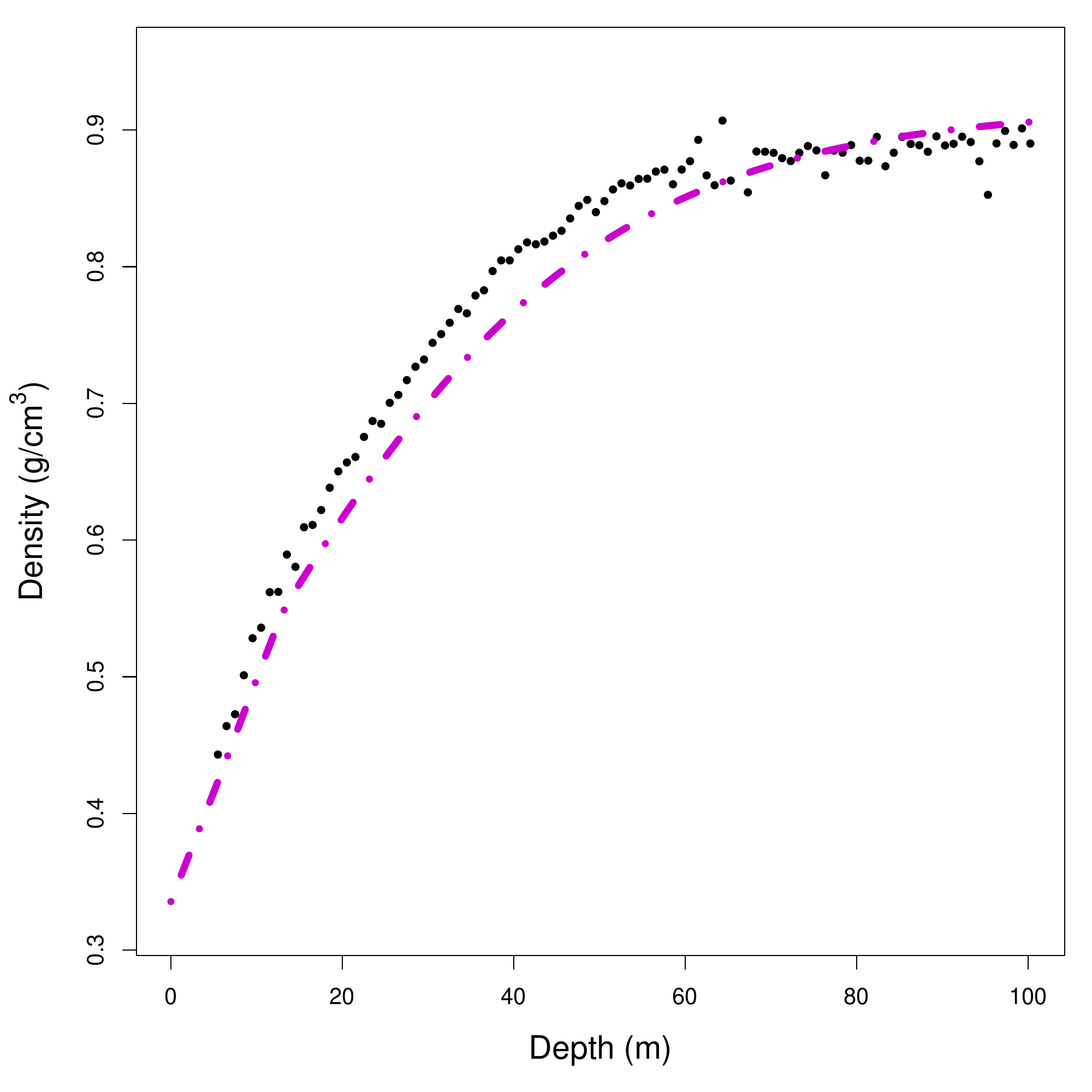}
\includegraphics[width=0.24\textwidth]{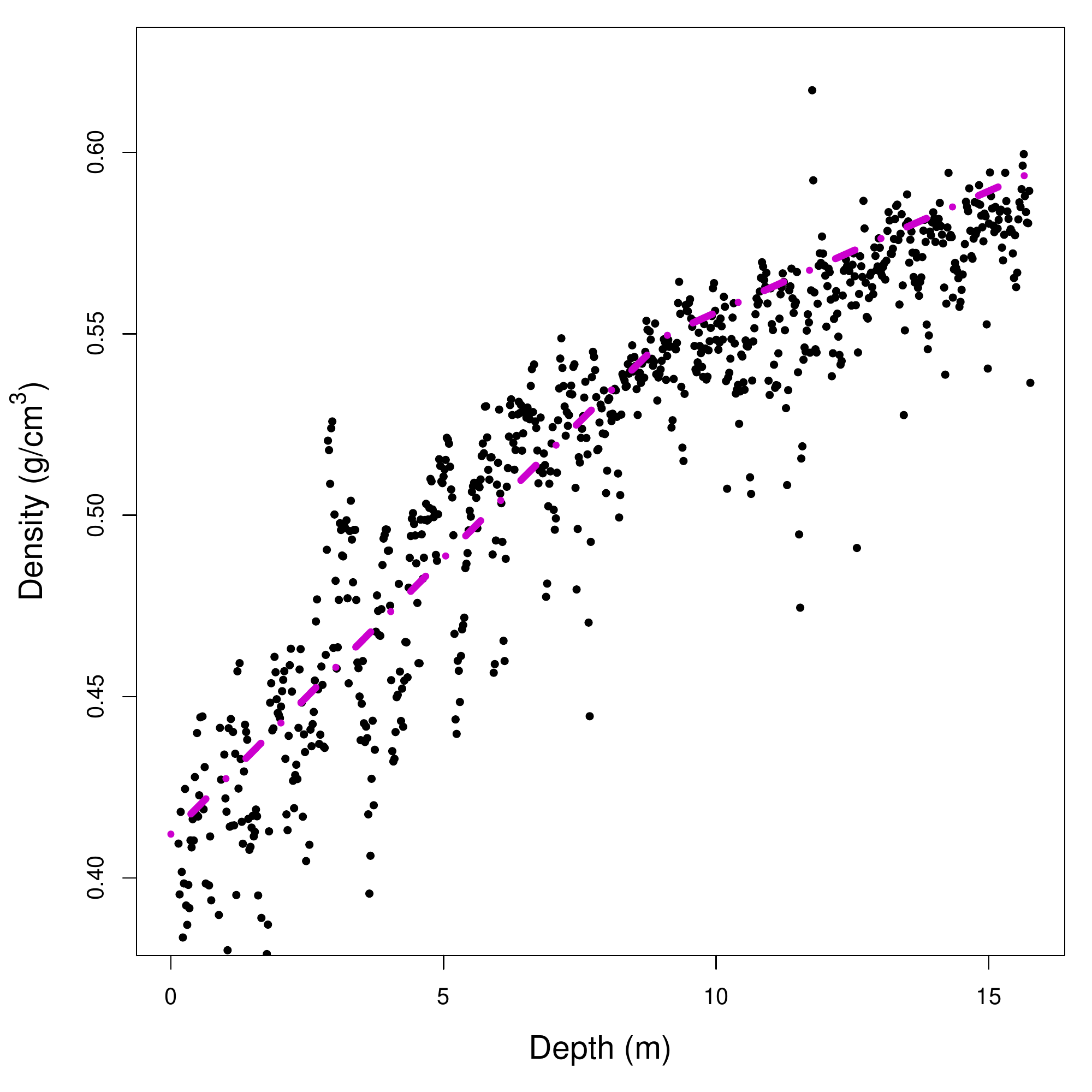}
\includegraphics[width=0.24\textwidth]{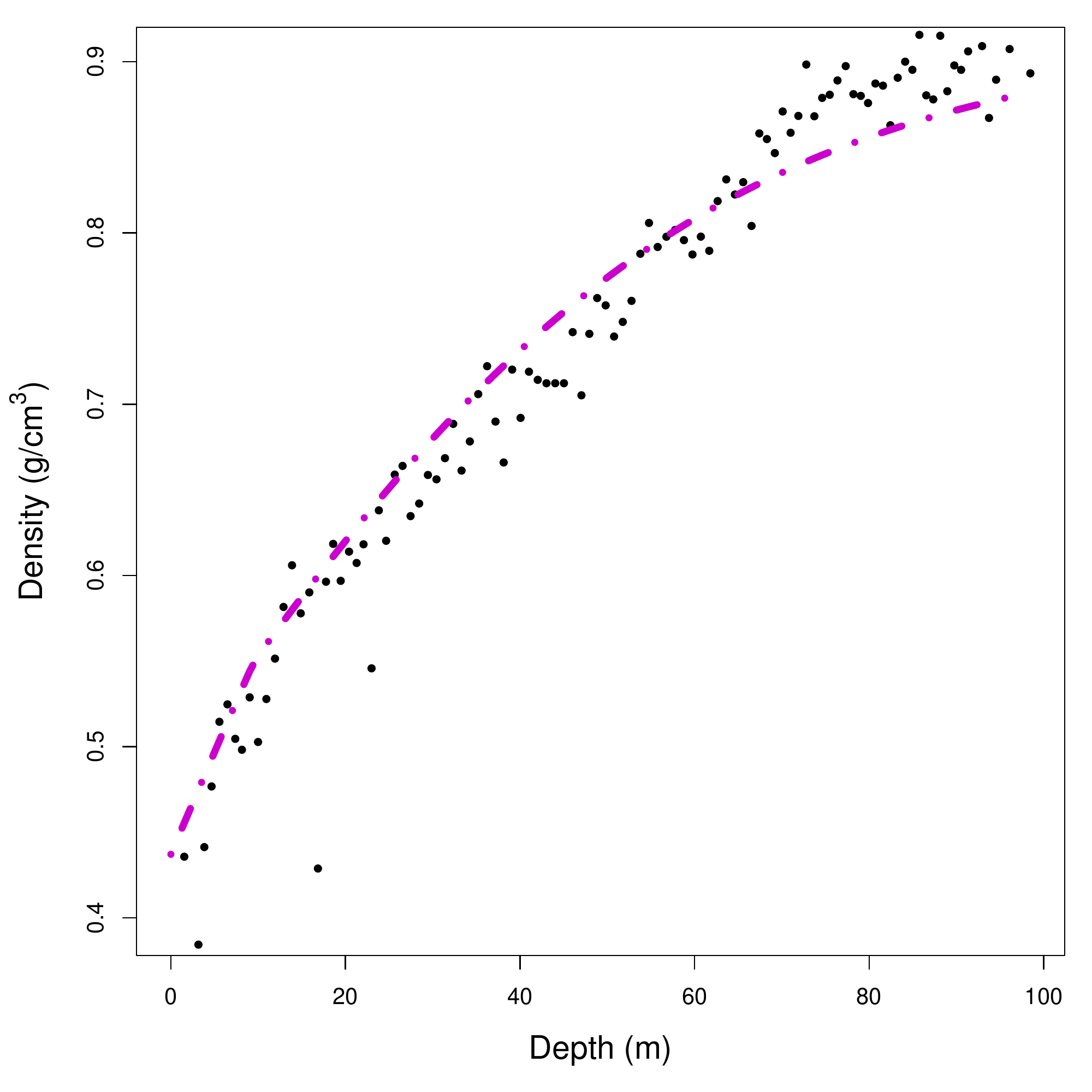}
\end{center}

        \vspace{-5mm}

\caption{HL model fit to core (Left) 4, (Middle Left) 8, (Middle Right) 12, (Right) 49.}\label{fig:herron_langway_fit}
\end{figure}

Using the posterior means for $T(\bs)$ and $\text{SMB}(\bs)$, plotted in Figure \ref{fig:temps}, we compare the HL model fit for each core to the observed field data. While the model matches the overall trend of density in general, it fails to accurately estimate snow density in many cases. To illustrate, we plot the fit of the HL model for two SDM cores (cores 4 and 8), one SEAT core (core 12), and one US-ITASE core (core 49) in Figure \ref{fig:herron_langway_fit}. The HL model fit is generally good for core 4, undershoots the observed values for core 8, overshoots density measurements for much of core 12, and under- and over-estimates densities in core 49. These model failures suggest that a single HL model for all cores is inappropriate and that the model should vary site-to-site. We revisit the fits of our proposed models for these same cores in Section \ref{sec:model_fit}.

\begin{figure}[H]

\begin{center}

\includegraphics[width=0.32\textwidth]{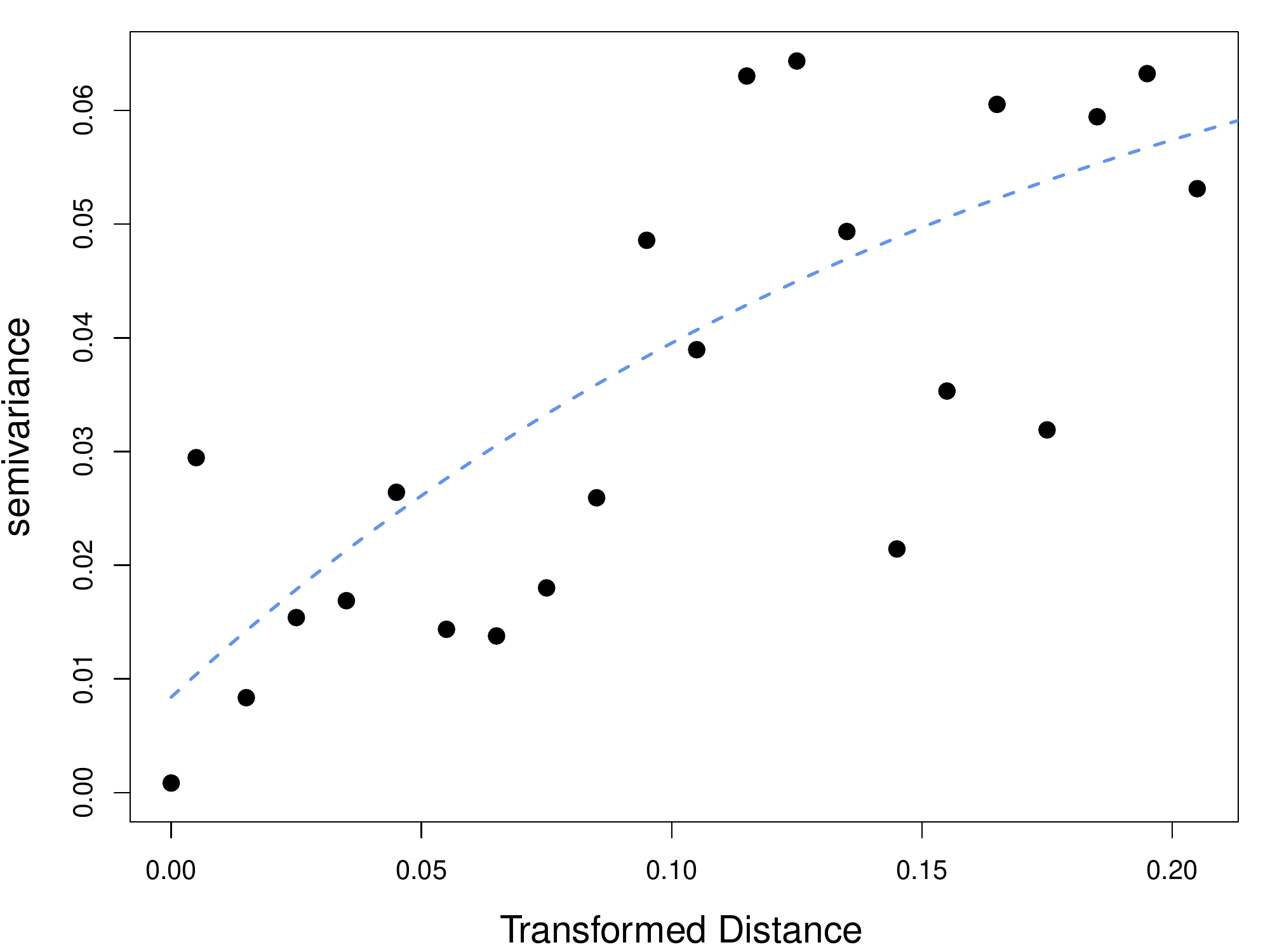}
\includegraphics[width=0.32\textwidth]{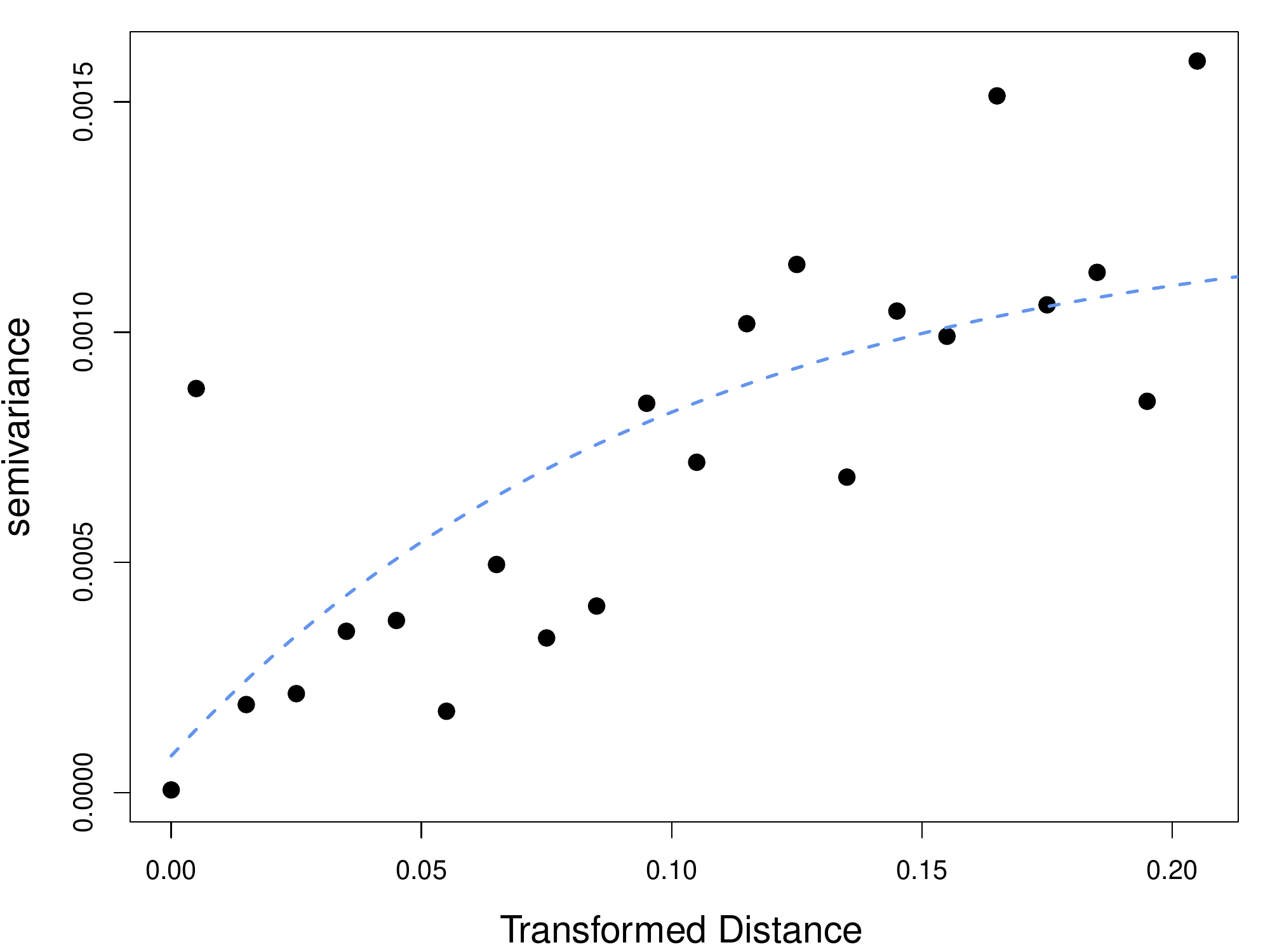}
\includegraphics[width=0.32\textwidth]{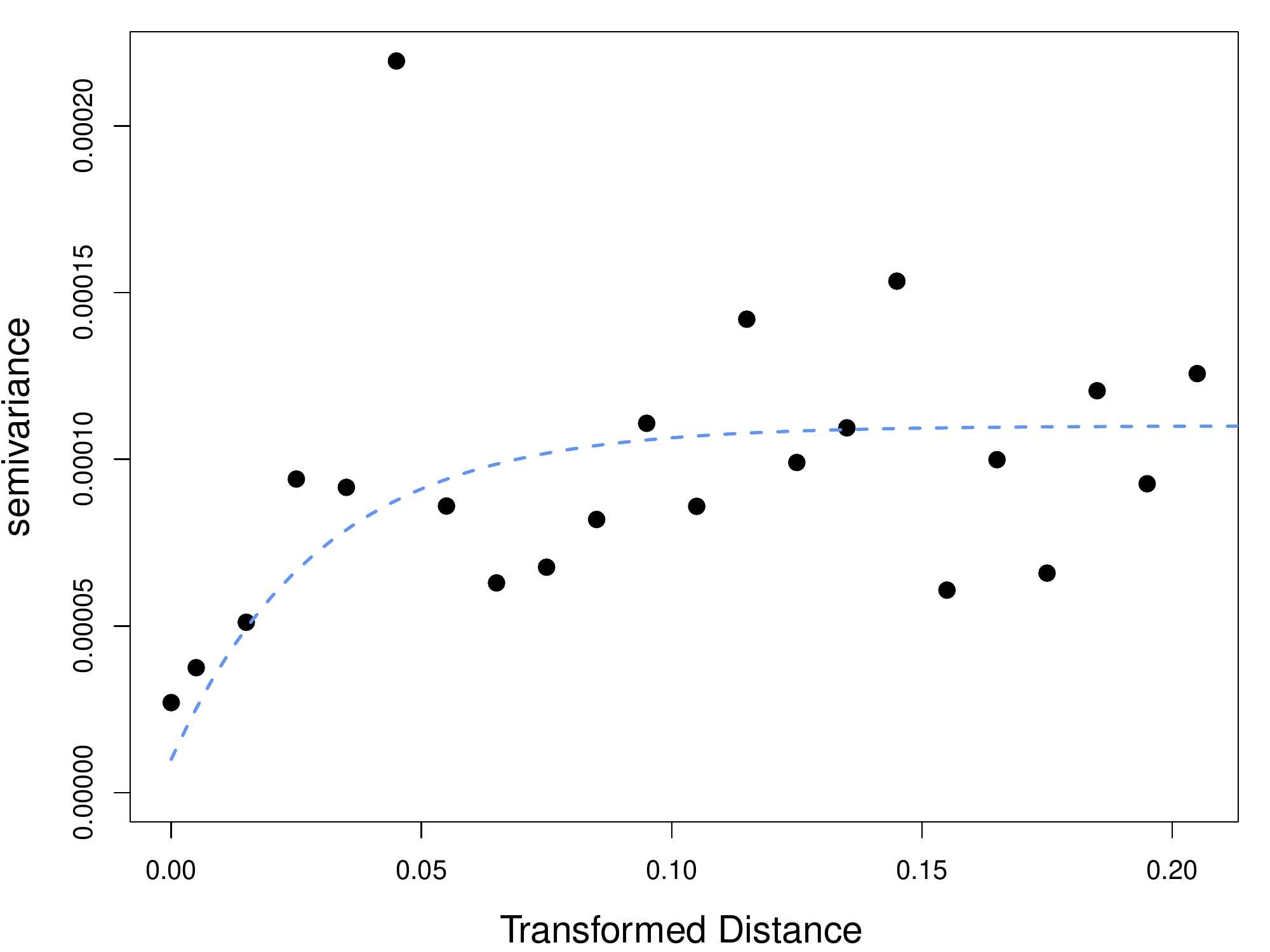}

\end{center}

        \vspace{-5mm}

\caption{Empirical binned semivariogram for site-specific (Left) intercepts, (Center) First-stage Arrhenius constant, and (Right) Second-stage Arrhenius constants of the HL model. All binned semivariograms are plotted with a fit to an exponential semivariogram.}\label{fig:variogram}

\end{figure}

We fit piecewise linear models with the same discontinuities as the HL model to explore whether there are evident spatial patterns in HL model fits. We fit these models on the logit-transformed density measurements less than $\rho_I$ at each core. For each coefficient, we plot the binned semivariogram with the fit for an exponential semivariogram in Figure \ref{fig:variogram} to explore spatial patterns in the estimated regression coefficients. These show spatial patterns motivating a spatial model for the parameters of the HL model.

\section{Spatially Varying Snow Density Models}\label{sec:arrhenius}

\subsection{Spatially Varying Snow Density Model}\label{sec:spatial_arrhenius}

Motivated by the systematic errors in the HL model (Figure \ref{fig:herron_langway_fit}) and the spatial autocorrelation of parameters in Figure \ref{fig:variogram}, we propose a joint hierarchical spatial model for the HL model parameters for each stage of our model and the critical densities $\rho_{1}$, $\rho_{2}$, and $\rho_{3}$. In the SVSD model, we denote the spatially varying densification rates for each stage of the SVSD model, indexed with the subscript $l = 1,2,3,4$, as $A_l(\bs)$, $E_l(\bs)$, where each stage is separated by critical densities $\rho_{1}(\bs)$, $\rho_{2}(\bs)$, and $\rho_{3}(\bs)$. 

We specify a joint spatial process for parameters $\alpha(\bs)$, $A_l(\bs)$, ${E}_l(\bs)$, $\rho_{1}(\bs)$, $\rho_{2}(\bs)$, and $\rho_{3}(\bs)$. We model $\alpha(\bs)$ using a Gaussian (spatial) process. Because $A_l(\bs)$ and ${E}_l(\bs)$ are positive, we model these parameters on the log scale. Similarly, we use generalized logistic transformations of $\rho_{1}(\bs)$, $\rho_{2}(\bs)$, and $\rho_{3}(\bs)$ to bound these quantities to physically-plausible quantities, between (0.42-0.68), (0.68-0.78), and (0.78-0.88) g/cm$^3$, respectively. Moreover, because each stage of our model only has one rate, we use a spatial prior distribution that enables information sharing between sites to make $A_l(\bs)$ and ${E}_l(\bs)$ jointly identifiable. Importantly, the SVSD model allows us to estimate how thermal processes vary over space.

Since snow density is bounded between $(0, \rho_I]$, we specify our mean function $\mu(\bs_i,x)$, rather than the observations, in terms of the SVSD model.
We write this as
\begin{equation}\label{eq:herron_spatial}
\log\left( \frac{\mu(\bs_i,x)}{ \rho_I - \mu(\bs_i,x)} \right) = \alpha(\bs_i) + \bz_\theta(\bs_i,x)^T \bk(\bs_i),
\end{equation} 
where $\bz_\theta(\bs_i,x)$ and $\bk(\bs_i)$ are made up of elements
\begin{equation*}
\begin{aligned}
z_1(\bs,x) &= \rho_I\min(x,\kappa_1(\bs)), \\
z_2(\bs,x) &= \frac{\rho_I\min(x - \kappa_1(\bs),\kappa_2(\bs)- \kappa_1(\bs))}{\sqrt{SMB(\bs)}} \bone\left( x > \kappa_1(\bs) \right),\\
z_3(\bs,x) &= \frac{\rho_I \min(x - \kappa_2(\bs),\kappa_3(\bs)- \kappa_2(\bs))}{\sqrt{SMB(\bs)}}  \bone\left( x > \kappa_2(\bs) \right), \\
z_4(\bs,x) &= \frac{\rho_I(x - \kappa_3(\bs))}{\sqrt{SMB(\bs)}}  \bone\left( x > \kappa_3(\bs) \right).
\end{aligned}
\qquad
\begin{aligned}
k_1(\bs) &= A_1(\bs) \exp \left( -\frac{{E}_1(\bs)}{RT(\bs)} \right)  \\
k_2(\bs) &= A_2(\bs) \exp \left( -\frac{{E}_2(\bs)}{RT(\bs)} \right)  \\
k_3(\bs) &=  A_3(\bs) \exp \left( -\frac{{E}_3(\bs)}{RT(\bs)} \right) \\
k_4(\bs) &=  A_4(\bs) \exp \left( -\frac{{E}_4(\bs)}{RT(\bs)} \right) 
\end{aligned}
\end{equation*}
Because they are parameter dependent, we write the space-depth covariates as $\bz_\theta(\bs_i,x)$. Although we do not model the depth change points $\kappa_1(\bs)$, $\kappa_2(\bs)$, and $\kappa_3(\bs)$ directly, we infer them from the critical densities $\rho_{1}(\bs)$, $\rho_{2}(\bs)$, and $\rho_{3}(\bs)$, as in \eqref{eq:change_depths}.

Somewhat as an aside, this model can be scaled to be expressed as a generalized linear I-spline model with unknown knots, where the critical depths $\kappa_1(\bs)$, $\kappa_2(\bs)$, and $\kappa_3(\bs)$ are equivalent to interior knots of the I-spline. I-splines are defined as the integral of M-spline, or, equivalently, as the integral of a scaled B-spline \citep[see][for more details]{ramsay1988,meyer2008}. \cite{white2020} use M-spline basis functions to construct integrated spatial processes for monotone function estimation.

\subsection{Functionally-Smoothed Snow Density Model}\label{sec:smoothed_arrhenius}

Because the HL and SVSD models are piecewise linear (on the transformed scale) and could thus miss important deviations from linearity, we present the functionally-smoothed SVSD model. Because parameter inference for the SVSD is our primary goal and the column space of the SVSD model changes with location, we constrain the smooth component of this model to be orthogonal to the SVSD model for each site (i.e., the functional component must lie in the orthogonal column (null) space of $\bz_\theta(\bs_i,x)$ for each site). If parameter inference is the only goal, then this extension may be unnecessary. However, if snow density prediction is of interest, this extension may boost the model's predictive power, depending on the dataset.

Because we orthogonalize the smooth component of the smoothed SVSD model with respect to $\bz_\theta(\bs_i,x)$ at observed depths, we define the projection into the orthogonal columns space of $\bZ(\bs_{i})$ as $$P^\perp_z(\bs_{i}) = \bI - \bZ(\bs_{i}) \left(\bZ(\bs_{i})^T \bZ(\bs_{i}) \right)^{-1} \bZ(\bs_{i})^T,$$ where $\bZ(\bs_{i})$ is a matrix with $\bz_\theta(\bs_{i},x)$ at all measured depths for core $i$. Note that we do not include a column of ones in $\bZ(\bs_{i})$ because we wish to preserve the simple correspondence between $\alpha(\bs)$ and the expected snow density at the surface (i.e., $x = 0$).  

If we let $\bh^\perp_\theta(\bs_{i},x) = P^\perp_z(\bs_{i}) \bh(x) $, where $\bh(x)$ are covariates specified through basis functions of smooth curves (e.g., polynomial splines), then the smoothed SVSD model is
\begin{equation}\label{eq:smooth_herron}
\log\left( \frac{\mu(\bs_{i},x)}{ \rho_I - \mu(\bs_{i},x)} \right) = \alpha(\bs_i) + \bz_\theta(\bs_{i},x)^T \bk(\bs_i) + \bh^\perp_\theta(\bs_{i},x)^T \bbeta(\bs_i),
\end{equation} 
where $\bbeta(\bs)$ are corresponding spatially varying mean-zero coefficients. We write this model in terms of $\mu(\bs_{i})$ rather than $\mu(\bs)$ because the projection is dependent on depths measured at core $i$. We note that because the parameters of the SVSD vary spatially, the projection in the null space of the SVSD model that defines $\bh^\perp_\theta(\bs_{i},x)$ depends on location. The smoothed SVSD model preserves inference on Arrhenius parameters but is more flexible, allowing deviations from piecewise linearity. Of course, $\bh^\perp_\theta(\bs_{i},x)^T \bbeta(\bs)$ could be replaced with a mean-zero Gaussian random effect with covariance $\bS$ augmented as $P^\perp_z(\bs_{i}) \bS P^\perp_z(\bs_{i})^T$, but, given the relatively simple shape of these data and the number of measurements, we use polynomials and splines instead of Gaussian processes.

Because the bases of the SVSD model $\bz_\theta(\bs_i,x)$ depend on the spatially varying parameters, the span of the SVSD model varies by site. Thus, we address confounding between the functional smoothing (spline terms) and the spatially varying snow density model at each site. Although not pursued in this manuscript, in the spatial confounding literature, it is conventional to use dimension reduction through the eigenvalue decomposition of the orthogonal basis and to scale the orthogonal projection with adjacency information \citep[see, e.g.,][]{griffith2003spatial, tiefelsdorf2007semiparametric,hughes2013dimension,murakami2015random}. However, because the SVSD model is different at every site, there is not a common or global projection into the null space of the SVSD model. Moreover, sites do not share common eigenvalue decompositions for the functional bases $\bh^\perp_\theta(\bs_i,x)$. Thus, it is not clear how we would coherently specify a joint prior distribution for spatially varying coefficients in this setting. Likewise, because we are addressing functional confounding at each site, it is unclear what benefit incorporating \emph{functional} adjacency in the orthogonal functional basis $\bh^\perp_\theta(\bs_i,x)$ would have. Although both of these approaches in the spatial confounding literature present possible areas of future exploration, we take a more direct approach to address functional confounding in this application.

\section{Methods and Models}\label{sec:models}

\subsection{Models Comparison}\label{sec:mod_comp}

In our model comparison, we address four modeling questions: (1) Which truncated error distribution best suits these data? (2) How should we account for group differences in measurement accuracy, including the core length used to obtain density measurements? (3) Which cross-covariance model is best for our spatially varying HL model? (4) Should our model include deviations from the piecewise linearity by including \textit{smoother} components to the model?  

We use the Watanabe–Akaike information criterion (WAIC) as our model selection criterion \citep{watanabe2010}, computed using each observation as a partition \citep{vehtari2017}. We use WAIC in favor of deviance information criterion (DIC) \citep{spiegelhalter2002} because WAIC is a generally more stable approximation to cross-validation than DIC, uses the entire posterior distribution, and is less prone to select overfit models.

We discuss model comparison results in detail in Appendix \ref{app:mod_comp}. To summarize the results, we find that using a truncated $t$-distribution is better than the corresponding truncated Normal models. Error distributions with hierarchically-specified scale parameters, weighted to account for the length of the core used to obtain density measurements, improve WAIC relative to homoscedastic models. We select a separable cross-covariance model for the spatially varying HL parameters because it has the lowest WAIC, but we did not find significant differences in model performance for most cross-covariance specifications. Lastly, we compare seven possible functional smoothings for the smoothed SVSD and select a quadratic spline with three knots to specify \eqref{eq:smooth_herron} because it has the lowest WAIC; however, several spline models perform comparably. We present our final model in Section \ref{sec:final_mod}.

\subsection{Hierarchical Model}\label{sec:final_mod}

Using the quantities defined in Sections \ref{sec:data} and \ref{sec:smoothed_arrhenius}, our hierarchical model is
\begin{equation}
\begin{aligned}
y(\bs_{i},x) &\sim t_\nu\left( \mu(\bs_{i},x) ,\tau^2_i,0,\infty \right) \\
\log\left( \frac{\mu(\bs_{i},x)}{ \rho_I - \mu(\bs_{i},x)} \right) &= \alpha(\bs_i) + \bz_\theta(\bs_i,x)^T \bk(\bs_i) + \bh^\perp_\theta(\bs_{i},x)^T \bbeta(\bs) \\
\log(\tau^2_i) &\sim \mathcal{N}\left\lbrace \log\left[ \tau^2_{m_i} \,  dx_i^{\eta_{m_i }}  \right], \sigma^2_\tau \right\rbrace,
\end{aligned}
\end{equation}
where $t_\nu(\mu(\bs_{i},x) ,\tau^2_i,0,\infty)$ is a $t$-distribution, truncated below by 0, with location $\mu(\bs_{i},x)$ and scale $\tau_i$. We use the truncated-$t$ distribution to give a coherent model, but the truncation has little practical effect on our analysis because snow densities are not close to zero. As a consequence, $\mu(\bs_i,x)$ is effectively a mean function rather than only a location function.
For the EAP and SEAT campaigns, there is no variation in $dx_i$; therefore, we fix $\eta_{m_i} = 0$ because this weighting parameter is not identifiable.

As discussed, we specify a joint spatial process for the smoothed SVSD model through a multivariate Gaussian process (MGP) model for the intercept, transformed Arrhenius constant parameters, and critical densities. We let $\btheta(\bs)$ be transformed spatial parameters,
$\alpha(\bs) $, $ \log(A_1(\bs))$, $ \log(A_2(\bs)) $, $ \log(A_3(\bs)) $, $ \log(A_4(\bs)) $, $ \log(E_1(\bs)) $, $ \log(E_2(\bs)) $, $ \log(E_3(\bs)) $, $ \log(E_4(\bs)) $, $ \log\left( \frac{\rho_1(\bs) - 0.42}{0.68 - \rho_1(\bs)} \right) $, $ \log\left( \frac{\rho_2(\bs) - 0.68}{0.78 - \rho_2(\bs)} \right) $, and $\log\left( \frac{\rho_3(\bs) - 0.78}{0.88 - \rho_3(\bs)} \right)$, and we specify $\btheta(\bs)$ using a multivariate GP. Thus, the transformed spatial parameters at observed sites $\btheta(\mathcal{S}) = \left(\btheta(\bs_1)^T ,...,\btheta(\bs_{n_s})^T  \right)^T$ follows a multivariate-normal distribution. That is,
\begin{equation}
\btheta(\mathcal{S}) \sim \mathcal{N}\left( \bM \bgamma, \bR \otimes \bV \right),
\end{equation}
where the hierarchical mean for $\btheta(\bs)$ is 
$$\bM \bgamma = \left(\gamma_\alpha, \gamma_{A_1}, \gamma_{A_2},  \gamma_{A_2},\gamma_{A_2}, \gamma_{E_1},\gamma_{E_2},\gamma_{E_2}, \gamma_{E_2}, \gamma_{\rho_1},\gamma_{\rho_2}, \gamma_{\rho_3} \right)^T,$$ 
$\bgamma =  \left(\gamma_\alpha, \gamma_{A_1}, \gamma_{A_2}, \gamma_{E_1},\gamma_{E_2}, \gamma_{\rho_1},\gamma_{\rho_2}, \gamma_{\rho_3} \right)^T$, and $\bM$ is a $12 \times 8$ matrix that repeats elements of $\bgamma$ to account for the weak transitions suggested by \cite{horhold_densification_2011}. We use repeated elements in our hierarchical mean because densification rates show little or weak transitions at the second and third critical densities \citep{horhold_densification_2011}. The $i$th row and $i'$th column of the correlation matrix $\bR$ is $\exp\left( \phi d(\bs_{i},\bs_{i'} )\right)$.

For the spatial-varying coefficients $\beta_k(\bs)$ of the orthogonalized spline basis functions, we assume that
\begin{equation}
\beta_k\left(\mathcal{S}\right) \sim \mathcal{N}\left(0,\sigma^2_{\beta_m} \bR_\beta \right),
\end{equation}
where the $i$th row and $i'$th column of the correlation matrix $(\bR)_{i,i'} = \exp\left( -\phi_\beta d(\bs_{i},\bs_{i'} )\right)$. Therefore, we assume that the spatially varying spline coefficients are centered on zero, share a common decay parameter, but have their own scale parameter.

\subsection{Prior Distributions, Model Fitting, and Prediction}\label{sec:model_fit}

We use the following prior distributions for our hierarchical model:
\begin{equation}
\begin{aligned}[c]
\gamma_\alpha &\sim \mathcal{N}\left( -0.5,0.5^2 \right) \\
\gamma_{A_1} &\sim \mathcal{N}\left( 2.4,0.2^2 \right) \\
\gamma_{A_2}  &\sim \mathcal{N}\left( 6.35,0.2^2 \right) \\
\gamma_{E_1} &\sim \mathcal{N}\left( 9.23,0.2^2 \right) \\
\gamma_{E_2} &\sim \mathcal{N}\left( 9.97,0.25^2 \right) \\
\end{aligned}
\hspace{5mm}
\begin{aligned}[c]
\gamma_{\rho_1} &\sim \mathcal{N}\left( 0,1 \right) \\
\gamma_{\rho_2} &\sim \mathcal{N}\left( 0,1 \right) \\
\gamma_{\rho_3} &\sim \mathcal{N}\left( 0,1 \right) \\
\nu &\sim \text{Unif}\left(4,30\right) \\
\log(\tau^2_{m_i}) &\sim \mathcal{N}\left(-7,4\right) \\
\eta_{m_i} &\sim \mathcal{N}\left(-8,4\right) \\
\end{aligned}
\hspace{5mm}
\begin{aligned}[c]
\sigma^2_\tau &\sim \mathcal{IG}\left(2.1,1/10 \right)\\
\bV &\sim \text{Inverse-Wishart}\left( 13 ,\bI \right) \\
\phi^{-1} &\sim \text{Unif}\left( 10, 1000 \right) \\
\sigma^2_{\beta_k} &\sim \mathcal{IG}\left(2.1,1/10 \right) \\
\phi^{-1}_\beta &\sim \text{Unif}\left( 10, 1000 \right) \\
\end{aligned}
\end{equation}
The hierarchical mean parameters are informative, and we use the parameters from the original HL parameters in \cite{herron_firn_1980} to specify posterior medians. Because the $t$-distribution is nearly Gaussian when $\nu \geq 30$, we set 30 degrees of freedom as the upper bound. We set the lower bound on $\nu$ to be four so that the first four moments of the error distribution are finite.  The prior median for $\log(\tau^2_{m_i})$ corresponds to a median standard deviation of about 0.03 (for large $\nu$) but allows much larger and smaller values. We choose this because, for a fixed depth, observations are quite concentrated about the center of the density curve. Because we do not know how much values of $\log(\tau^2_i)$ may vary, we use a somewhat diffuse prior distribution on $\sigma^2_\tau$.  We choose the prior distribution for $\eta_{m_i}$ because using more of the core to obtain a density measurement yields a smoother curve (i.e., less variance).

We use uniform prior distributions for the range parameters of our spatially varying HL parameters $\phi^{-1}$ and cubic spline coefficients $\phi^{-1}_\beta$. For the exponential correlation function, our prior distribution bounds the effective range of our spatially varying parameters between 30 and 3,000 km. Because we expect the spatially varying parameters of our model not to differ much from the hierarchical means, we use flexible prior distributions for $\bV$ and $\sigma^2_{\beta_k}$ that heavily weight small values. For exponential covariance models, only the product of scale and decay parameters is identifiable, and model-based interpolations are equal if the product of scale and decay parameters is equal \citep{zhang2004}. Thus, we are not able to make good inference on $\phi$, $\phi_\beta$, $\bV$, on $\sigma^2_{\beta_k}$ individually.

We use an adaptive Markov chain Monte Carlo (MCMC) model fitting approach to obtain posterior samples using Metropolis-within-Gibbs and Gibbs updates. We sample HL parameters $\theta(\bs)$ site-wise using the multivariate Metropolis algorithm with multivariate Normal proposal distributions centered on the current parameter values and covariance equal to the scaled empirical covariance of the site-specific parameters \citep{haario2001}. We use the univariate Metropolis algorithm updates with Normal random walk proposal distributions to sample from the posterior distributions of $\beta_k(\bs)$ and $\nu$, while we use the univariate Metropolis-Hastings algorithm with log-Normal random walk proposal distributions to sample from the posterior distributions of $\tau^2(\bs_{i})$, $\phi$, and $\phi_\beta$. During the burn-in part of our MCMC, we tune the candidate variances so that acceptance rates are between 0.2-0.6 for univariate updates and  0.15-0.5 for multivariate updates. Because spatially varying parameters mix slowly due to high correlation, we follow the recommendation of \cite{neal1998}, updating the spatially varying parameters several times, five, in this case, each iteration of the sampler. Because posterior conditional distributions for $\bgamma$, $\log(\tau^2_{m_i})$, $\eta_{m_i}$, $\sigma^2_\tau$, and $\sigma^2_{\beta_k}$ can be found in closed form, we update these parameters using Gibbs updates. 

Using our model, we can estimate model parameters and mean density curves. Our primary goal is estimating how various parameters (or functions of parameters) vary spatially. To estimate spatially varying parameters at new locations $\bs_{new}$, we simply sample the transformed parameters $\btheta(\bs_{new})$ using the appropriate conditional normal distribution. If estimating the mean $\mu(\bs,x)$ or observed density $y(\bs,x)$ at a location outside of $\mathcal{S}$, then we follow the same process for interpolating regression spline coefficients as we use for spatially varying HL parameters and use composition sampling to sample from the posterior predictive distribution \citep{tanner1996}.

\section{Results}\label{sec:res}

We run our MCMC model fitting for 250,000 iterations, discard the first 50,000, and keep every 20th sample. Our results are based on the remaining 10,000 posterior samples. Before presenting results, we again emphasize that the truncated-$t$ distribution gives a coherent model, but the truncation has no practical effect on our analysis because snow densities are not close to zero. For this reason, the location function $\mu(\bs,x)$ is essentially a mean function.
 
In Section \ref{sec:res_hier}, we summarize the global hierarchical parameters of the smooth SVSD model and compare them to previous studies. Then, we explore the differences in scale parameters by core, discussing campaign-specific effects. We discuss the spatial variability in the physical model's parameters in Section \ref{sec:res_spat_vary}. Because our primary focus is understanding snow density's response to temperature and other climatic variables, we give particular attention to $E_l(\bs)$. In this section, we also follow up on the hypothesis that the second, third, and fourth stages of densification show no significant difference. Lastly, in Section \ref{sec:comp_curve}, we compare the simple HL, SVSD, and smoothed SVSD models for several cores to illustrate the benefits of our proposed framework. 

\subsection{Model Summaries and Comparisons to Previous Studies}\label{sec:res_hier}

In this subsection, we present the posterior summaries of the hierarchical parameters and compare the hierarchical mean to the parameter estimates of the HL model, as well as other scientific assertions about snow densification (Discussed in Section \ref{sec:snow}). We defer discussion of several parameters to the Supplementary Material. We present posterior summaries (posterior mean, median, standard deviation, and a 90\% central credible interval) in Table \ref{tab:post_sum_hier}. To compare previous parameter estimates to our posterior distribution, we calculate the quantile of our posterior distribution corresponding to the previous estimates (See Table \ref{tab:post_sum_hier}). 

\begin{table}[H]
\centering
\scriptsize
\begin{tabular}{|l|l|l|lllll|}
  \hline
   Estimate  & Function of & Previous Estimate &  &  & &  &  \\ 
 Description  & Parameters & Posterior Quantile & Mean & Median & Std. dev. & 5\% & 95\% \\ 
  \hline
 Surface Density & $\frac{\rho_I \exp(\gamma_\alpha)}{1 + \exp(\gamma_\alpha)}$&  ------ & 0.365 & 0.364 & 0.046 & 0.291 & 0.442 \\
1st-stage $A$ &   $\exp(\gamma_{A_1})$ & 0.577 & 10.810 & 10.650 & 1.892 & 7.978 & 14.150 \\ 
  1st-stage $E$  &  $\exp(\gamma_{E_1})$ &  0.756 & 9.46e3 & 9.38e3 & 1.07e3 & 7.78e3 & 1.13e4 \\ 
 2nd-stage $A$&   $\exp(\gamma_{A_2})$ & 0.511 & 578.700 & 572.800 & 80.540 & 457.900 & 719.700 \\
2nd-stage $E$ &  $\exp(\gamma_{E_2})$ &  0.505  & 2.14e4 &2.14e4 & 1.47e3 & 1.91e4 & 2.39e4 \\ 
 1st Critical Density &  $f_1(\rho_1)$ & 0.835 & 0.524 & 0.522 & 0.026 & 0.484 & 0.571 \\ 
  2nd Critical Density &  $f_2(\rho_2)$ & 0.267 & 0.739 & 0.741 & 0.014 & 0.714 & 0.760 \\ 
 3rd Critical Density &   $f_3(\rho_3)$ & 0.071 & 0.846 & 0.847 & 0.010 & 0.828 & 0.860 \\ 
   \hline
\end{tabular}
\caption{Posterior summaries of untransformed hierarchical parameters.}\label{tab:post_sum_hier}
\end{table}

Parameter estimates from the original HL model correspond to quantiles between (0.365 and 0.756). The first critical density suggested by \cite{herron_firn_1980}, 0.55 g/cm$^3$, corresponds to the 84\% of our posterior distribution, suggesting that the snow cores in our dataset suggest a lower critical density than that in \cite{herron_firn_1980}. As discussed in \cite{horhold_densification_2011}, several studies suggest critical densities at 0.73 and 0.83 g/cm$^3$, which are, respectively, the 0.27 and 0.07 quantile of the posterior distribution. Although the hierarchical mean of our smoothed SVSD model is compatible with previous scientific estimates for snow densification, a global model for snow densification fails to accurately capture site-specific density patterns. 
Given the sparseness of density observations, improved constraints on these critical densities and their spatial patterns are crucial for statistically interpolating and physically modeling snow densification over ice sheets.
We explore how these patterns change over space in Section \ref{sec:res_spat_vary}.

\begin{figure}[H]
\begin{center}
\includegraphics[width=\textwidth]{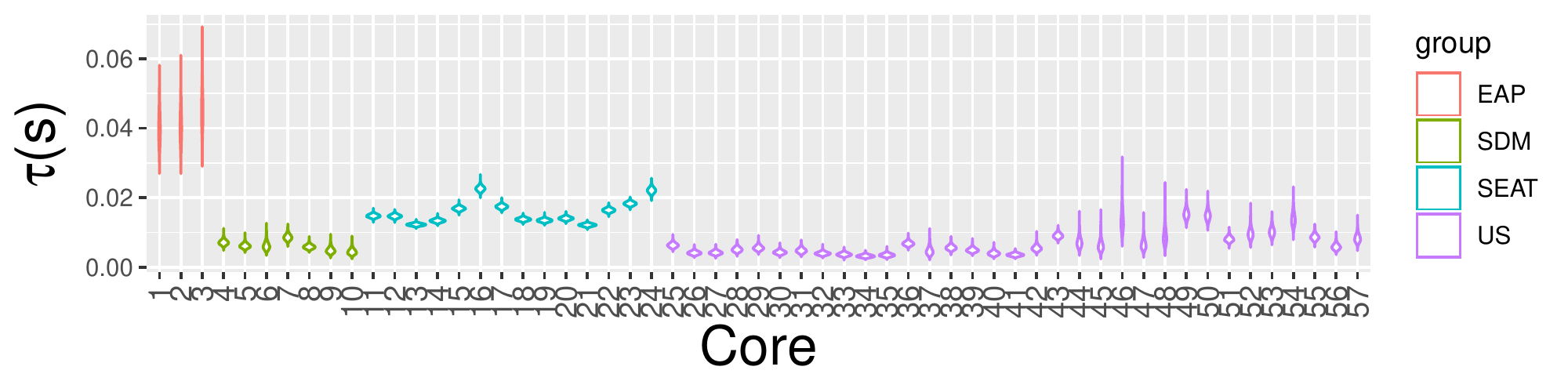}
\end{center}
\caption{Posterior distributions for core-specific scale parameters.}\label{fig:tau_sd}
\end{figure}

In Figure \ref{fig:tau_sd}, we give violin plots of the posterior distributions for the core-specific scale parameters $\sqrt{\tau^2_{i}}$. Recall that $\log(\tau^2_{i})$ are hierarchically-specified and incorporate expedition-specific scale and weighting parameters to account for the length of the core used to obtain density measurements, as well as campaign-specific differences. The differences between expeditions are apparent in Figure \ref{fig:tau_sd}. EAP snowpits and SEAT cores generally have shorter averaging lengths $dx_i$; thus, unsurprisingly, EAP snowpits have the highest estimated standard deviation, followed by SEAT cores.  SDM and US-ITASE cores have smaller estimated standard deviations, on average, but the US-ITASE cores show high variability in standard deviation. 

\subsection{Spatially Varying Parameters}\label{sec:res_spat_vary}

In this section, we focus on the spatially varying component of our model. Specifically, we interpolate all spatially varying HL parameters on a grid of about 2500 points over the convex hull of our data. Importantly, the model's Arrhenius constants depend on temperature 10-m below the surface. We rely on model-based spatial interpolations presented in Section \ref{sec:data} to estimate $T(\bs)$ which, together with $A_l(\bs)$, $E_l(\bs)$, define the Arrhenius constant for the $l$th stage of densification.  

\begin{figure}[H]
\begin{center}
\includegraphics[width=0.24\textwidth]{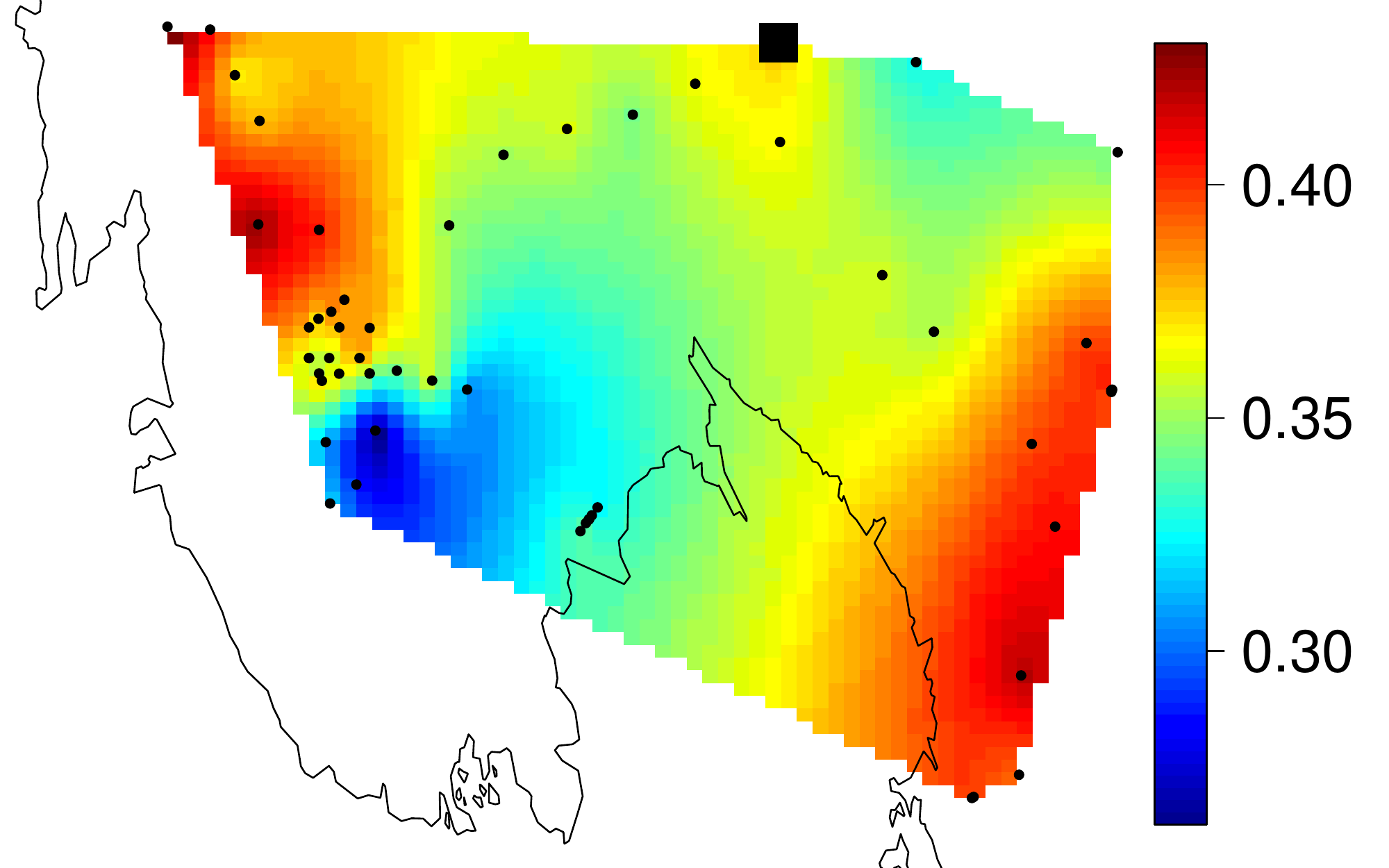}
\includegraphics[width=0.24\textwidth]{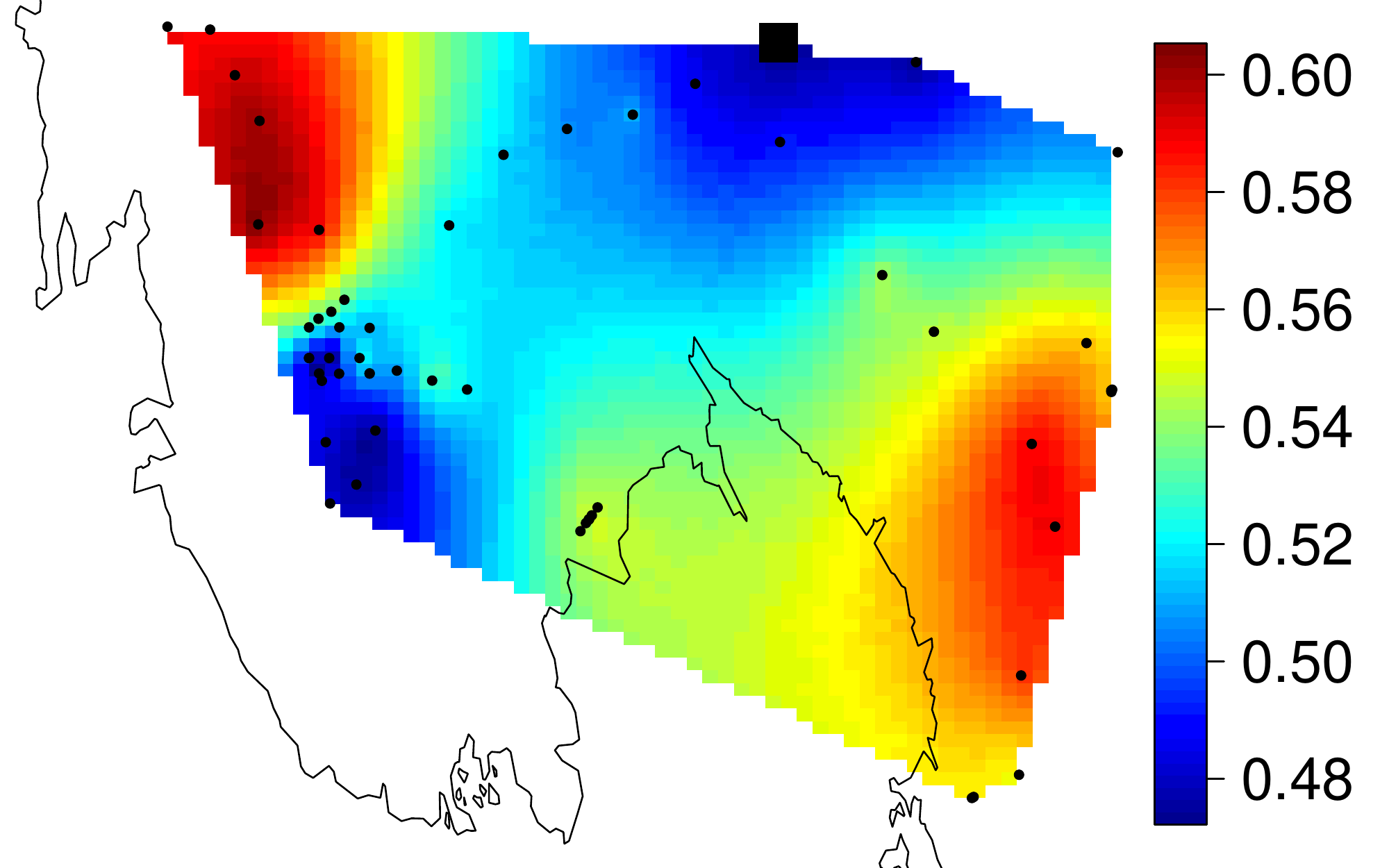}
\includegraphics[width=0.24\textwidth]{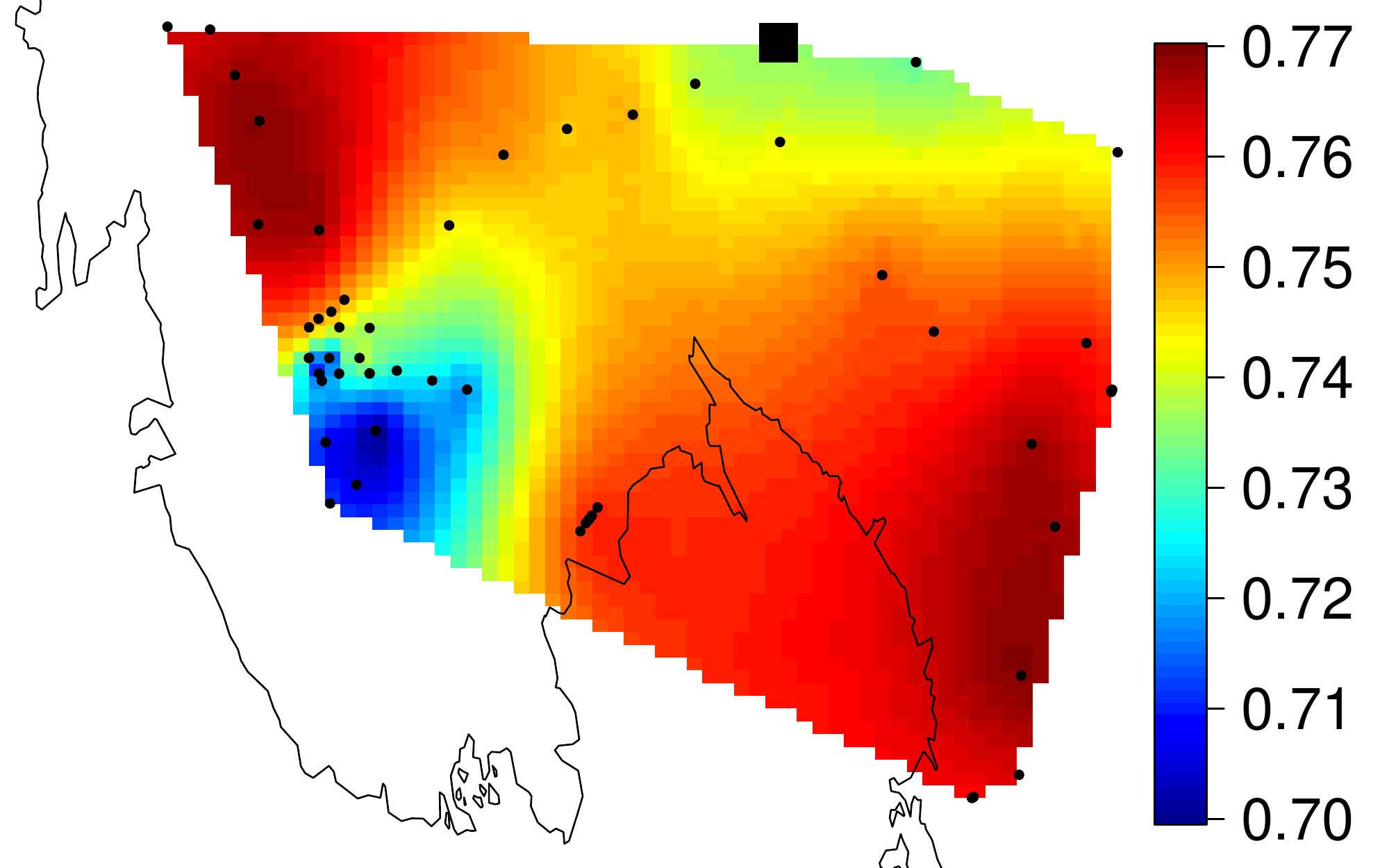}
\includegraphics[width=0.24\textwidth]{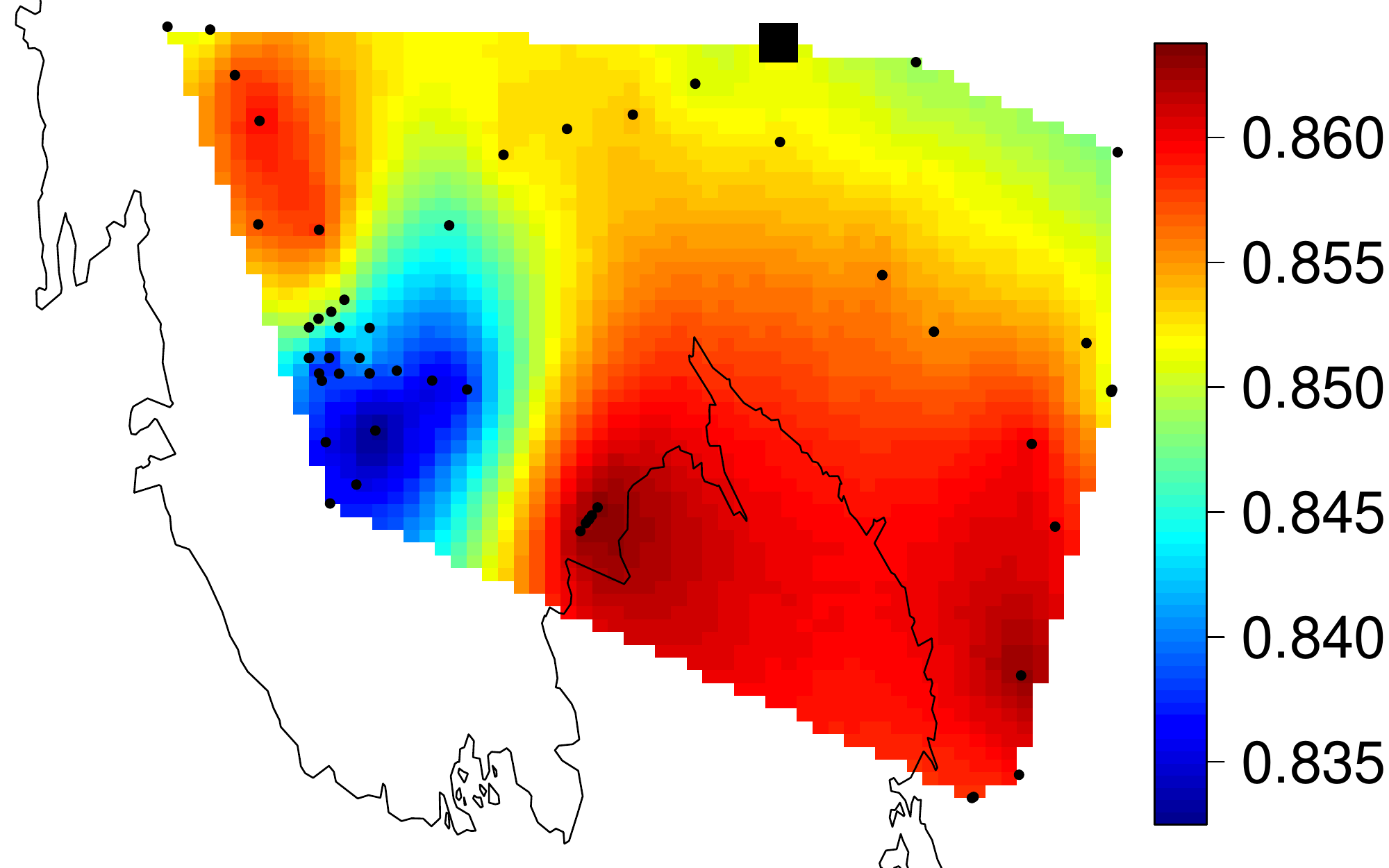} \\
\includegraphics[width=0.24\textwidth]{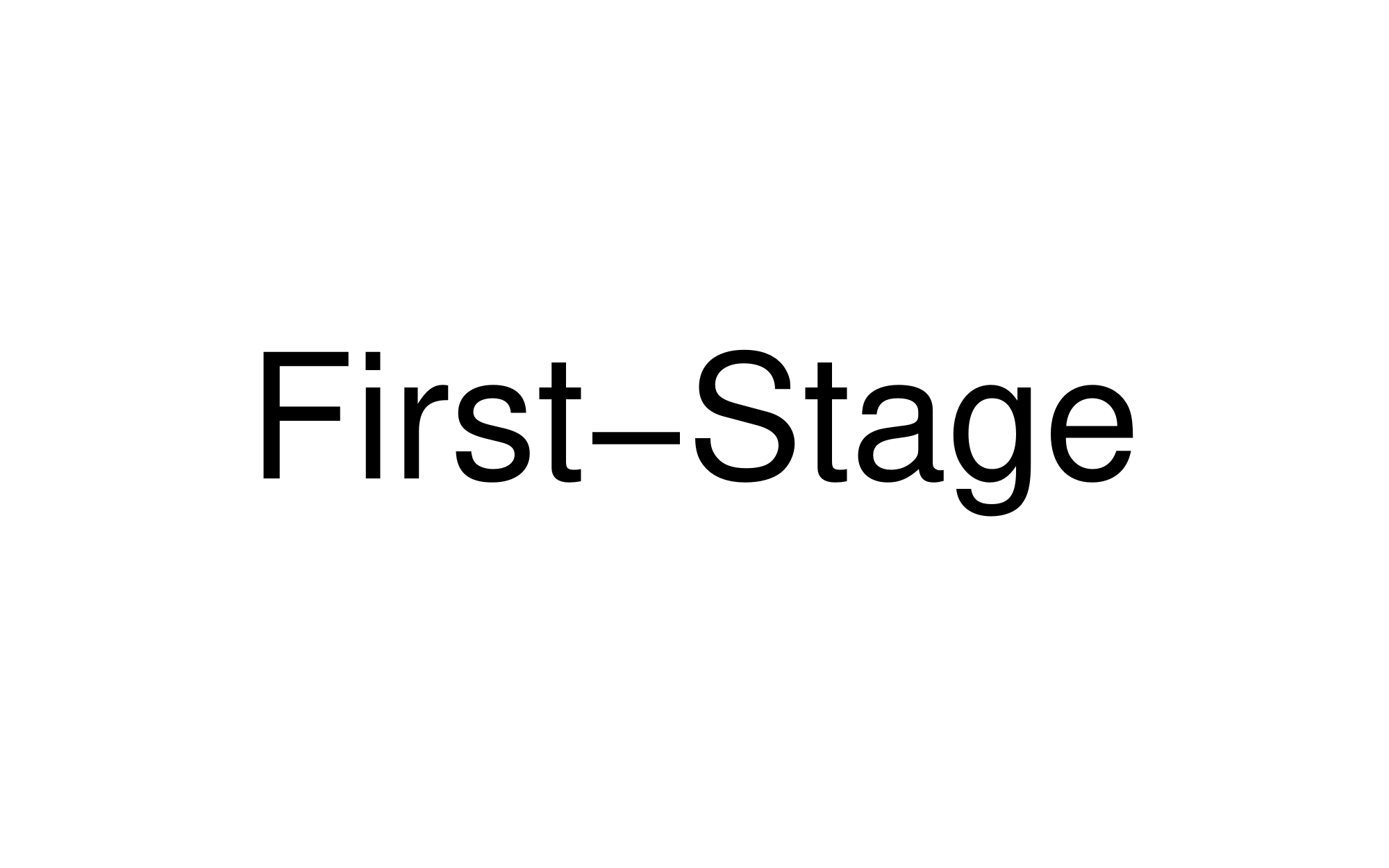}
\includegraphics[width=0.24\textwidth]{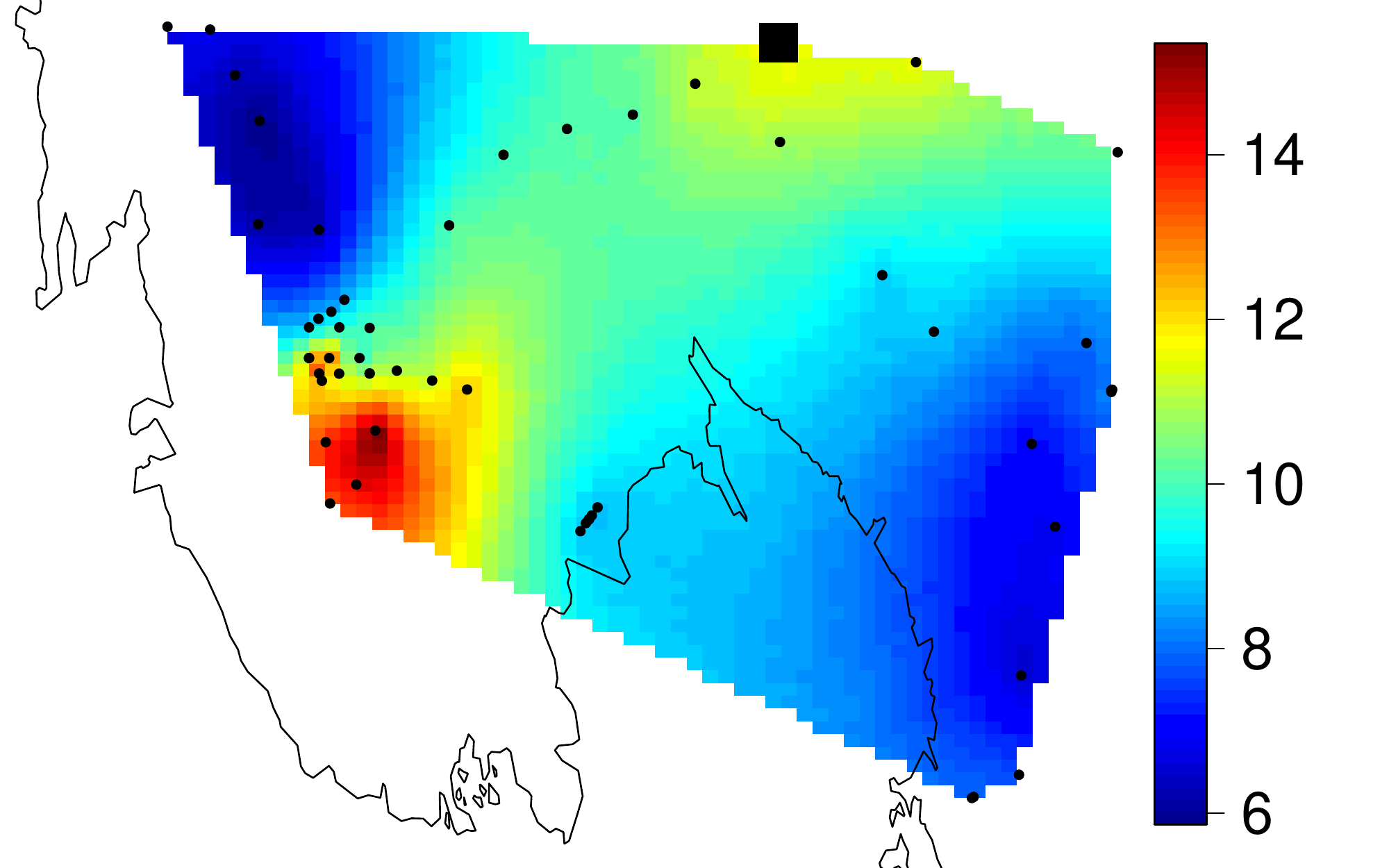}
\includegraphics[width=0.24\textwidth]{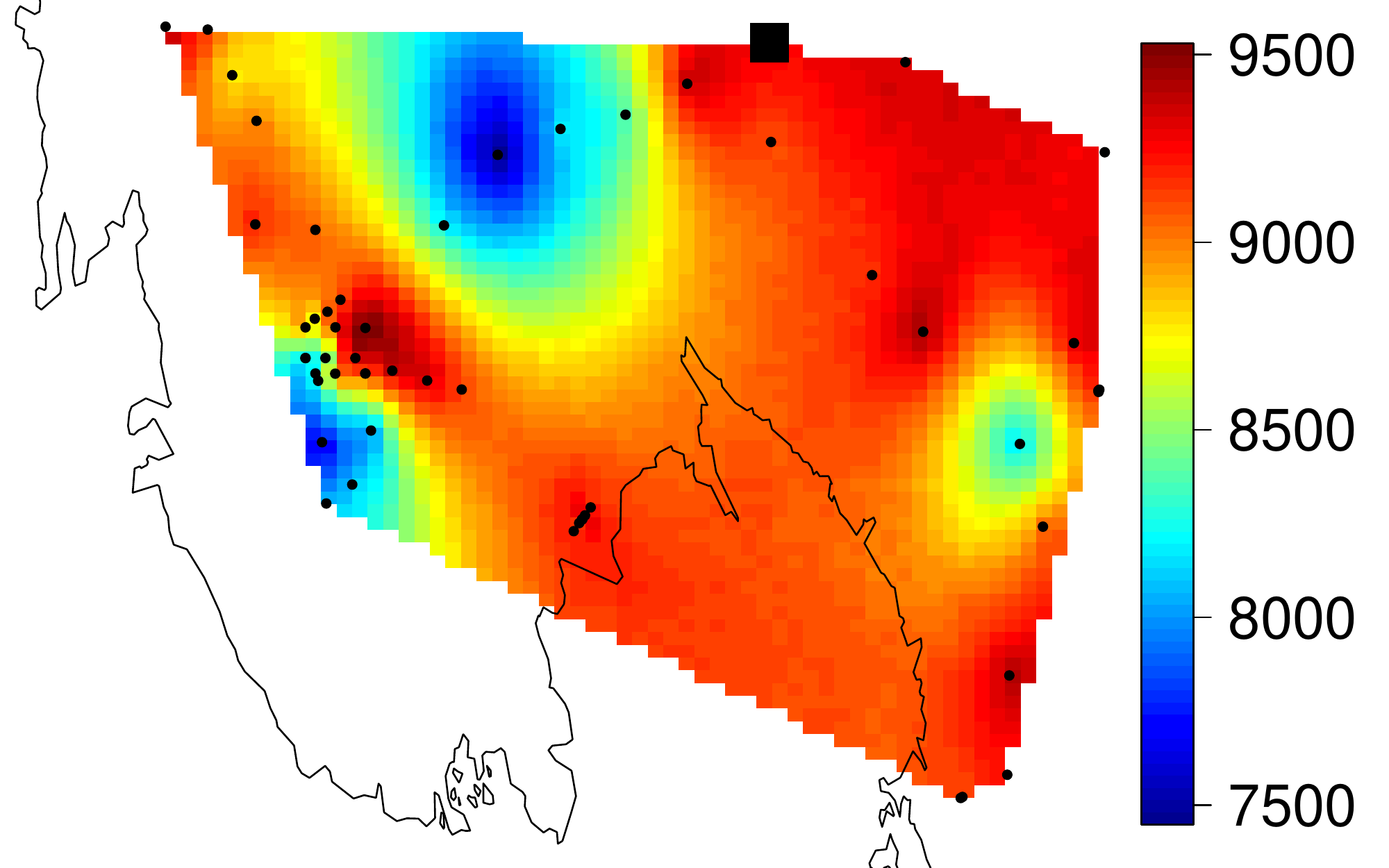}
\includegraphics[width=0.24\textwidth]{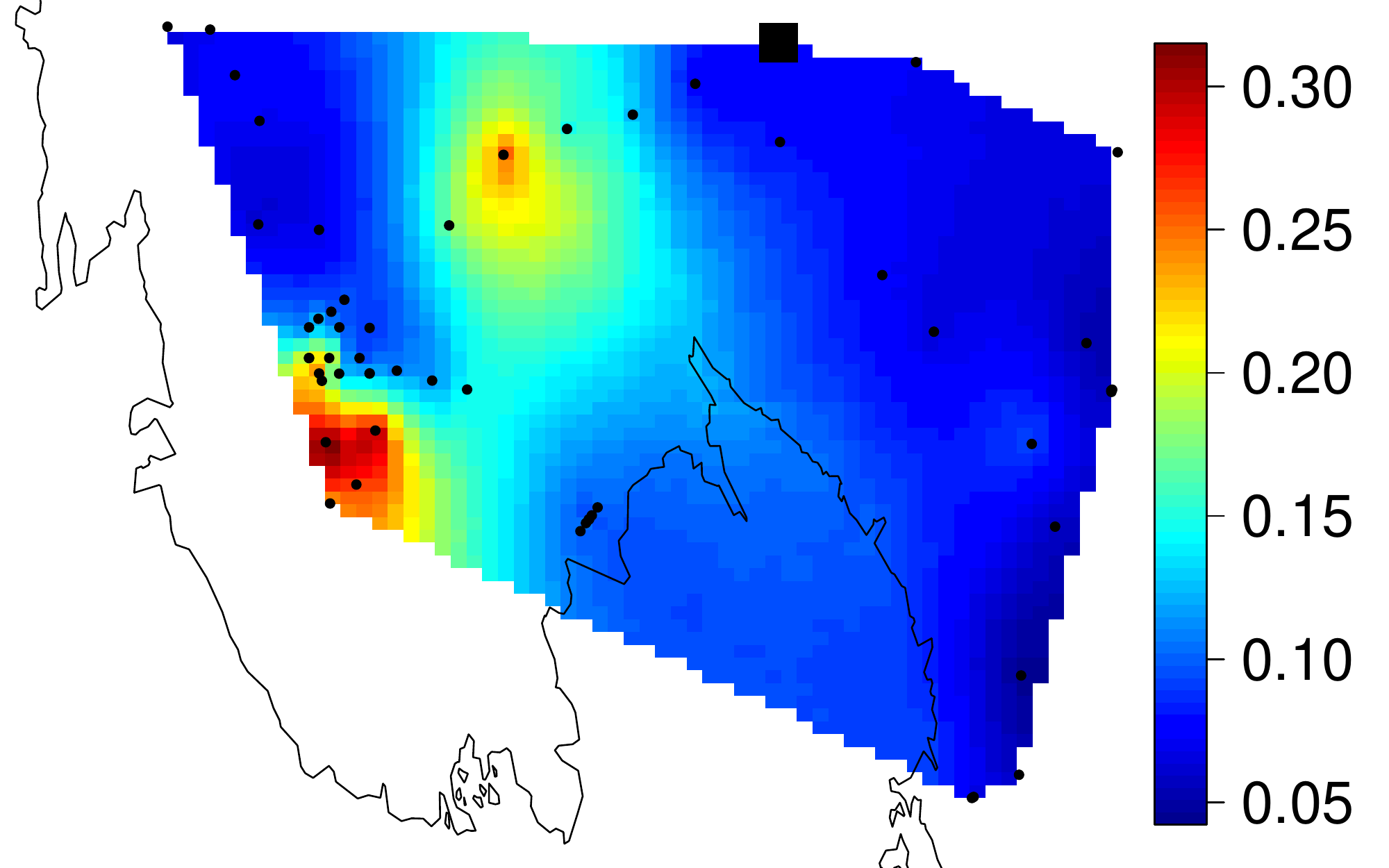}\\
\includegraphics[width=0.24\textwidth]{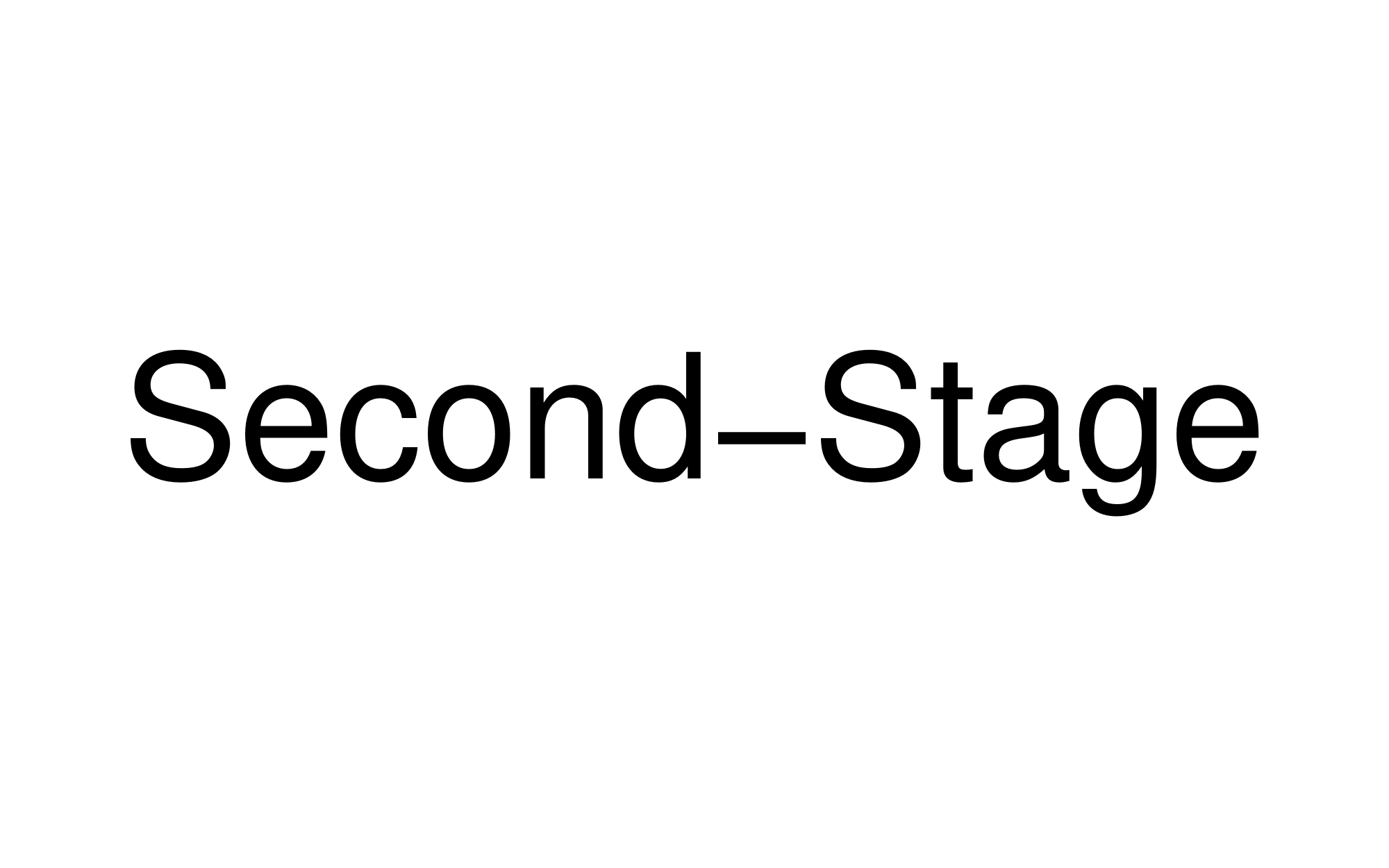}
\includegraphics[width=0.24\textwidth]{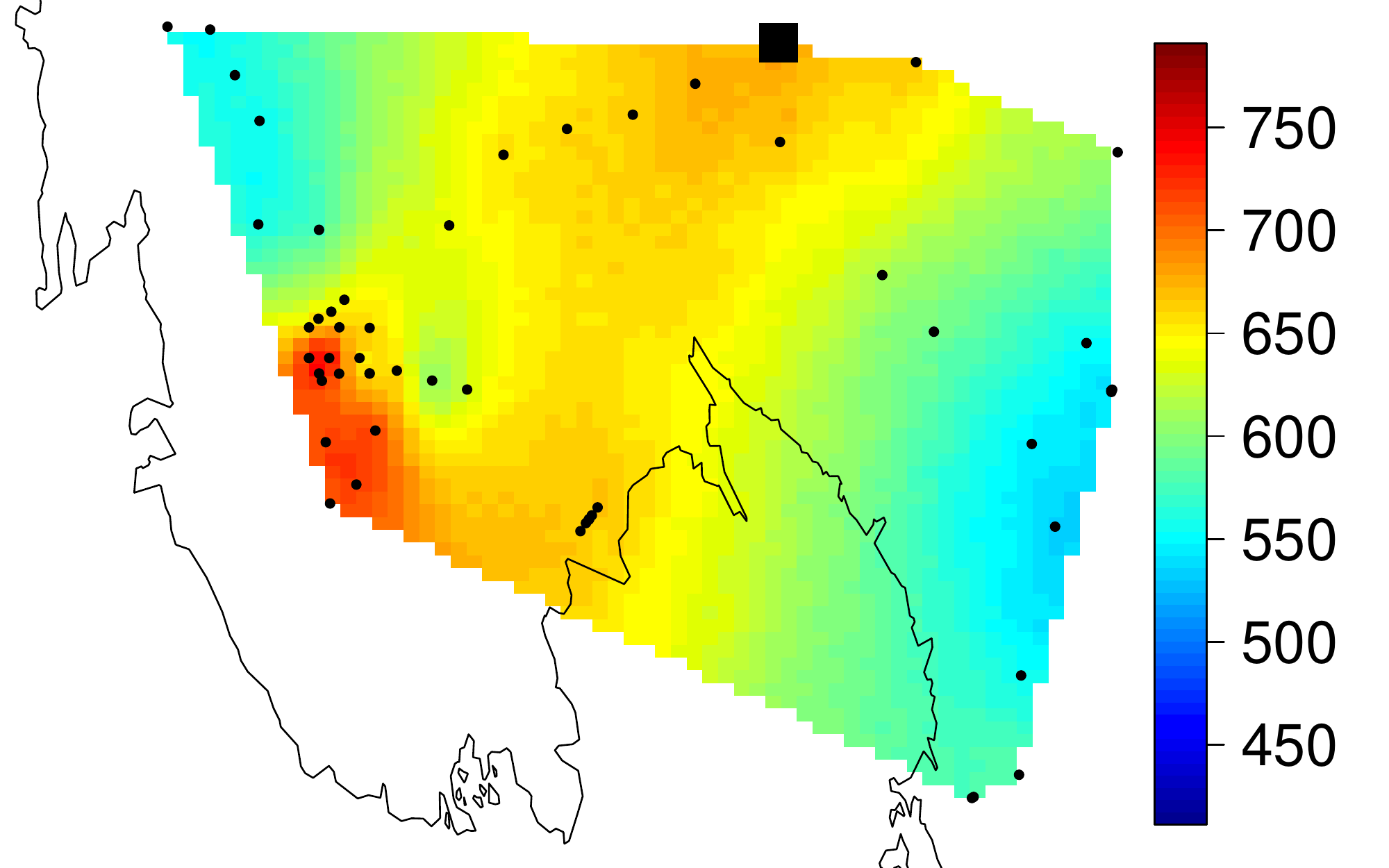}
\includegraphics[width=0.24\textwidth]{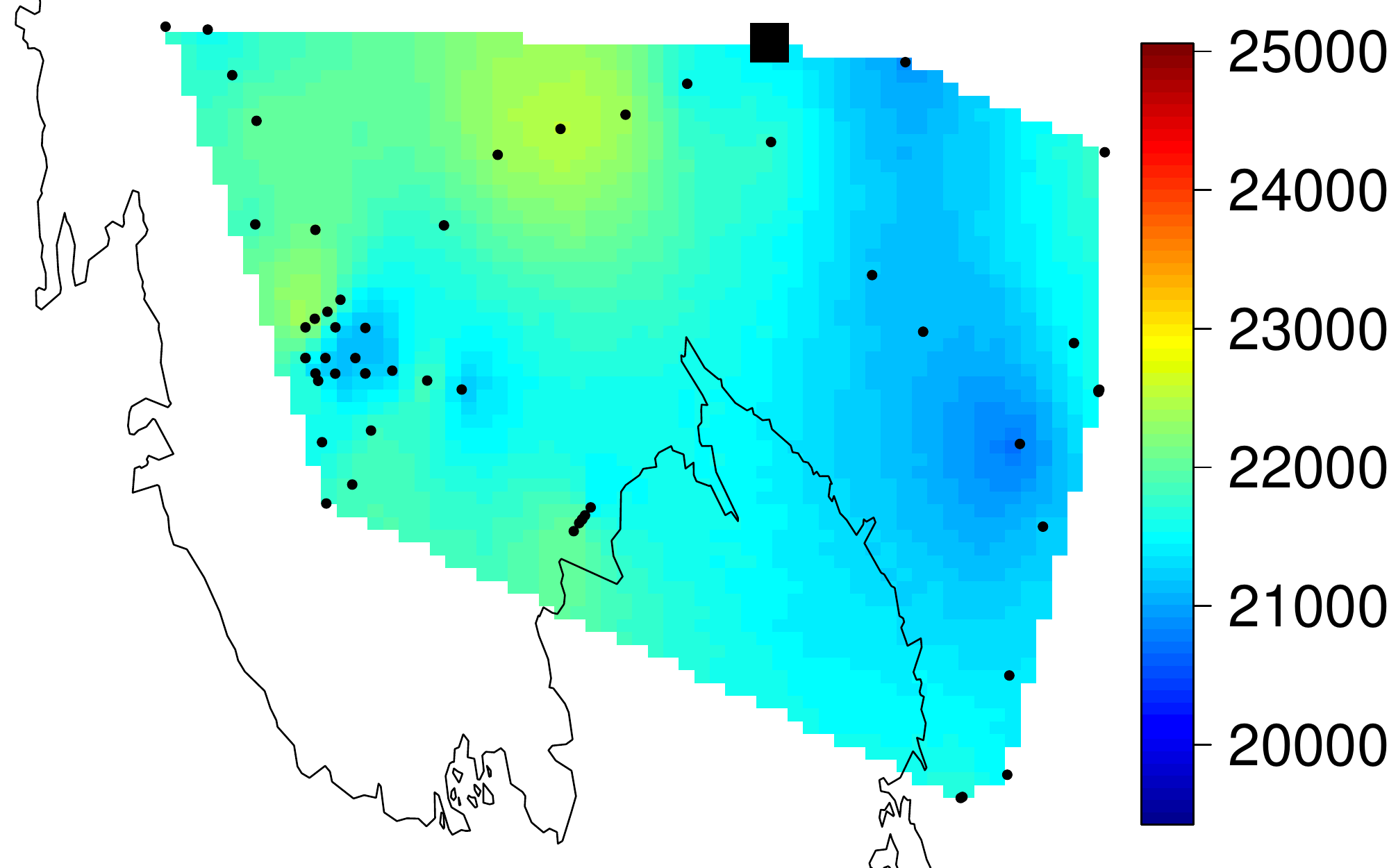}
\includegraphics[width=0.24\textwidth]{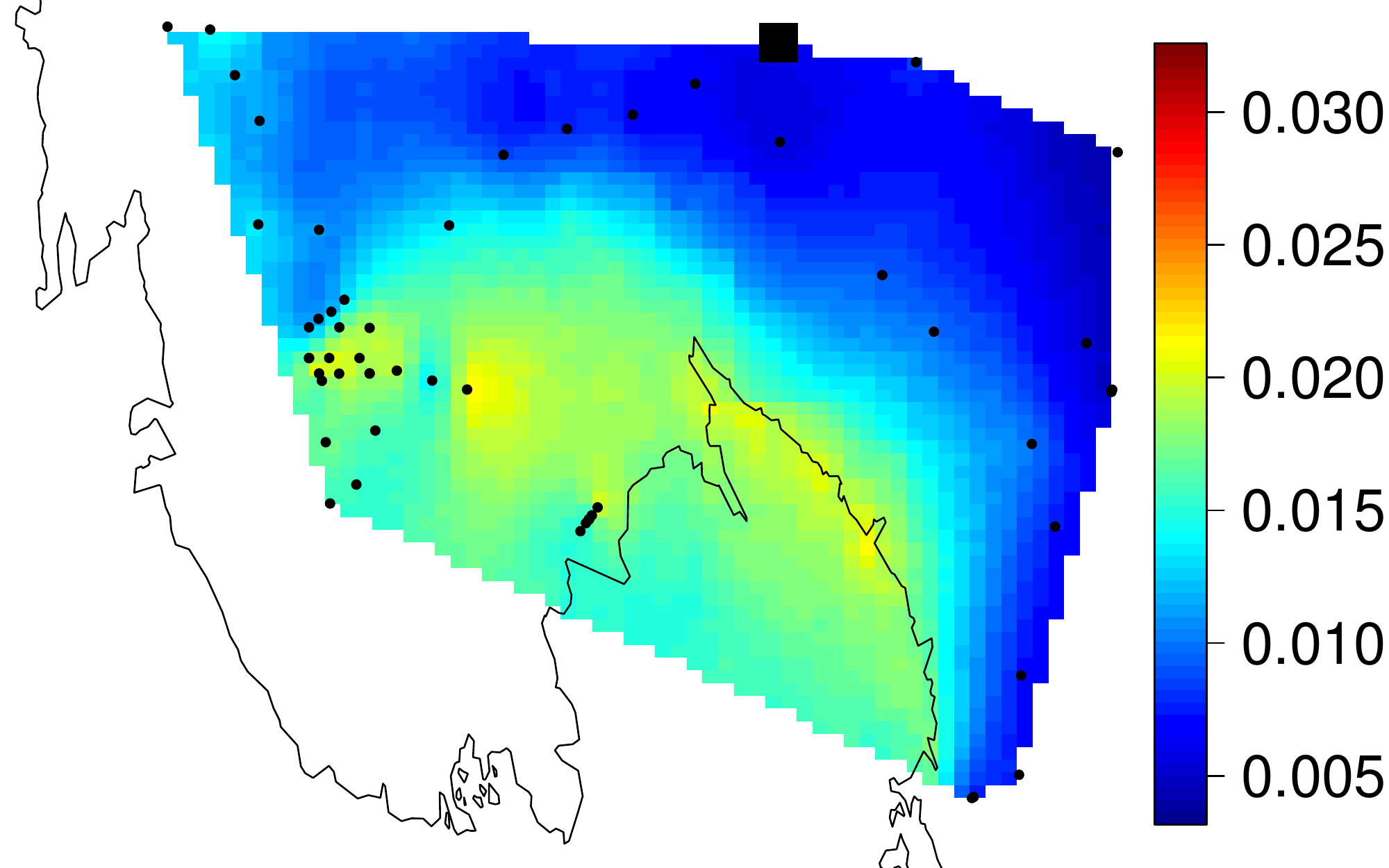}\\
\includegraphics[width=0.24\textwidth]{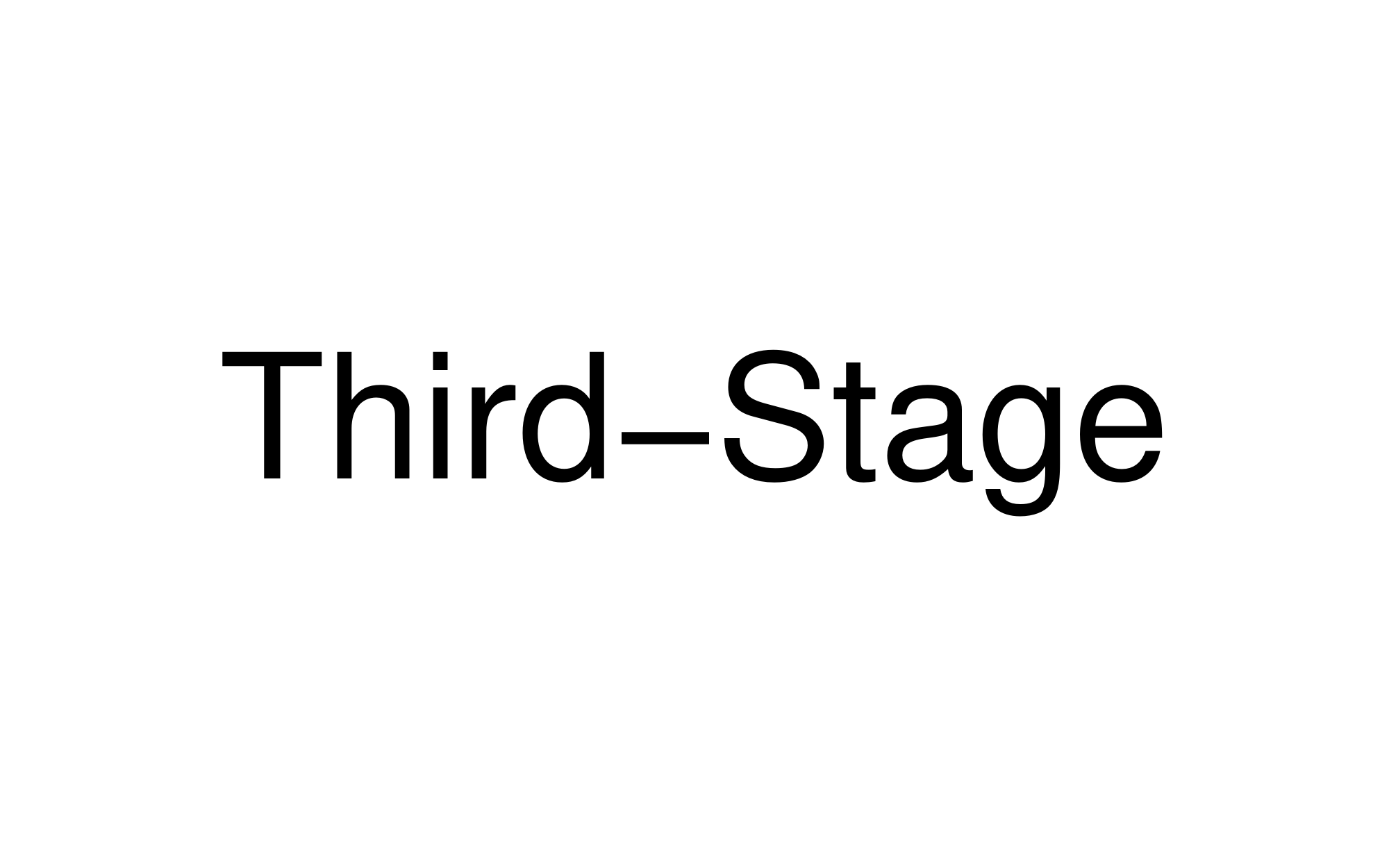}
\includegraphics[width=0.24\textwidth]{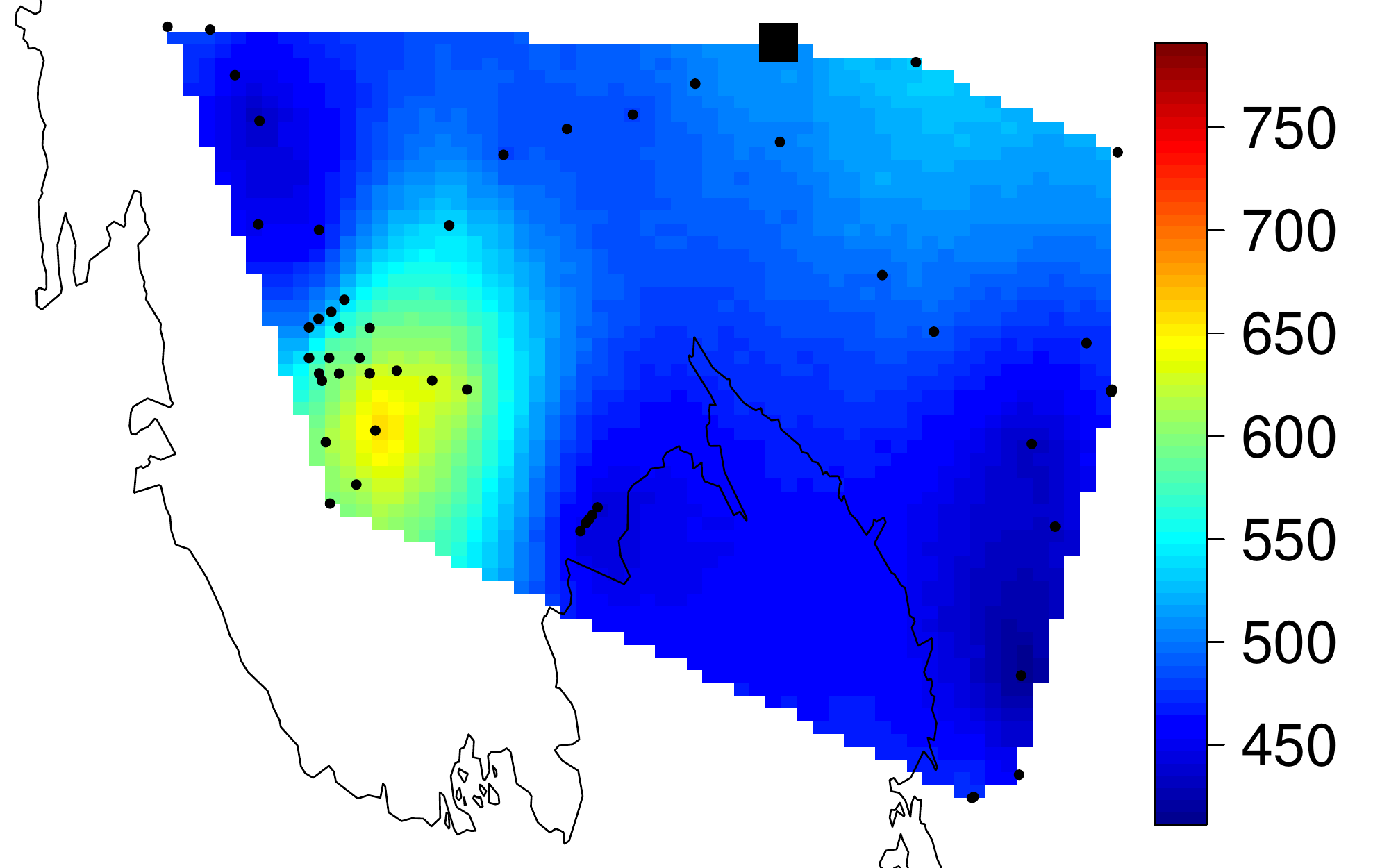}
\includegraphics[width=0.24\textwidth]{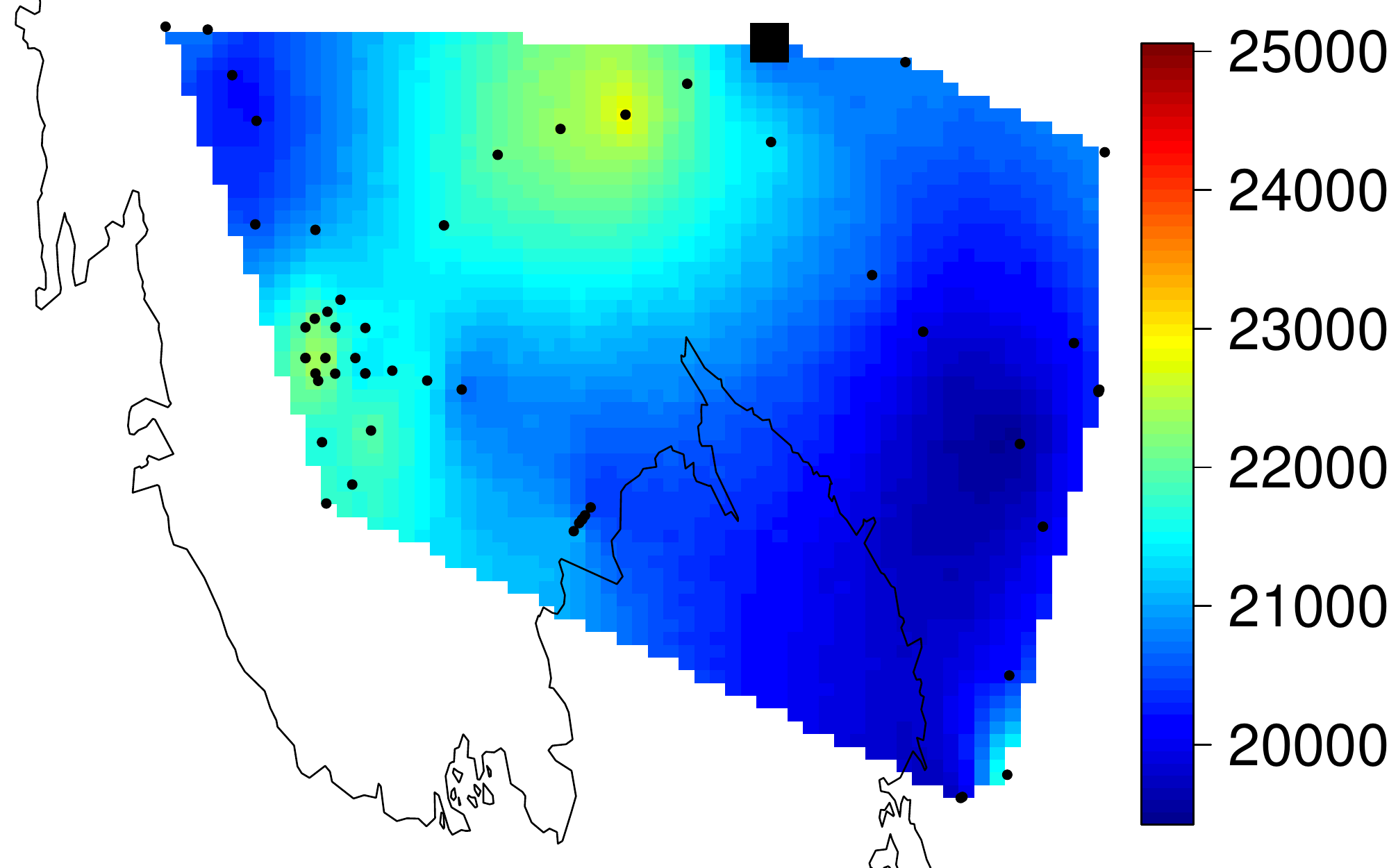}
\includegraphics[width=0.24\textwidth]{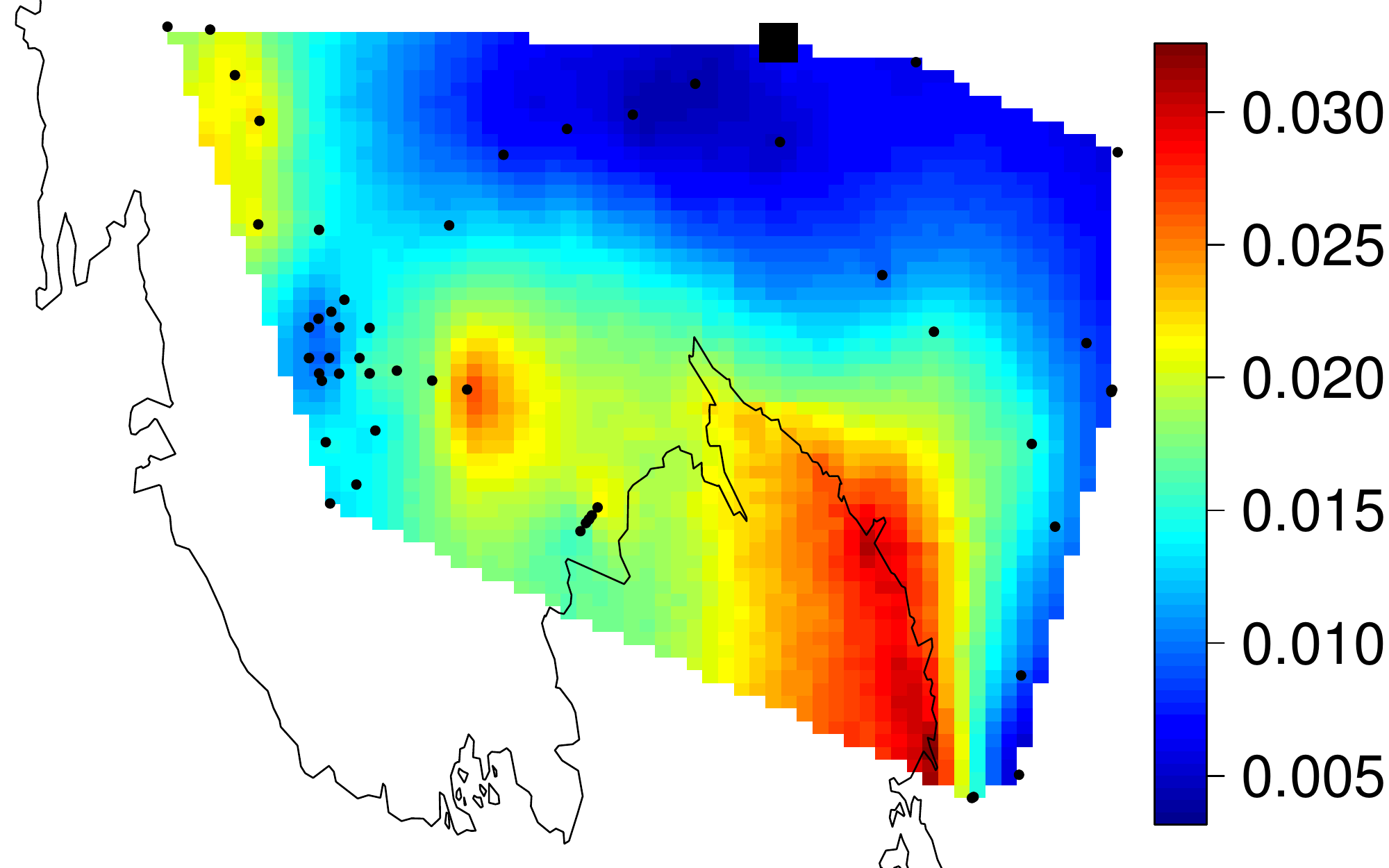}\\
\includegraphics[width=0.24\textwidth]{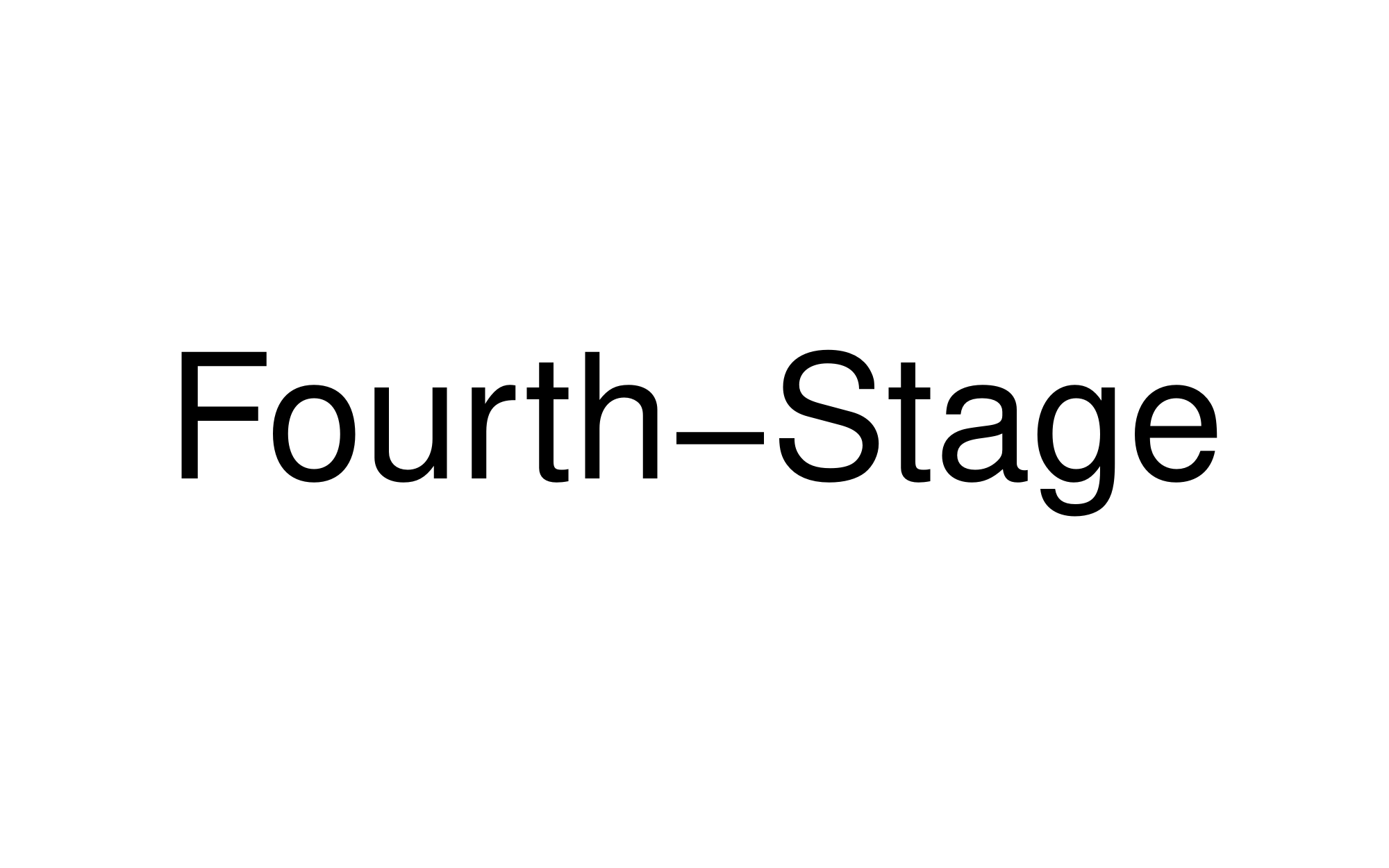}
\includegraphics[width=0.24\textwidth]{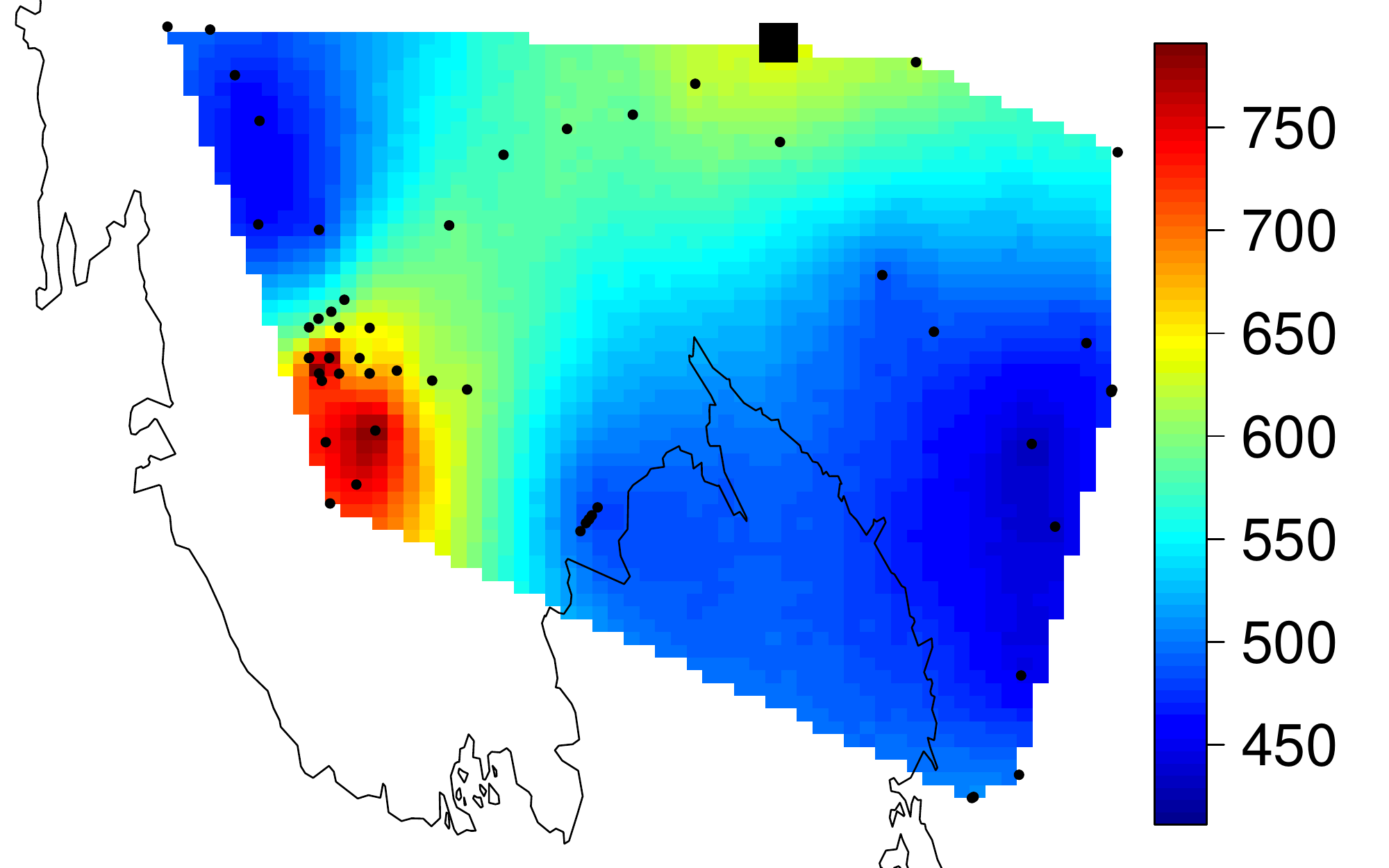}
\includegraphics[width=0.24\textwidth]{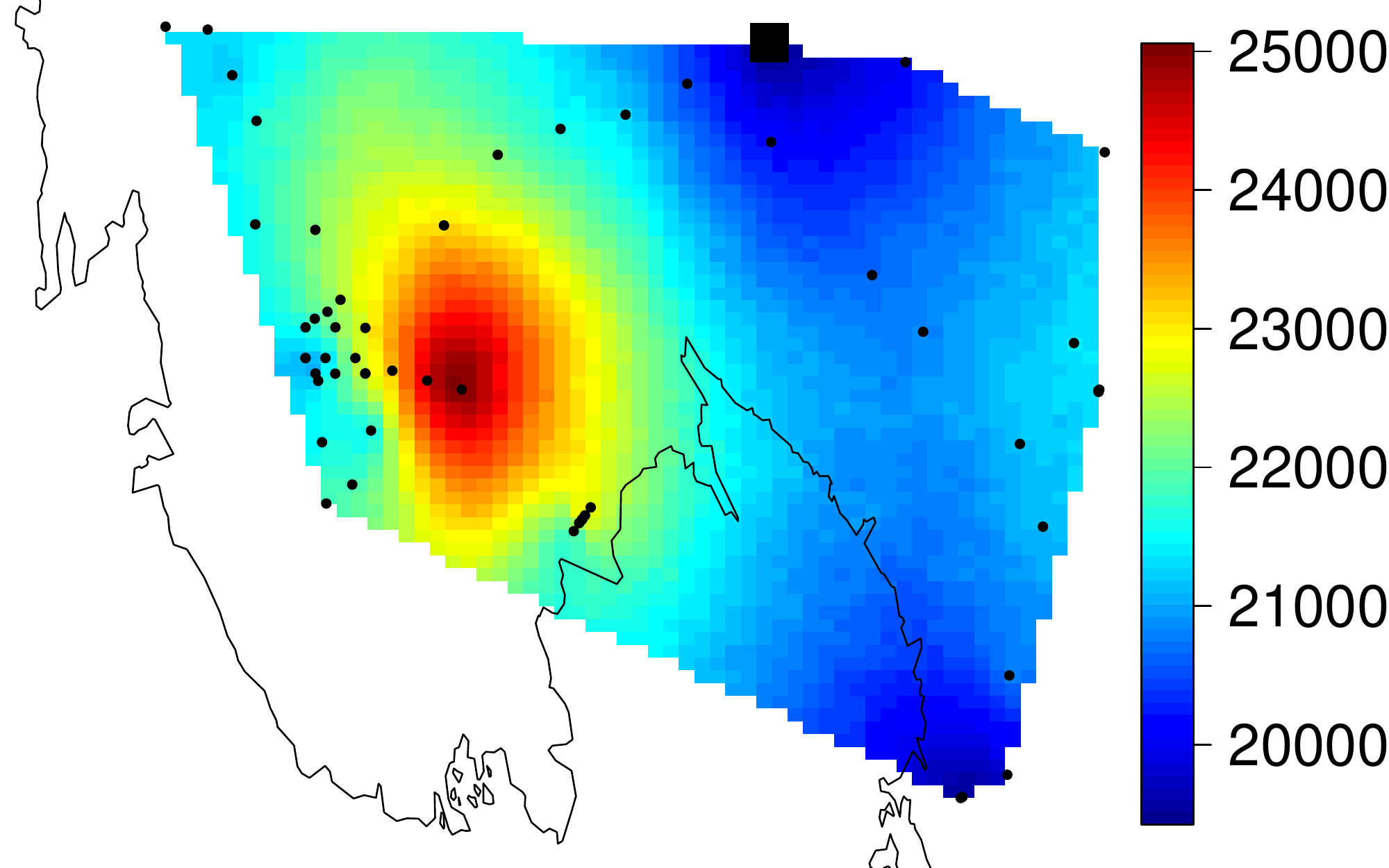}
\includegraphics[width=0.24\textwidth]{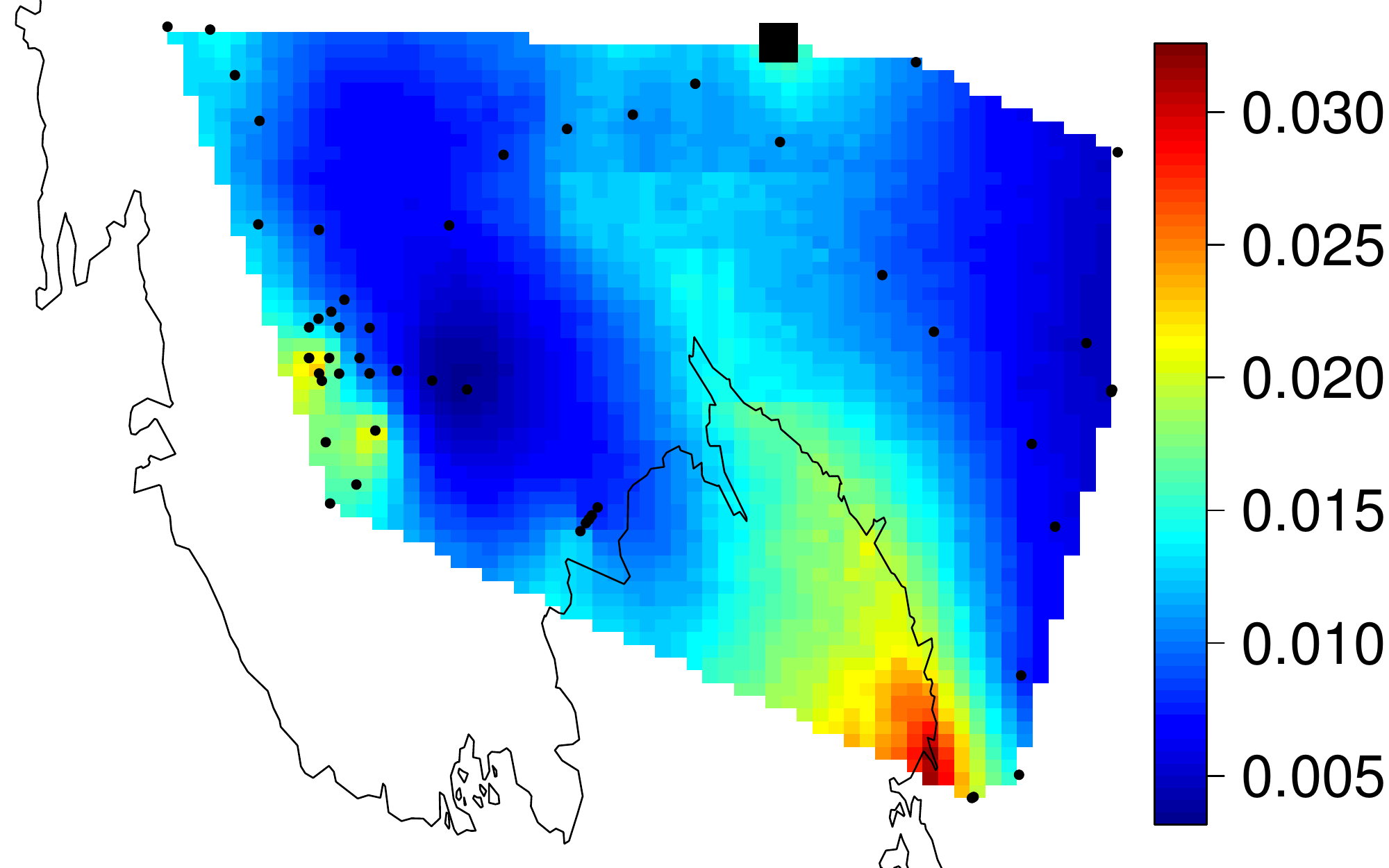}
\end{center}
\caption{(Row 1) Posterior median for surface density and the three critical densities, all in g/cm$^3$, from left to right. (Rows 2-5) Posterior medians for $A_l(\bs)$, $E_l(\bs)$, and $k_l(\bs) = A_l(\bs) \exp\left(-\frac{E_l(\bs)}{RT(\bs)} \right)$, for stages $l = 1,2,3,4$. Second, third, and fourth stages share the same scale to allow visual comparison. In all plots, the square shows the south pole.}\label{fig:spat_pars}
\end{figure}

In Figure \ref{fig:spat_pars}, we plot the posterior median of surface density, three critical densities, and the Arrhenius constants and parameters for the first two densification stages. We use the median because the untransformed parameters have some extreme values. In each plot, we plot the location of observations and note that the uncertainty in our estimates increases as the distance to observations increases. Altogether, these plots show high heterogeneity in the spatially varying parameters. We give measures of uncertainty for all spatially varying parameters in the Supplementary Material. From these uncertainty plots, we see the trends we would expect using a spatial model: uncertainty is higher (i) as the distance to observations increases and (ii) at the boundaries of our data. The spatial patterns in surface and critical densities have distinctly geographical distributions. For example, there is an interesting dipole in surface density centered on West Antarctica, that may be the result of regional patterns in snow accumulation rates and temperatures. Similarly interesting spatial patterns in the critical densities are expressed across the region. Thus, the spatial variability in model parameters has important physical implications not captured in prior models.

In our model, the energies of activation $E_l(\bs)$ quantify the snow density response to temperature for stage $l$, where lower temperature sensitivity is signaled by relatively high values of $E_l(\bs)$. The first and fourth stages of densification appear to be most spatially heterogeneous, looking at $E_1(\bs)$ and $E_4(\bs)$ in Figure \ref{fig:spat_pars}.
In the first stage (near the ice sheet's surface), the strongest estimated snow density responses to temperature (lowest $E_1(\bs)$) appear in three regions: left of the south pole, bottom-left, and middle-right (note that cardinal directions near the South Pole are not intuitive; we therefore refer to locations in Antarctica with reference to the figure orientation---e.g. top/bottom/left/right). The second and third stages appear most similar, showing the highest values of $E_l(\bs)$ (lowest temperature sensitivity) left of the south pole. On the other hand, the fourth stage of densification shows only one regional peak in $E_4(\bs)$ in the left-bottom region. Altogether, our estimates suggest that the West Antarctic Ice Sheets' snow density response to temperature varies over space. The HL model, as a thermal model, assumes temperature dependence as the primary climatic control on snow densification. The spatial variability demonstrated in our results indicates that there are other physical and/or climatic parameters (other than temperature) that influence the densification. These results can potentially be used to determine which physical/climatic variables should be accounted for to improve firn densification models.

\begin{figure}[H]
\begin{center}
\includegraphics[width=0.32\textwidth]{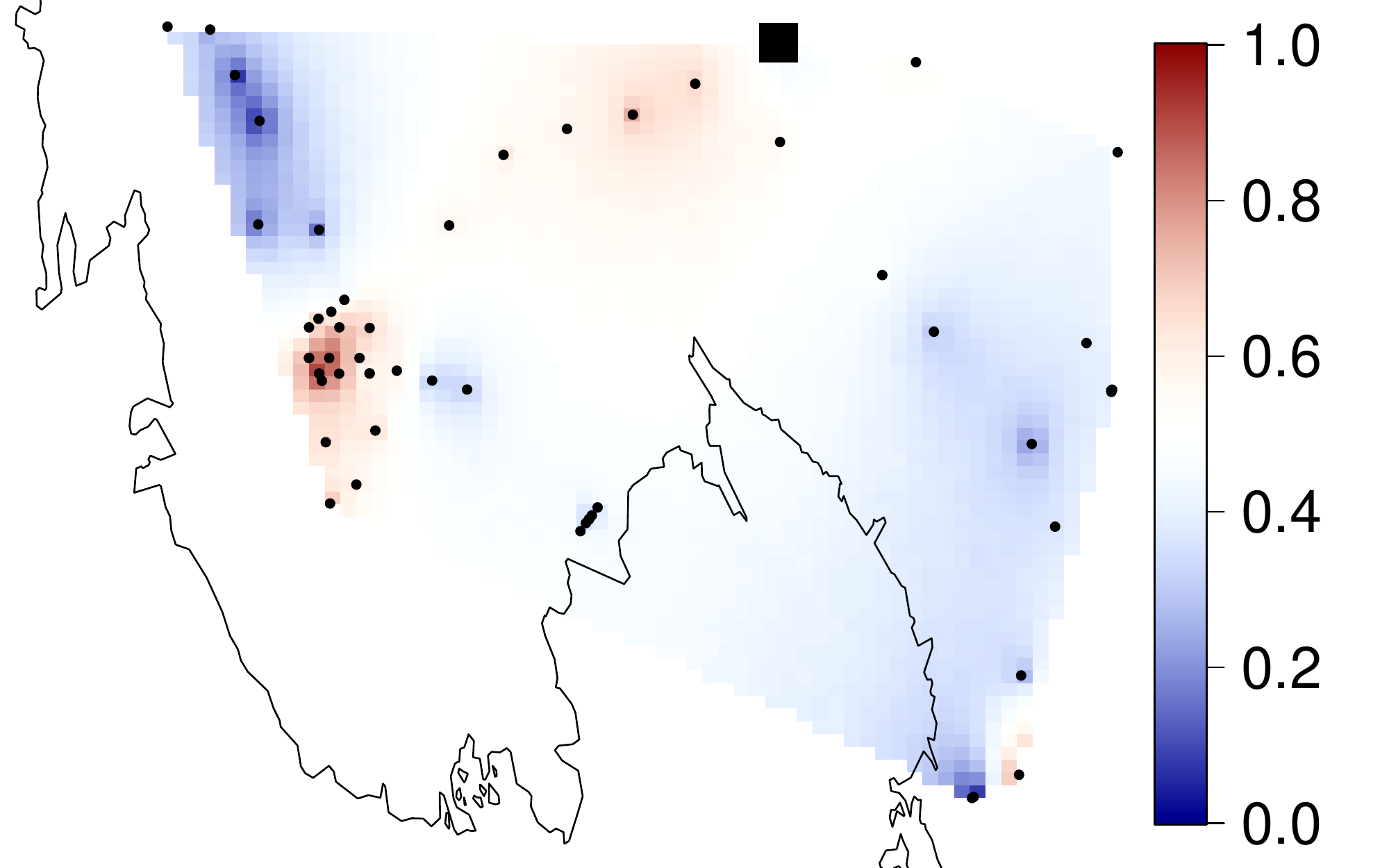}
\includegraphics[width=0.32\textwidth]{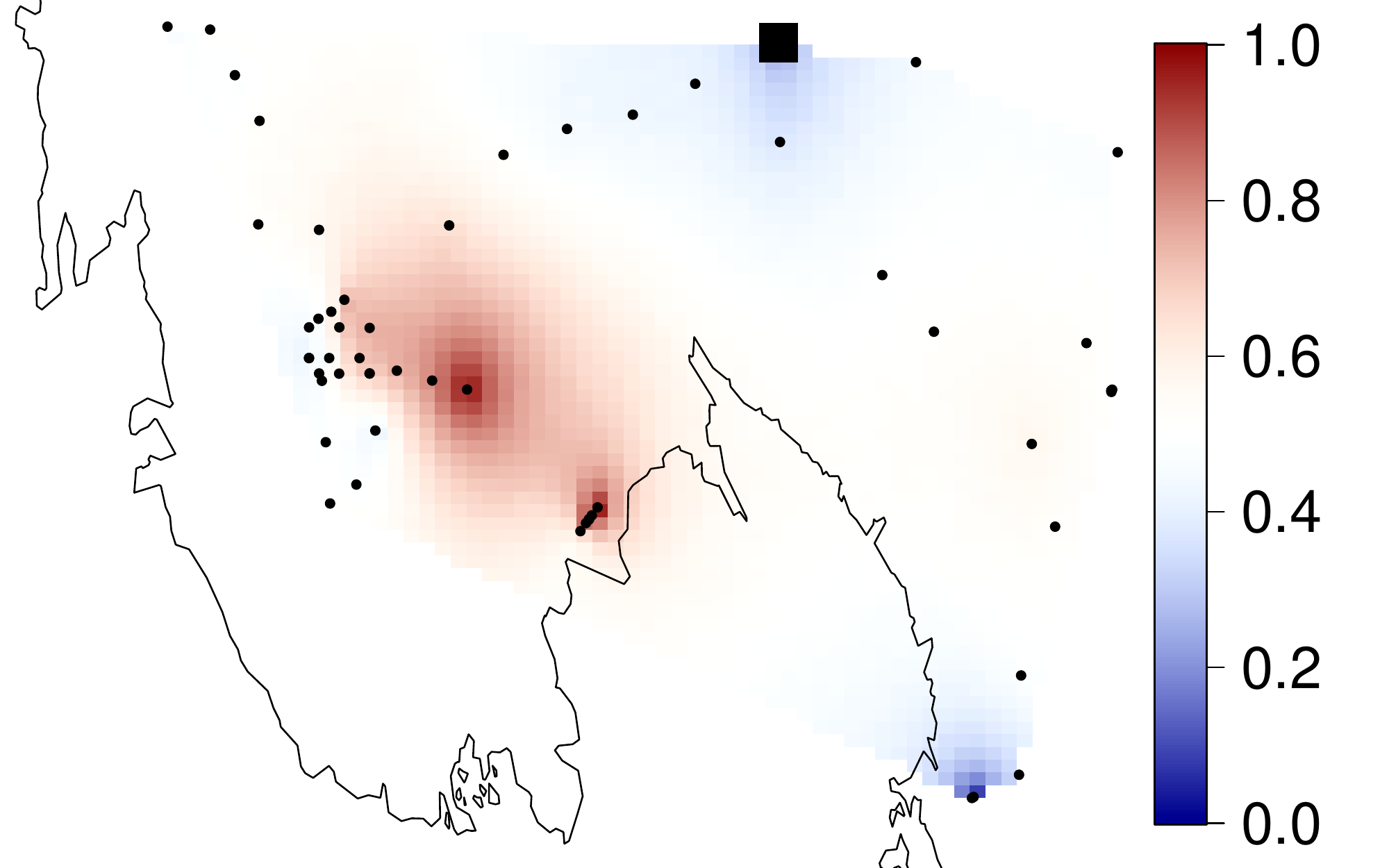}
\includegraphics[width=0.32\textwidth]{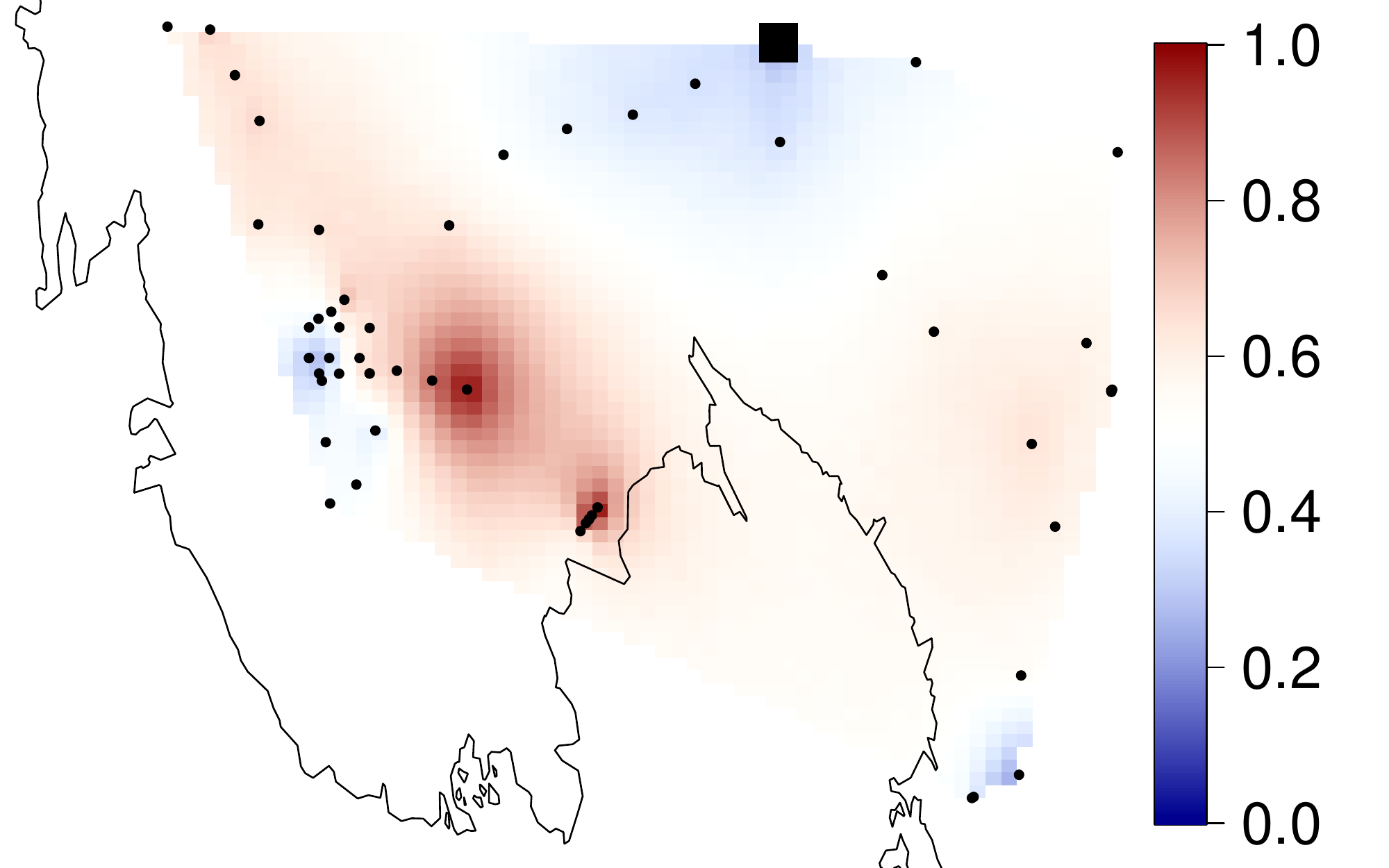}
\end{center}
\caption{Posterior probability of (Left) $k_2(\bs) > k_3(\bs)$, (Center) $k_2(\bs) > k_4(\bs)$, and (Right) $k_3(\bs) > k_4(\bs)$. Both small and large probabilities indicate differences between adjacent stages. The dotted line is the convex hull of our data. In all plots, the square shows the south pole.}\label{fig:post_prob}
\end{figure}

We also follow up on the assertion that stages two, three, and four are not significantly different \citep{horhold_densification_2011} which we assume is true \emph{a priori} in the mean of the hierarchical model. To do this, we compute and plot the posterior probability that $k_2(\bs) > k_3(\bs)$, $k_2(\bs) > k_4(\bs)$, and $k_3(\bs) > k_4(\bs)$ in Figure \ref{fig:post_prob}. In general, the estimated differences appear to be somewhat weak, confirming the suggestion of weak transitions between densification stages at many sites \citep{horhold_densification_2011}. However, we estimate that there are many regions where the densification rates differ across the last three stages of the densification model. This indicates that there is stage-specific spatial heterogeneity in snow densification processes. 
These observations are noteworthy for two reasons.
First, the results indicate that a global assumption of either strong or weak transitions at deeper critical densities is inaccurate for specific locations, resulting in decreased confidence in the final density estimates. 
Second, this stage-specific heterogeneity indicates deficiencies in the current understanding of snow densification mechanisms.
These results therefore provide an effective target for improved understanding and physical modeling capabilities of snow densification in polar regions.

\subsection{Comparison of Snow Density Estimates}\label{sec:comp_curve}

\begin{figure}[H]
\begin{center}
\includegraphics[width=0.4\textwidth]{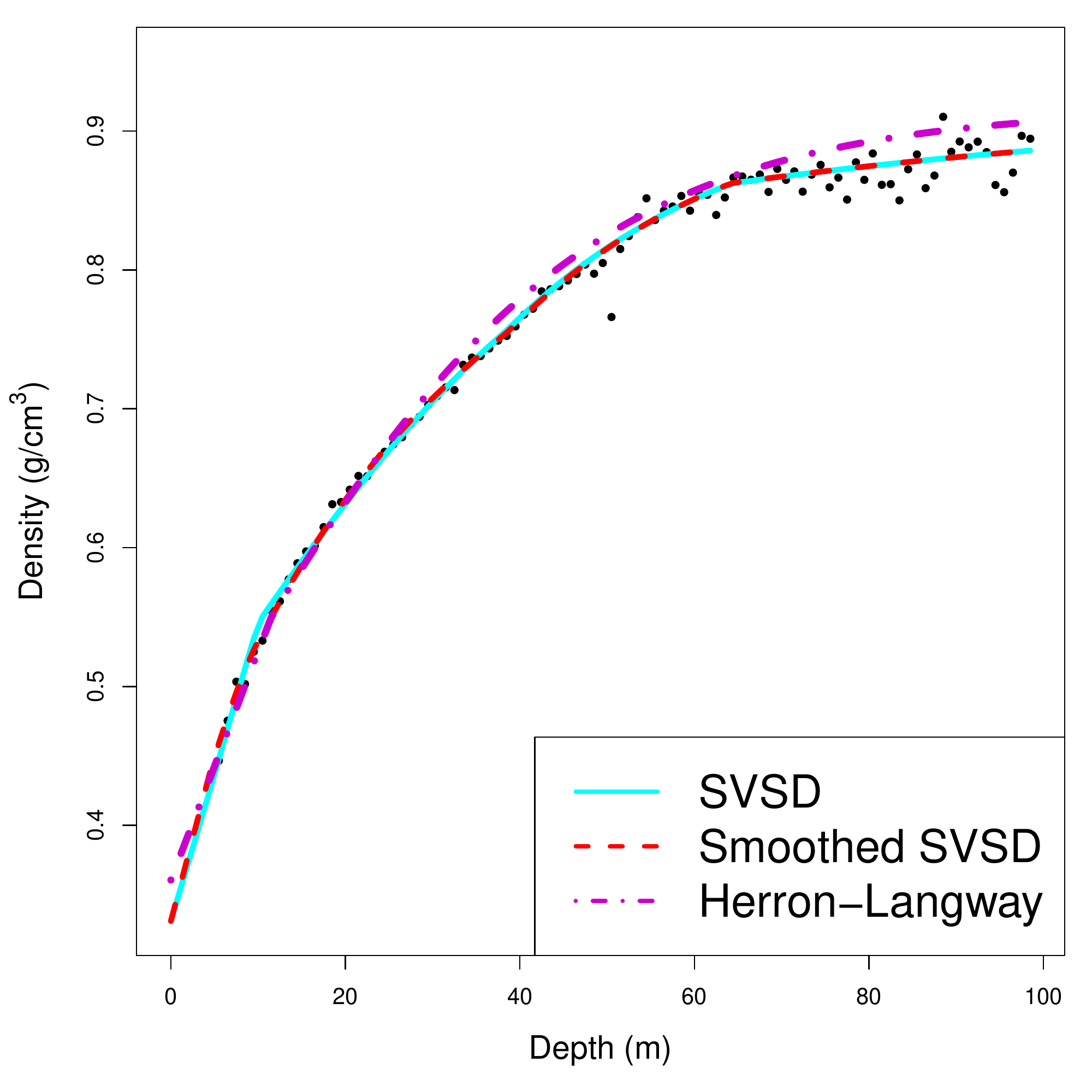}
\includegraphics[width=0.4\textwidth]{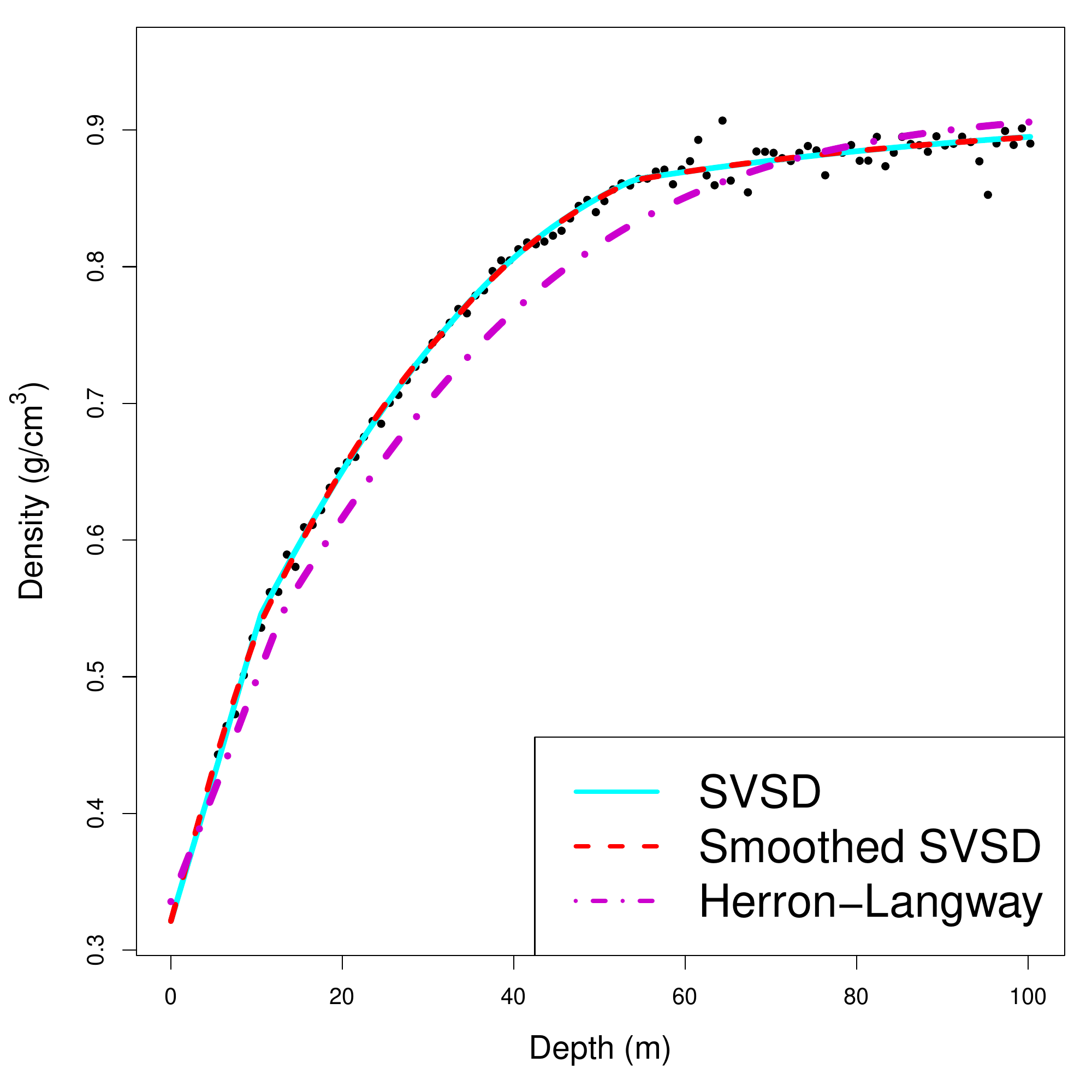}
\includegraphics[width=0.4\textwidth]{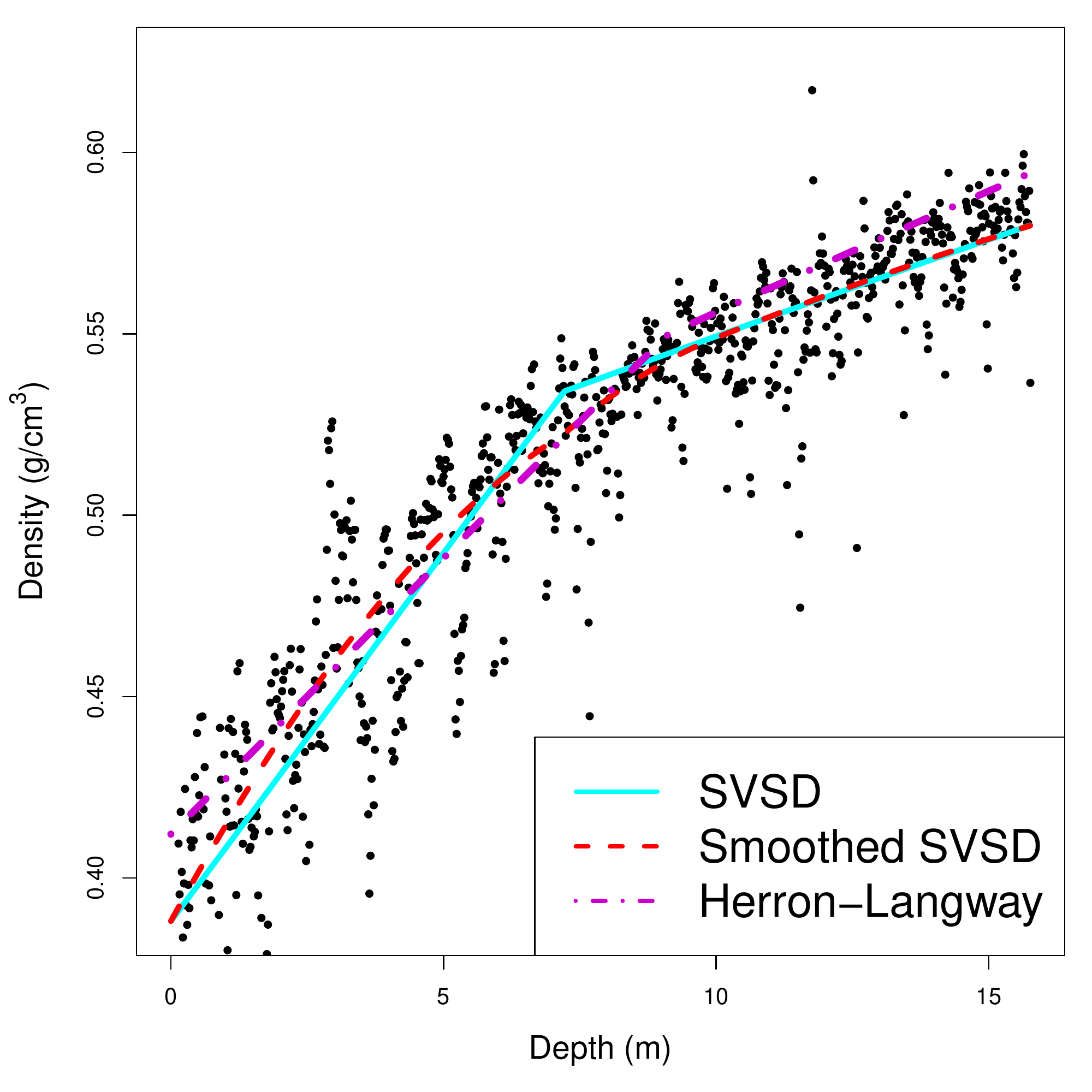}
\includegraphics[width=0.4\textwidth]{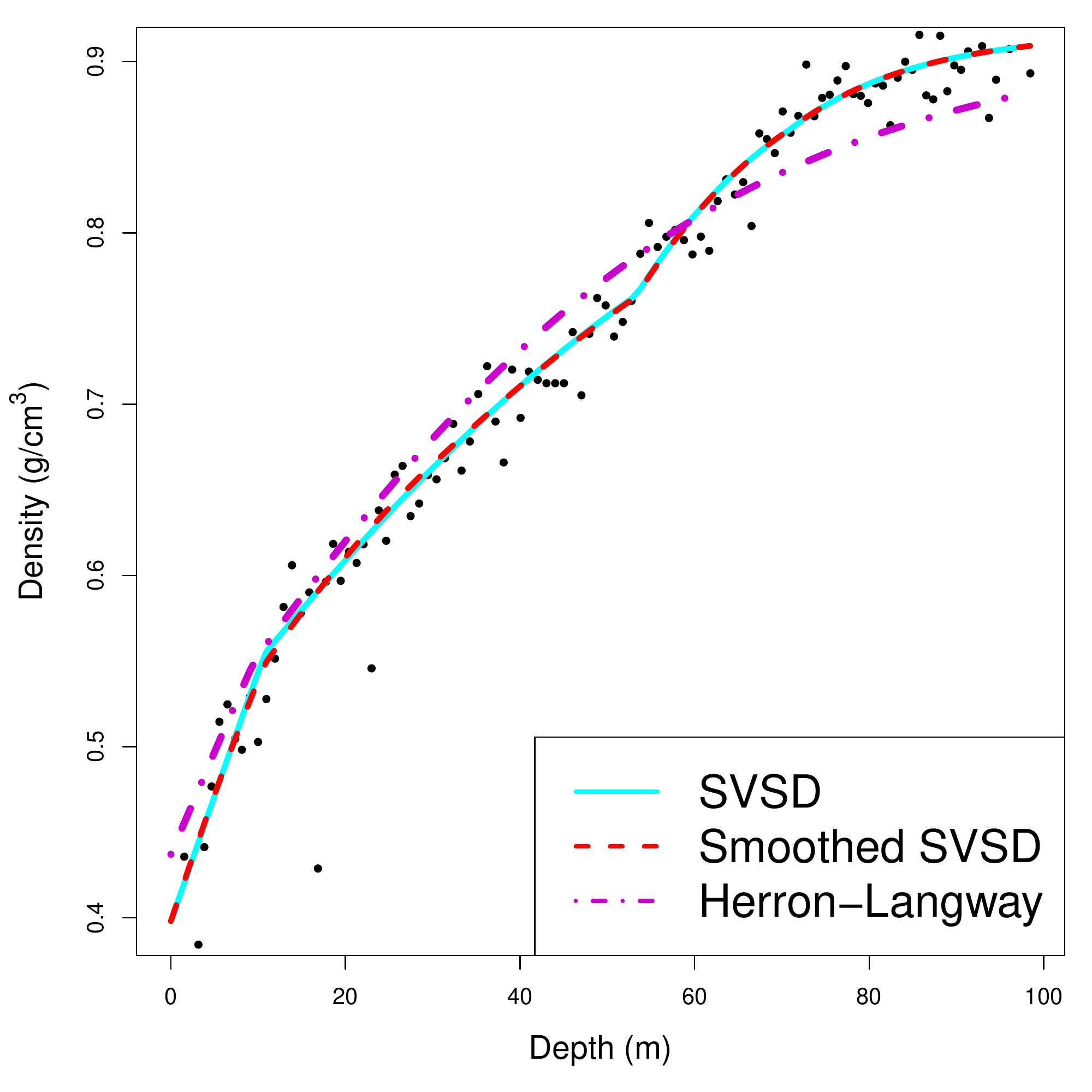}
\end{center}
\caption{A comparison of the HL, SVSD, and the smoothed SVSD models for cores 4, 8, 12, and 49, from top-left to bottom-right. For the SVSD and the smoothed SVSD models, we illustrate the model fit using the posterior mean.}\label{fig:fits}
\end{figure}

To highlight the differences between the original HL model \eqref{eq:herron}, the SVSD model \eqref{eq:herron_spatial}, and the smoothed SVSD model \eqref{eq:smooth_herron}, we revisit the cores discussed in Section \ref{sec:ourdata} and plotted in Figure \ref{fig:herron_langway_fit}. For these four cores, we plot the three model fits in Figure \ref{fig:fits}.  For all cores, there is a clear improvement using both of our proposed models. The smoothed SVSD model is only apparently better for some cores. For the two SDM cores, 4 and 8, there is no evident improvement using the more complicated smoothed SVSD model. However, for core 12, the smoothed SVSD model is much better than the SVSD model, while it is slightly better for core 49. Overall, we find that the smoothed SVSD model was best for the high-resolution, data-rich cores and provided only marginal improvement for low-resolution cores with relatively fewer measurements.
This is likely an effect of the inherent smoothing present in the low-resolution cores compared to the increased variability exhibited by the high-resolution cores.

\section{Conclusions and Future Work}\label{sec:conc}

We analyzed a dataset of 14,844 snow density measurements taken from 57 snow cores drilled at 56 sites in West Antarctica. We proposed a model that allowed physical constants to vary spatially. Our model allowed for deviations from the simple physical model through functional smoothing. To preserve inference on the spatially varying physical constants, we orthogonalized the functional smoothing with respect to the spatially varying span of the physical model. Our model allowed us to explore how important physical quantities vary over West Antarctica while not imposing an overly simple model for these data. Using this model, we explored how estimated snow densification patterns change over space and compared our model to previous studies. 

The proposed model provides important contributions to the polar and ice sheet research communities.
Perhaps most importantly, the model performs spatial modeling of snow densification rate constants, allowing for improved inference on the regional variability of physical constants. 
Specifically, this modeling includes flexibility in the densification rates and density/depth cutoffs between different stages of densification, better capturing the true variability in such parameters.
The model also allows for non-linear deviations in densification rate parameters and is thus able to better handle noisy data, either from measurement errors or physical disequilibrium effects.
Finally, our model includes quantitative bounds of estimate uncertainty.
All of these attributes are valuable for the scientific community, while most current observation-based snow density models include few if any of the above points.
The proposed model, therefore, represents an important step forward in providing spatially-distributed and error-constrained estimates of density with depth in Antarctic snow from \textit{in-situ} data.

This research opens several paths to further advance the state of polar cryospheric science.
Increased understanding of the spatial patterns in density model parameters informs our fundamental knowledge of snow densification processes and can help improve physical modeling efforts in the future.
Several areas of the discipline also rely on spatially-distributed estimates of snow density with depth.
The proposed density modeling framework should be immediately applicable to estimates of annual surface mass balance using radar data, which require snow density as input \citep{keeler_probabilistic_2020}.
Additionally, depth-density estimates are necessary for the conversion of ice sheet thickness changes to mass changes \citep{csatho_laser_2014}. Ice sheet thickness change observations are becoming more prevalent, and at increasingly higher spatial resolution, but fully leveraging new thickness change time series is hampered by the lack of high resolution density data \citep{shepherd_trends_2019}.
Recent investigations of Antarctica also increasingly rely on the outputs of coupled climate models and climate reanalyses to investigate SMB, mass loss, and similar climate changes \citep[see, e.g.][]{lenaerts_new_2012,wessem_modelling_2018}.
The improved prediction of snow/ice densities from the results of this study can provide data for radar inference, converting ice sheet thickness changes to mass changes, and assessment of  climate model performance across West Antarctica.

Due to the large commitment and expense required for \textit{in-situ} data collection in Antarctica, the careful selection of ice/snow coring sites is essential for expedition planning and execution. 
We, therefore, plan to pursue a follow-up project proposing a spatial design method for selecting where and how deep cores should be drilled.
As more data become available, we speculate that richer spatial models will become increasingly beneficial. More recently, snow cores are measured using high-resolution approaches that yield $\approx$50,000 density measurements per core. With such cores, smoothed HL models become computationally prohibitive; thus, we suggest exploring appropriate approximations for this framework. Future extensions could also explore stochastic differential equation models for snow density.

Although the work here is motivated by snow density modeling, our approach could be applied in other domains that make use of interpretable piecewise constant or linear models to infer trends in, for example, climate parameters \citep[see, e.g.,][]{pawson1998,tome2004piecewise,seidel2004assessment,tebaldi2009joint,schofield_model-based_2016,banesh_comparison_2019}.
Additional examples of piecewise constant or linear models are present in soil water content modeling \citep{chen_using_2014,koster_using_2020}, estimating effects of climate change on food production \citep{katz1977assessing,tebaldi2008towards,ortiz2018another}, understanding human impacts on landscape ecology \citep{horner2009mortality,howarth2012nitrogen}, hydrological predictions and risk assessments \citep{das2018streamflow,dai_flood-risk_2019}, and quantifying environmental impacts on economies \citep{engle1986semiparametric,du2017impact}.


\section*{Acknowledgment}

We thank the two anonymous reviewers, as well as the Associate Editor, for their comments that have helped improve the manuscript. Summer Rupper acknowledges funding from NASA grant NNX16AJ72G. 

\section*{Supplementary Material}

In the Supplementary Material, we present extended posterior analysis, including between-parameter coefficient correlations, violin plots for all spatially varying parameters, and interquartile ranges for the spatially varying parameters plotted over space. Code and data have also been included as Supplementary Material. 

\bibliographystyle{apalike}
\bibliography{ref}

\begin{thebibliography}{}

\bibitem[Albert et~al., 2004]{albert_extreme_2004}
Albert, M., Shuman, C., Courville, Z., Bauer, R., Fahnestock, M., and Scambos,
  T. (2004).
\newblock Extreme firn metamorphism: Impact of decades of vapor transport on
  near-surface firn at a low-accumulation glazed site on the {East} {Antarctic}
  plateau.
\newblock {\em Annals of Glaciology}, 39:73--78.

\bibitem[Alley et~al., 1982]{alley1982}
Alley, R.~B., Bolzan, J.~F., and Whillans, I.~M. (1982).
\newblock Polar firn densification and grain growth.
\newblock {\em Annals of Glaciology}, 3:7--11.

\bibitem[{\'A}lvarez and Lawrence, 2011]{alvarez2011}
{\'A}lvarez, M.~A. and Lawrence, N.~D. (2011).
\newblock Computationally efficient convolved multiple output {G}aussian
  processes.
\newblock {\em Journal of Machine Learning Research}, 12(May):1459--1500.

\bibitem[Arthern et~al., 2010]{arthern_situ_2010}
Arthern, R.~J., Vaughan, D.~G., Rankin, A.~M., Mulvaney, R., and Thomas, E.~R.
  (2010).
\newblock In situ measurements of {Antarctic} snow compaction compared with
  predictions of models.
\newblock {\em Journal of Geophysical Research: Earth Surface}, 115(F3).
\newblock \_eprint:
  https://agupubs.onlinelibrary.wiley.com/doi/pdf/10.1029/2009JF001306.

\bibitem[Banerjee et~al., 2014]{banerjee2014}
Banerjee, S., Carlin, B.~P., and Gelfand, A.~E. (2014).
\newblock {\em Hierarchical modeling and analysis for spatial data}.
\newblock CRC press.

\bibitem[Banesh et~al., 2019]{banesh_comparison_2019}
Banesh, D., Petersen, M., Wendelberger, J., Ahrens, J., and Hamann, B. (2019).
\newblock Comparison of piecewise linear change point detection with
  traditional analytical methods for ocean and climate data.
\newblock {\em Environmental Earth Sciences}, 78(21):623.

\bibitem[Bohlander and Scambos, 2001]{bohlander2001}
Bohlander, J. and Scambos, T. (2001).
\newblock Thermap {A}ntarctic ice sheet temperature data.
\newblock {\em Boulder, CO: National Snow and Ice Data Centre}.

\bibitem[Burgener et~al., 2013]{burgener_observed_2013}
Burgener, L., Rupper, S., Koenig, L., Forster, R., Christensen, W.~F.,
  Williams, J., Koutnik, M., Miege, C., Steig, E.~J., Tingey, D., Keeler, D.,
  and Riley, L. (2013).
\newblock An observed negative trend in {West} {Antarctic} accumulation rates
  from 1975 to 2010: {Evidence} from new observed and simulated records.
\newblock {\em Journal of Geophysical Research-Atmospheres},
  118(10):4205--4216.

\bibitem[Chen et~al., 2014]{chen_using_2014}
Chen, T., de~Jeu, R. A.~M., Liu, Y.~Y., van~der Werf, G.~R., and Dolman, A.~J.
  (2014).
\newblock Using satellite based soil moisture to quantify the water driven
  variability in {NDVI}: {A} case study over mainland {Australia}.
\newblock {\em Remote Sensing of Environment}, 140:330--338.

\bibitem[Cressie and Wikle, 2015]{cressie2015statistics}
Cressie, N. and Wikle, C.~K. (2015).
\newblock {\em Statistics for spatio-temporal data}.
\newblock John Wiley \& Sons.

\bibitem[Csatho et~al., 2014]{csatho_laser_2014}
Csatho, B.~M., Schenk, A.~F., Veen, C. J. v.~d., Babonis, G., Duncan, K.,
  Rezvanbehbahani, S., Broeke, M. R. v.~d., Simonsen, S.~B., Nagarajan, S., and
  Angelen, J. H.~v. (2014).
\newblock Laser altimetry reveals complex pattern of {Greenland} {Ice} {Sheet}
  dynamics.
\newblock {\em Proceedings of the National Academy of Sciences},
  111(52):18478--18483.
\newblock Publisher: National Academy of Sciences Section: Physical Sciences.

\bibitem[Dai et~al., 2019]{dai_flood-risk_2019}
Dai, L., Zhou, J., Chen, L., Huang, K., Wang, Q., and Zha, G. (2019).
\newblock Flood-risk analysis based on a stochastic differential equation
  method.
\newblock {\em Journal of Flood Risk Management}, 12(S1):e12515.
\newblock \_eprint: https://onlinelibrary.wiley.com/doi/pdf/10.1111/jfr3.12515.

\bibitem[Das~Bhowmik et~al., 2018]{das2018streamflow}
Das~Bhowmik, R., Seo, S.~B., and Sahoo, S. (2018).
\newblock Streamflow simulation using bayesian regression with multivariate
  linear spline to estimate future changes.
\newblock {\em Water}, 10(7):875.

\bibitem[DeConto and Pollard, 2016]{deconto2016contribution}
DeConto, R.~M. and Pollard, D. (2016).
\newblock Contribution of antarctica to past and future sea-level rise.
\newblock {\em Nature}, 531(7596):591--597.

\bibitem[Du et~al., 2017]{du2017impact}
Du, D., Zhao, X., and Huang, R. (2017).
\newblock The impact of climate change on developed economies.
\newblock {\em Economics Letters}, 153:43--46.

\bibitem[Eisen et~al., 2008]{eisen_ground-based_2008}
Eisen, O., Frezzotti, M., Genthon, C., Isaksson, E., Magand, O., van~den
  Broeke, M.~R., Dixon, D.~A., Ekaykin, A., Holmlund, P., Kameda, T., Karlof,
  L., Kaspari, S., Lipenkov, V.~Y., Oerter, H., Takahashi, S., and Vaughan,
  D.~G. (2008).
\newblock Ground-based measurements of spatial and temporal variability of snow
  accumulation in {East} {Antarctica}.
\newblock {\em Reviews of Geophysics}, 46(1):RG2001.
\newblock WOS:000255079700001.

\bibitem[Engle et~al., 1986]{engle1986semiparametric}
Engle, R.~F., Granger, C.~W., Rice, J., and Weiss, A. (1986).
\newblock Semiparametric estimates of the relation between weather and
  electricity sales.
\newblock {\em Journal of the American statistical Association},
  81(394):310--320.

\bibitem[Freitag et~al., 2004]{freitag2004}
Freitag, J., Wilhelms, F., and Kipfstuhl, S. (2004).
\newblock Microstructure-dependent densification of polar firn derived from
  x-ray microtomography.
\newblock {\em Journal of Glaciology}, 50(169):243--250.

\bibitem[Gelfand et~al., 2003]{gelfand2003}
Gelfand, A.~E., Kim, H.-J., Sirmans, C., and Banerjee, S. (2003).
\newblock Spatial modeling with spatially varying coefficient processes.
\newblock {\em Journal of the American Statistical Association},
  98(462):387--396.

\bibitem[Genton and Kleiber, 2015]{genton2015}
Genton, M.~G. and Kleiber, W. (2015).
\newblock Cross-covariance functions for multivariate geostatistics.
\newblock {\em Statistical Science}, pages 147--163.

\bibitem[Gneiting et~al., 2010]{gneiting2010}
Gneiting, T., Kleiber, W., and Schlather, M. (2010).
\newblock Mat{\'e}rn cross-covariance functions for multivariate random fields.
\newblock {\em Journal of the American Statistical Association},
  105(491):1167--1177.

\bibitem[Gow, 1975]{gow1975}
Gow, A.~J. (1975).
\newblock Time-temperature dependence of sintering in perennial isothermal
  snowpacks.
\newblock {\em International Association of Hydrological Sciences
  Publications}, 114:25--41.

\bibitem[Griffith, 2003]{griffith2003spatial}
Griffith, D.~A. (2003).
\newblock {\em Spatial autocorrelation and spatial filtering: gaining
  understanding through theory and scientific visualization}.
\newblock Springer Science \& Business Media.

\bibitem[Grzebyk and Wackernagel, 1994]{grzebyk1994}
Grzebyk, M. and Wackernagel, H. (1994).
\newblock Multivariate analysis and spatial/temporal scales: real and complex
  models.
\newblock In {\em Proceedings of the XVIIth International Biometrics
  Conference}, volume~1, pages 19--33. Citeseer.

\bibitem[Haario et~al., 2001]{haario2001}
Haario, H., Saksman, E., and Tamminen, J. (2001).
\newblock An adaptive {M}etropolis algorithm.
\newblock {\em Bernoulli}, 7(2):223--242.

\bibitem[Helsen et~al., 2008]{helsen_elevation_2008}
Helsen, M.~M., Broeke, M. R. v.~d., Wal, R. S. W. v.~d., Berg, W. J. v.~d.,
  Meijgaard, E.~v., Davis, C.~H., Li, Y., and Goodwin, I. (2008).
\newblock Elevation {Changes} in {Antarctica} {Mainly} {Determined} by
  {Accumulation} {Variability}.
\newblock {\em Science}, 320(5883):1626--1629.
\newblock Publisher: American Association for the Advancement of Science
  Section: Report.

\bibitem[Herron and Langway, 1980]{herron_firn_1980}
Herron, M.~M. and Langway, C.~C. (1980).
\newblock Firn densification: An empirical model.
\newblock {\em Journal of Glaciology}, 25(93):373--385.

\bibitem[Hock et~al., 2009]{hock2009mountain}
Hock, R., de~Woul, M., Radi{\'c}, V., and Dyurgerov, M. (2009).
\newblock Mountain glaciers and ice caps around antarctica make a large
  sea-level rise contribution.
\newblock {\em Geophysical Research Letters}, 36(7).

\bibitem[Hodges and Reich, 2010]{hodges2010}
Hodges, J.~S. and Reich, B.~J. (2010).
\newblock Adding spatially-correlated errors can mess up the fixed effect you
  love.
\newblock {\em The American Statistician}, 64(4):325--334.

\bibitem[H{\"o}rhold et~al., 2011]{horhold_densification_2011}
H{\"o}rhold, M.~W., Kipfstuhl, S., Wilhelms, F., Freitag, J., and Frenzel, A.
  (2011).
\newblock The densification of layered polar firn.
\newblock {\em Journal of Geophysical Research: Earth Surface}, 116(F1).

\bibitem[Horner et~al., 2009]{horner2009mortality}
Horner, G.~J., Baker, P.~J., Mac~Nally, R., Cunningham, S.~C., Thomson, J.~R.,
  and Hamilton, F. (2009).
\newblock Mortality of developing floodplain forests subjected to a drying
  climate and water extraction.
\newblock {\em Global Change Biology}, 15(9):2176--2186.

\bibitem[Howarth et~al., 2012]{howarth2012nitrogen}
Howarth, R., Swaney, D., Billen, G., Garnier, J., Hong, B., Humborg, C.,
  Johnes, P., M{\"o}rth, C.-M., and Marino, R. (2012).
\newblock Nitrogen fluxes from the landscape are controlled by net
  anthropogenic nitrogen inputs and by climate.
\newblock {\em Frontiers in Ecology and the Environment}, 10(1):37--43.

\bibitem[Hughes and Haran, 2013]{hughes2013dimension}
Hughes, J. and Haran, M. (2013).
\newblock Dimension reduction and alleviation of confounding for spatial
  generalized linear mixed models.
\newblock {\em Journal of the Royal Statistical Society: Series B (Statistical
  Methodology)}, 75(1):139--159.

\bibitem[{IPCC}, 2013]{ipcc_climate_2013}
{IPCC} (2013).
\newblock {\em Climate {Change} 2013: {The} {Physical} {Science} {Basis}.
  {Contribution} of {Working} {Group} {I} to the {Fifth} {Assessment} {Report}
  of the {Intergovernmental} {Panel} on {Climate} {Change}}.
\newblock Cambridge University Press, Cambridge, United Kingdom and New York,
  NY, USA.

\bibitem[Johnson, 1998]{johnson1998}
Johnson, J.~B. (1998).
\newblock A preliminary numerical investigation of the micromechanics of snow
  compaction.
\newblock {\em Annals of Glaciology}, 26:51--54.

\bibitem[Katz, 1977]{katz1977assessing}
Katz, R.~W. (1977).
\newblock Assessing the impact of climatic change on food production.
\newblock {\em Climatic Change}, 1(1):85--96.

\bibitem[Keeler et~al., 2020]{keeler_probabilistic_2020}
Keeler, D.~G., Rupper, S.~B., Forster, R., and Miège, C. (2020).
\newblock A {Probabilistic} {Automated} {Isochrone} {Picking} {Routine} to
  {Derive} {Annual} {Surface} {Mass} {Balance} {From} {Radar} {Echograms}.
\newblock {\em IEEE Transactions on Geoscience and Remote Sensing}, pages
  1--14.

\bibitem[Keenan et~al., 2021]{keenan2021physics}
Keenan, E., Wever, N., Dattler, M., Lenaerts, J., Medley, B., Kuipers~Munneke,
  P., and Reijmer, C. (2021).
\newblock Physics-based snowpack model improves representation of near-surface
  antarctic snow and firn density.
\newblock {\em The Cryosphere}, 15(2):1065--1085.

\bibitem[Kleiber and Nychka, 2012]{kleiber2012}
Kleiber, W. and Nychka, D. (2012).
\newblock Nonstationary modeling for multivariate spatial processes.
\newblock {\em Journal of Multivariate Analysis}, 112:76--91.

\bibitem[Koenig et~al., 2016]{koenig_annual_2016}
Koenig, L.~S., Ivanoff, A., Alexander, P.~M., MacGregor, J.~A., Fettweis, X.,
  Panzer, B., Paden, J.~D., Forster, R.~R., Das, I., McConnell, J.~R., Tedesco,
  M., Leuschen, C., and Gogineni, P. (2016).
\newblock Annual {Greenland} accumulation rates (2009-2012) from airborne snow
  radar.
\newblock {\em Cryosphere}, 10(4):1739--1752.
\newblock WOS:000381218000016.

\bibitem[Koster et~al., 2020]{koster_using_2020}
Koster, R.~D., Schubert, S.~D., DeAngelis, A.~M., Molod, A.~M., and Mahanama,
  S.~P. (2020).
\newblock Using a {Simple} {Water} {Balance} {Framework} to {Quantify} the
  {Impact} of {Soil} {Moisture} {Initialization} on {Subseasonal}
  {Evapotranspiration} and {Air} {Temperature} {Forecasts}.
\newblock {\em Journal of Hydrometeorology}, 21(8):1705--1722.
\newblock Publisher: American Meteorological Society.

\bibitem[Lamorey and Cooper, 2002]{lamorey_waiscores}
Lamorey, G. and Cooper, T. (2002).
\newblock Siple dome waiscores density data for shallow cores.

\bibitem[Lenaerts et~al., 2012]{lenaerts_new_2012}
Lenaerts, J. T.~M., van~den Broeke, M.~R., van~de Berg, W.~J., van Meijgaard,
  E., and Munneke, P.~K. (2012).
\newblock A new, high-resolution surface mass balance map of {Antarctica}
  (1979-2010) based on regional atmospheric climate modeling.
\newblock {\em Geophysical Research Letters}, 39:L04501.
\newblock WOS:000300794400004.

\bibitem[Lewis et~al., 2019]{lewis_recent_2019}
Lewis, G., Osterberg, E., Hawley, R., Marshall, H.~P., Meehan, T., Graeter, K.,
  McCarthy, F., Overly, T., Thundercloud, Z., and Ferris, D. (2019).
\newblock Recent precipitation decrease across the western {Greenland} ice
  sheet percolation zone.
\newblock {\em The Cryosphere}, 13(11):2797--2815.

\bibitem[Li and Zwally, 2011]{li_modeling_2011}
Li, J. and Zwally, H.~J. (2011).
\newblock Modeling of firn compaction for estimating ice-sheet mass change from
  observed ice-sheet elevation change.
\newblock {\em Annals of Glaciology}, 52(59):1--7.

\bibitem[Lindgren et~al., 2011]{lindgren2011explicit}
Lindgren, F., Rue, H., and Lindstr{\"o}m, J. (2011).
\newblock An explicit link between gaussian fields and gaussian markov random
  fields: the stochastic partial differential equation approach.
\newblock {\em Journal of the Royal Statistical Society: Series B (Statistical
  Methodology)}, 73(4):423--498.

\bibitem[Maeno and Ebinuma, 1983]{maeno1983}
Maeno, N. and Ebinuma, T. (1983).
\newblock Pressure sintering of ice and its implication to the densification of
  snow at polar glaciers and ice sheets.
\newblock {\em The Journal of Physical Chemistry}, 87(21):4103--4110.

\bibitem[Martinerie et~al., 1992]{martinerie1992}
Martinerie, P., Raynaud, D., Etheridge, D.~M., Barnola, J.-M., and Mazaudier,
  D. (1992).
\newblock Physical and climatic parameters which influence the air content in
  polar ice.
\newblock {\em Earth and Planetary Science Letters}, 112(1-4):1--13.

\bibitem[Matheron, 1982]{matheron1982}
Matheron, G. (1982).
\newblock Pour une analyse krigeante des donn{\'e}es r{\'e}gionalis{\'e}es.
\newblock {\em Centre de G{\'e}ostatistique, Report N-732, Fontainebleau}.

\bibitem[Mayewski et~al., 2005]{mayewski_international_2005}
Mayewski, P.~A., Frezzotti, M., Bertler, N., Ommen, T.~V., Hamilton, G., Jacka,
  T.~H., Welch, B., Frey, M., Dahe, Q., Jiawen, R., Simões, J., Fily, M.,
  Oerter, H., Nishio, F., Isaksson, E., Mulvaney, R., Holmund, P., Lipenkov,
  V., and Goodwin, I. (2005).
\newblock The {International} {Trans}-{Antarctic} {Scientific} {Expedition}
  ({ITASE}): An overview.
\newblock {\em Annals of Glaciology}, 41:180--185.

\bibitem[Medley et~al., 2014]{medley_constraining_2014}
Medley, B., Joughin, I., Smith, B.~E., Das, S.~B., Steig, E.~J., Conway, H.,
  Gogineni, S., Lewis, C., Criscitiello, A.~S., McConnell, J.~R., van~den
  Broeke, M.~R., Lenaerts, J. T.~M., Bromwich, D.~H., Nicolas, J.~P., and
  Leuschen, C. (2014).
\newblock Constraining the recent mass balance of {Pine} {Island} and
  {Thwaites} glaciers, {West} {Antarctica}, with airborne observations of snow
  accumulation.
\newblock {\em The Cryosphere}, 8(4):1375--1392.

\bibitem[Meyer, 2008]{meyer2008}
Meyer, M.~C. (2008).
\newblock Inference using shape-restricted regression splines.
\newblock {\em The Annals of Applied Statistics}, 2(3):1013--1033.

\bibitem[Molteni et~al., 1996]{molteni1996}
Molteni, F., Buizza, R., Palmer, T.~N., and Petroliagis, T. (1996).
\newblock The {ECMWF} ensemble prediction system: Methodology and validation.
\newblock {\em Quarterly Journal of the Royal Meteorological Society},
  122(529):73--119.

\bibitem[Murakami and Griffith, 2015]{murakami2015random}
Murakami, D. and Griffith, D.~A. (2015).
\newblock Random effects specifications in eigenvector spatial filtering: a
  simulation study.
\newblock {\em Journal of Geographical Systems}, 17(4):311--331.

\bibitem[Neal, 1998]{neal1998}
Neal, R. (1998).
\newblock Regression and classification using {G}aussian process priors.
\newblock {\em Bayesian Statistics}, 6:475.

\bibitem[Ortiz-Bobea and Tack, 2018]{ortiz2018another}
Ortiz-Bobea, A. and Tack, J. (2018).
\newblock Is another genetic revolution needed to offset climate change impacts
  for us maize yields?
\newblock {\em Environmental Research Letters}, 13(12):124009.

\bibitem[Paciorek, 2010]{paciorek2010importance}
Paciorek, C.~J. (2010).
\newblock The importance of scale for spatial-confounding bias and precision of
  spatial regression estimators.
\newblock {\em Statistical Science}, 25(1):107.

\bibitem[Pawson et~al., 1998]{pawson1998}
Pawson, S., Labitzke, K., and Leder, S. (1998).
\newblock Stepwise changes in stratospheric temperature.
\newblock {\em Geophysical research letters}, 25(12):2157--2160.

\bibitem[Ramsay, 1988]{ramsay1988}
Ramsay, J.~O. (1988).
\newblock Monotone regression splines in action.
\newblock {\em Statistical Science}, 3(4):425--441.

\bibitem[Reich et~al., 2006]{reich2006effects}
Reich, B.~J., Hodges, J.~S., and Zadnik, V. (2006).
\newblock Effects of residual smoothing on the posterior of the fixed effects
  in disease-mapping models.
\newblock {\em Biometrics}, 62(4):1197--1206.

\bibitem[Salamatin et~al., 2009]{salamatin2009}
Salamatin, A.~N., Lipenkov, V.~Y., Barnola, J.~M., Hori, A., Duval, P., and
  Hondoh, T. (2009).
\newblock Snow/firn densification in polar ice sheets.
\newblock {\em Low Temperature Science}, 68(Supplement):195--222.

\bibitem[Schofield et~al., 2016]{schofield_model-based_2016}
Schofield, M.~R., Barker, R.~J., Gelman, A., Cook, E.~R., and Briffa, K.~R.
  (2016).
\newblock A {Model}-{Based} {Approach} to {Climate} {Reconstruction} {Using}
  {Tree}-{Ring} {Data}.
\newblock {\em Journal of the American Statistical Association},
  111(513):93--106.
\newblock Publisher: Taylor \& Francis \_eprint:
  https://doi.org/10.1080/01621459.2015.1110524.

\bibitem[Seidel and Lanzante, 2004]{seidel2004assessment}
Seidel, D.~J. and Lanzante, J.~R. (2004).
\newblock An assessment of three alternatives to linear trends for
  characterizing global atmospheric temperature changes.
\newblock {\em Journal of Geophysical Research: Atmospheres}, 109(D14).

\bibitem[Shepherd et~al., 2019]{shepherd_trends_2019}
Shepherd, A., Gilbert, L., Muir, A.~S., Konrad, H., McMillan, M., Slater, T.,
  Briggs, K.~H., Sundal, A.~V., Hogg, A.~E., and Engdahl, M.~E. (2019).
\newblock Trends in {Antarctic} {Ice} {Sheet} {Elevation} and {Mass}.
\newblock {\em Geophysical Research Letters}, 46(14):8174--8183.
\newblock \_eprint:
  https://agupubs.onlinelibrary.wiley.com/doi/pdf/10.1029/2019GL082182.

\bibitem[Smith et~al., 2020]{smith_pervasive_2020}
Smith, B., Fricker, H.~A., Gardner, A.~S., Medley, B., Nilsson, J., Paolo,
  F.~S., Holschuh, N., Adusumilli, S., Brunt, K., Csatho, B., Harbeck, K.,
  Markus, T., Neumann, T., Siegfried, M.~R., and Zwally, H.~J. (2020).
\newblock Pervasive ice sheet mass loss reflects competing ocean and atmosphere
  processes.
\newblock {\em Science}, 368(6496):1239--1242.
\newblock Publisher: American Association for the Advancement of Science
  Section: Report.

\bibitem[Spiegelhalter et~al., 2002]{spiegelhalter2002}
Spiegelhalter, D.~J., Best, N.~G., Carlin, B.~P., and Van Der~Linde, A. (2002).
\newblock Bayesian measures of model complexity and fit.
\newblock {\em Journal of the Royal Statistical Society: Series B (Statistical
  Methodology)}, 64(4):583--639.

\bibitem[Tanner, 1996]{tanner1996}
Tanner, M.~A. (1996).
\newblock {\em Tools for Statistical Inference}.
\newblock Springer-Verlag New York, 3 edition.

\bibitem[Tebaldi and Lobell, 2008]{tebaldi2008towards}
Tebaldi, C. and Lobell, D. (2008).
\newblock Towards probabilistic projections of climate change impacts on global
  crop yields.
\newblock {\em Geophysical Research Letters}, 35(8).

\bibitem[Tebaldi and Sans{\'o}, 2009]{tebaldi2009joint}
Tebaldi, C. and Sans{\'o}, B. (2009).
\newblock Joint projections of temperature and precipitation change from
  multiple climate models: a hierarchical bayesian approach.
\newblock {\em Journal of the Royal Statistical Society: Series A (Statistics
  in Society)}, 172(1):83--106.

\bibitem[Teh et~al., 2005]{teh2005}
Teh, Y.~W., Seeger, M., and Jordan, M.~I. (2005).
\newblock Semiparametric latent factor models.
\newblock {\em AISTATS 2005}, page 333.

\bibitem[Thomas et~al., 2004]{thomas2004accelerated}
Thomas, R., Rignot, E., Casassa, G., Kanagaratnam, P., Acu{\~n}a, C., Akins,
  T., Brecher, H., Frederick, E., Gogineni, P., Krabill, W., et~al. (2004).
\newblock Accelerated sea-level rise from west antarctica.
\newblock {\em Science}, 306(5694):255--258.

\bibitem[Tiefelsdorf and Griffith, 2007]{tiefelsdorf2007semiparametric}
Tiefelsdorf, M. and Griffith, D.~A. (2007).
\newblock Semiparametric filtering of spatial autocorrelation: the eigenvector
  approach.
\newblock {\em Environment and Planning A}, 39(5):1193--1221.

\bibitem[Tom{\'e} and Miranda, 2004]{tome2004piecewise}
Tom{\'e}, A. and Miranda, P. (2004).
\newblock Piecewise linear fitting and trend changing points of climate
  parameters.
\newblock {\em Geophysical Research Letters}, 31(2).

\bibitem[Vehtari et~al., 2017]{vehtari2017}
Vehtari, A., Gelman, A., and Gabry, J. (2017).
\newblock Practical {B}ayesian model evaluation using leave-one-out
  cross-validation and {WAIC}.
\newblock {\em Statistics and computing}, 27(5):1413--1432.

\bibitem[Verjans et~al., 2020]{verjans2020bayesian}
Verjans, V., Leeson, A.~A., Nemeth, C., Stevens, C.~M., Kuipers~Munneke, P.,
  No{\"e}l, B., and van Wessem, J.~M. (2020).
\newblock Bayesian calibration of firn densification models.
\newblock {\em The Cryosphere Discussions}, pages 1--23.

\bibitem[Watanabe, 2010]{watanabe2010}
Watanabe, S. (2010).
\newblock Asymptotic equivalence of {B}ayes cross validation and widely
  applicable information criterion in singular learning theory.
\newblock {\em Journal of Machine Learning Research}, 11(Dec):3571--3594.

\bibitem[Wessem et~al., 2018]{wessem_modelling_2018}
Wessem, J. M.~v., Berg, W. J. v.~d., Noël, B. P.~Y., Meijgaard, E.~v., Amory,
  C., Birnbaum, G., Jakobs, C.~L., Krüger, K., Lenaerts, J. T.~M., Lhermitte,
  S., Ligtenberg, S. R.~M., Medley, B., Reijmer, C.~H., Tricht, K.~v., Trusel,
  L.~D., Ulft, L. H.~v., Wouters, B., Wuite, J., and Broeke, M. R. v.~d.
  (2018).
\newblock Modelling the climate and surface mass balance of polar ice sheets
  using {RACMO2} – {Part} 2: {Antarctica} (1979–2016).
\newblock {\em The Cryosphere}, 12(4):1479--1498.
\newblock Publisher: Copernicus GmbH.

\bibitem[White et~al., 2020]{white2020}
White, P.~A., Keeler, D.~G., and Rupper, S. (2020).
\newblock Hierarchical integrated spatial process modeling of monotone {W}est
  {A}ntarctic snow density curves.
\newblock {\em arXiv preprint arXiv:2001.05520}.

\bibitem[White et~al., 2019]{white2019}
White, P.~A., Reese, C.~S., Christensen, W.~F., and Rupper, S. (2019).
\newblock A model for {A}ntarctic surface mass balance and ice core site
  selection.
\newblock {\em Environmetrics}, 30(8):e2579.

\bibitem[Wikle and Hooten, 2010]{wikle2010general}
Wikle, C.~K. and Hooten, M.~B. (2010).
\newblock A general science-based framework for dynamical spatio-temporal
  models.
\newblock {\em Test}, 19(3):417--451.

\bibitem[Zhang, 2004]{zhang2004}
Zhang, H. (2004).
\newblock Inconsistent estimation and asymptotically equal interpolations in
  model-based geostatistics.
\newblock {\em Journal of the American Statistical Association},
  99(465):250--261.

\end{thebibliography}

\appendix

\section{Model Comparison}\label{app:mod_comp}

Because the density measurements are bounded below, we consider truncated Normal and truncated Student's $t$ error distributions for these data. To account for expedition differences, as well as the length of the core used to obtain each density measurement, we consider hierarchical and weighted models for the scale of the error distribution. Our model comparison question here is which components of this model are beneficial. For this comparison, we find that the truncated Student's $t$ distribution with weighted, hierarchical scale parameters has the lowest WAIC (See Table \ref{tab:error}).

We consider eight possible cross-covariance specifications for the 12 spatially varying parameters used to model these data that can be written in the form \citep[see][for review]{alvarez2011,banerjee2014}:
\begin{equation}\label{eq:cov_mod}
\bSigma = \left(\bLam \otimes \bI \right) \text{BlockDiag}(\bR_1,...,\bR_{r}) \left(  \bLam^T \otimes \bI \right).
\end{equation} 
In this model, $\bLam$ controls between-parameter dependence and is $12 \times r$, where $r \leq 12$. To govern spatial dependence, we use correlation matrices $\bR_1,...,\bR_{12}$, where the $i$th row and $j$th column of the $k$th correlation matrix is $(\bR_l)_{i,j} = \exp\left( \phi_l d(\bs_{i},\bs_l )\right)$. Here, we use $d(\bs_{i},\bs_l )$ as the great-circle distance; however, other distance metrics may be just as effective in this application (e.g., the chordal distance). We consider the following possibilities:
\begin{description}
\item[Independent:] If (a) $\bLam$ is $12\times 12$ identity matrix, then we assume no between-parameter dependence and $\bSigma = \text{BlockDiag}(\bR_1,...,\bR_{12})$.
\item[Separable:] If we let $r = 12$ and $\bR = \bR_1 = \bR_2 =\cdots =\bR_{12} $, then we get a separable model $\bSigma = (\bLam \bLam^T) \otimes \bR$. This model is often called the intrinsic model of coregionalization \citep[see][]{matheron1982}. In this model, we assume that the spatial patterns for all parameters is similar and can be represented using a single correlation function .
\item[Latent Factor/Coregionalization:] If $r < 12$, then we get latent factor model and we cannot simplify \eqref{eq:cov_mod}. For a nice discussion of these models, see \cite{teh2005}. This model assumes that the we can describe the 12-dimensional spatial process using a lower-rank space. If we let $r = 12$ and let $\bLam$ be lower-triangular, then we get the linear model of coregionalization and we cannot simplify \eqref{eq:cov_mod} \citep[see][for early reference]{grzebyk1994}. In this model, we do not assume any rank-reduction.
\end{description}

Richer cross-covariance functions \citep[see, e.g.,][]{gneiting2010,genton2015} or non-stationary cross-covariance functions \citep[see, e.g.,][]{kleiber2012} are often beneficial. However, with $n_s = 56$ sites, we have limited ability to estimate complicated spatial models and find that more complicated spatial models perform worse for this data. Moreover, many models have limited parameter identifiability; therefore, we argue that these cross-covariance functions offer sufficient flexibility.

Because the spatially varying HL model \eqref{eq:herron_spatial} is piecewise linear, we explore various specification of the smoothed HL model \eqref{eq:smooth_herron} to see whether they improve upon \eqref{eq:herron_spatial}. Specifically, we use a cubic polynomial, quadratic splines (with one, two, or three knots), and cubic splines (with one, two, or three knots). The results of the model comparison are given in Table \ref{tab:smooth}. We use the cubic spline with two knots because it has the lowest WAIC; however, all spline models with two or more knots are comparable in terms of WAIC.



\begin{table}[H]
\centering
\scriptsize

\begin{tabular}{lllrrr}
  \hline
Truncated &  &  & &  &  \\ 
Distribution & Weighting & Hierarchical & WAIC& Relative WAIC & WAIC SE \\ 
  \hline
Normal & Yes & Yes & -73234.89 & 3410.35 & 338.17 \\ 
  Student's-$t$ & No & No & -73740.73 & 2904.52 & 272.37 \\ 
  Student's-$t$ & Yes & No & -75297.33 & 1347.91 & 275.06 \\ 
  Student's-$t$& No & Yes & -75775.72 & 869.53 & 272.71 \\ 
  {\bf Student's-$t$} & {\bf Yes} & {\bf Yes} & {\bf -75791.80} & {\bf  853.44} & {\bf 272.33} \\  
   \hline
\end{tabular}
\caption{WAIC for models differing in their error distribution.}\label{tab:error}
\end{table}


\begin{table}[H]
\centering
\scriptsize

\begin{tabular}{lrrr}
  \hline
Cross-Covariance Function  & WAIC & Relative WAIC & WAIC SE  \\ 
  \hline
\textbf{Separable GP} & \textbf{-75791.80} & {\bf  853.44} & {\bf 272.33} \\ 
Independent GP  & -75766.30 & 878.94 & 272.82 \\ 
2 Latent Factors  & -74572.49 & 2072.76 & 267.38 \\ 
4 Latent Factors  & -75654.19 & 991.06  & 268.43 \\  
6 Latent Factors  & -75600.94 & 1044.31 & 270.75 \\ 
8 Latent Factors  & -75606.62 & 1038.62 & 270.62 \\  
10 Latent Factors & -75776.52 & 868.72 & 271.53 \\
Coregionalization (12 Latent Factors) & -75690.65 & 954.59 & 270.70 \\ 
   \hline
\end{tabular}
\caption{WAIC for models differing by the cross-covariance of the transformed parameters.}\label{tab:spatial}
\end{table}


\begin{table}[H]
\centering
\scriptsize
\begin{tabular}{llrrr}
  \hline
Model for $\bh^\perp_\theta(\bs_{i},x)$ & Number of Knots & WAIC & Relative WAIC& WAIC SE  \\ 
  \hline
None & --- &-75791.80 & 853.44 & 272.33 \\  
  Cubic Polynomial & --- & -75822.69 & 822.56 & 272.95 \\  
  Quadratic Spline & 1 & -76383.12 & 262.13 & 278.51 \\ 
  \textbf{Quadratic Spline} & {\bf 2} &  {\bf -76645.25} & {\bf 0.00} & {\bf 278.69} \\ 
  Quadratic Spline & 3 &  -76601.84 & 43.41 & 279.04 \\ 
  Cubic Spline & 1 & -76541.38 & 103.87 & 278.77 \\ 
  Cubic Spline & 2 & -76625.65 & 19.60 & 280.73 \\ 
  Cubic Spline & 3 & -76617.85 & 27.40 & 280.09 \\ 
   \hline
\end{tabular}
\caption{WAIC for models differing by the smooth nonlinear adjustment to the piecewise linear model.}\label{tab:smooth}
\end{table}


        \vspace{-3mm}

\end{document}


\title{Supplemental Material for Improving Piecewise Linear Snow Density Models through Hierarchical Spatial and Functional Smoothing}
\author[1]{Philip A. White \thanks{Corresponding Author: pwhite@stat.byu.edu}}
\author[2]{Durban G. Keeler \thanks{durban.keeler@gmail.com}}
\author[1]{Daniel Sheanshang\thanks{ danielsheanshang@gmail.com }}
\author[2]{Summer Rupper \thanks{ summer.rupper@geog.utah.edu}}

\affil[1]{Department of Statistics, Brigham Young University, USA}
\affil[2]{Department of Geography, University of Utah, USA}

\maketitle

\section{Extended Posterior Analysis}\label{app:analysis}

\begin{figure}[H]
\begin{center}
\includegraphics[width=0.8\textwidth]{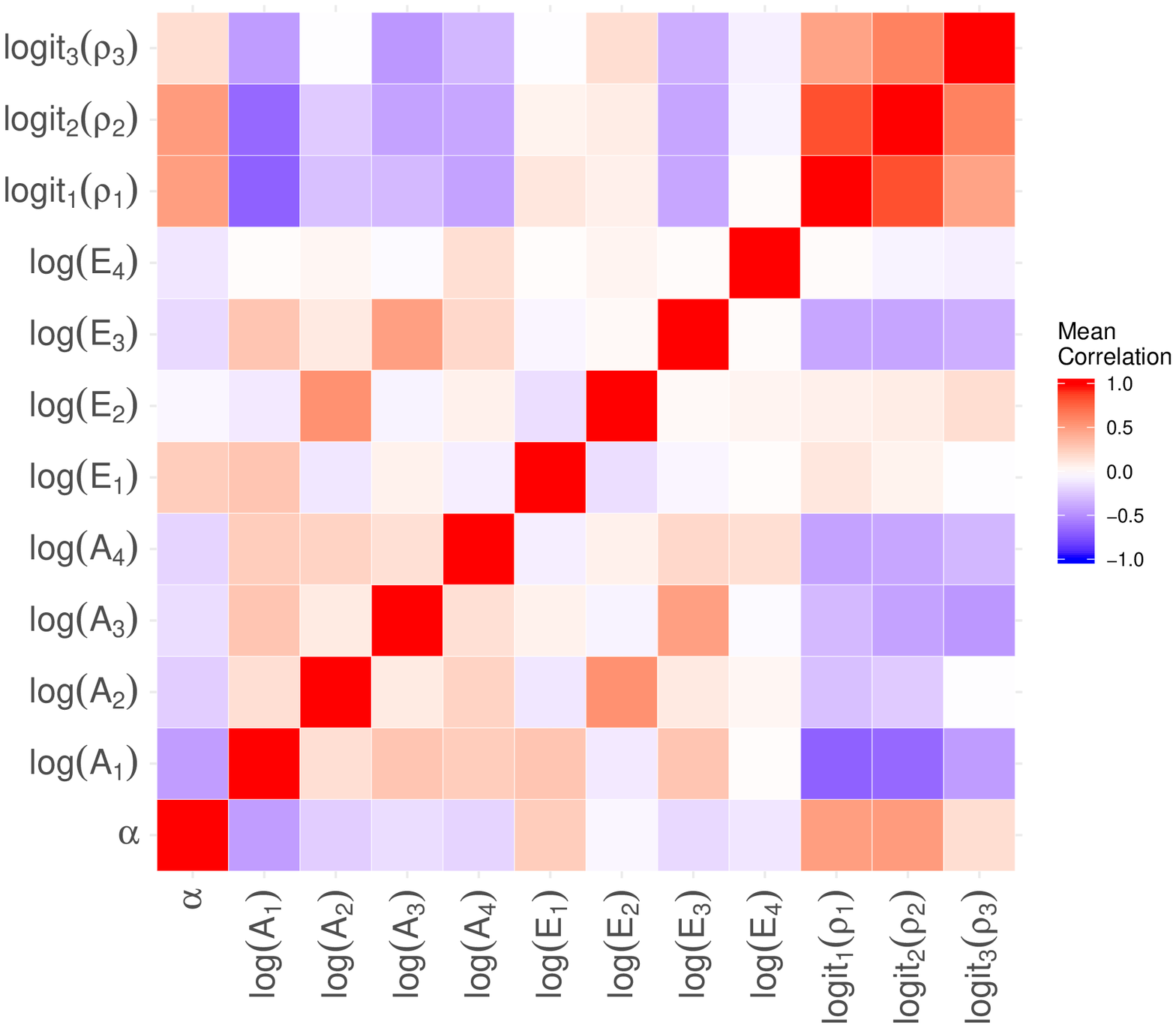}
\end{center}
\vspace{-12mm}
\caption{Empirical average between-parameter correlation estimated from 56 site-specific parameters.}\label{fig:cor}
\end{figure}

In Figure \ref{fig:cor}, we plot the mean empirical between-parameter correlation taken from the 56 sites. We examine the correlation on the transformed parameter scale to decrease the effects of large values. There are many strong correlations. We note only a few. The intercept and critical densities are all positively correlated. Interestingly, the log-transformed pre-exponentiation factors for the first and fourth stages have a strong negative correlation with the critical densities.

\begin{figure}[H]
\begin{center}
\includegraphics[width=\textwidth]{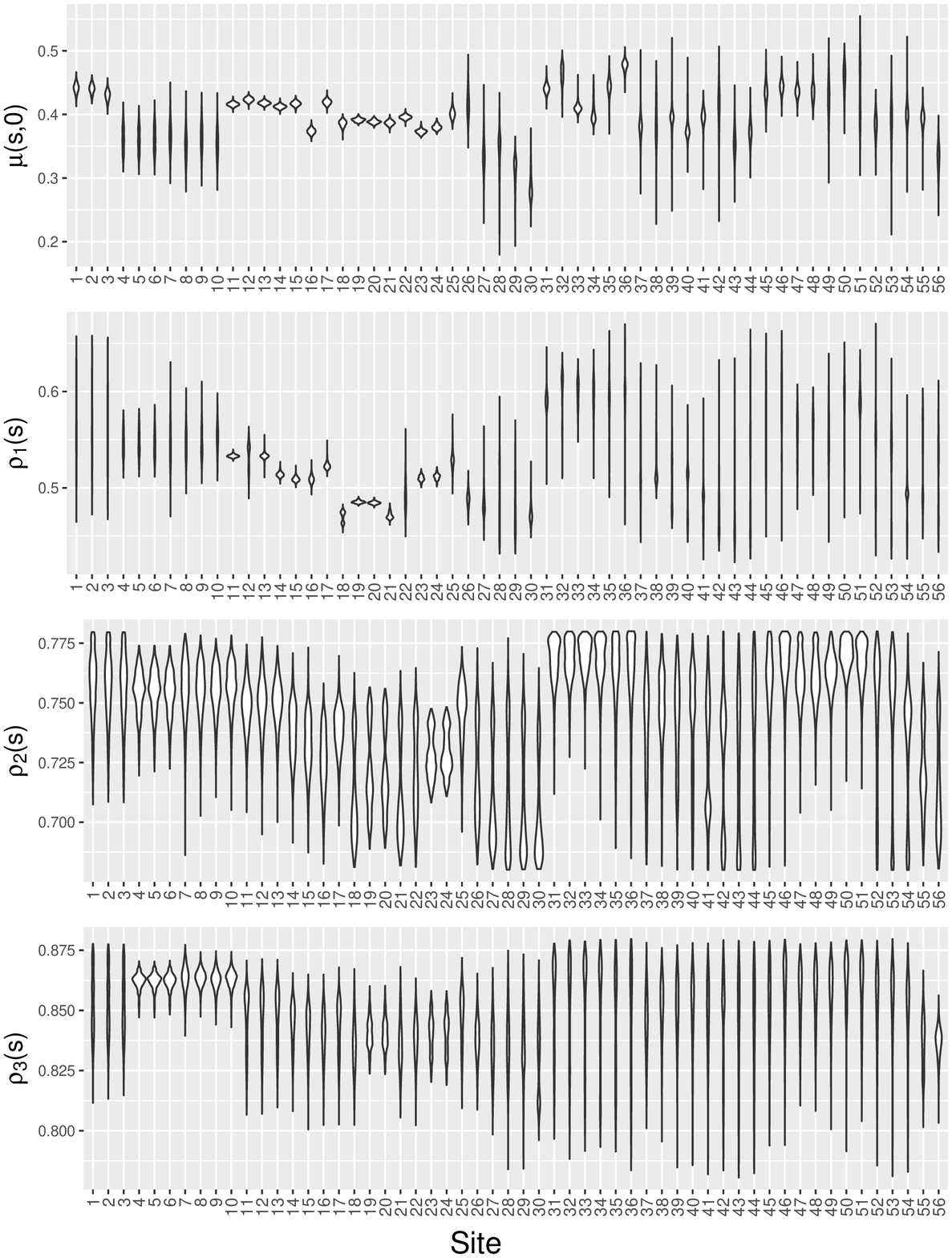}
\end{center}
\vspace{-4mm}

\caption{Posterior distributions for site-specific intercepts and critical densities.}\label{fig:dens_bonus}
\end{figure}

\begin{figure}[H]
\begin{center}
\includegraphics[width=\textwidth]{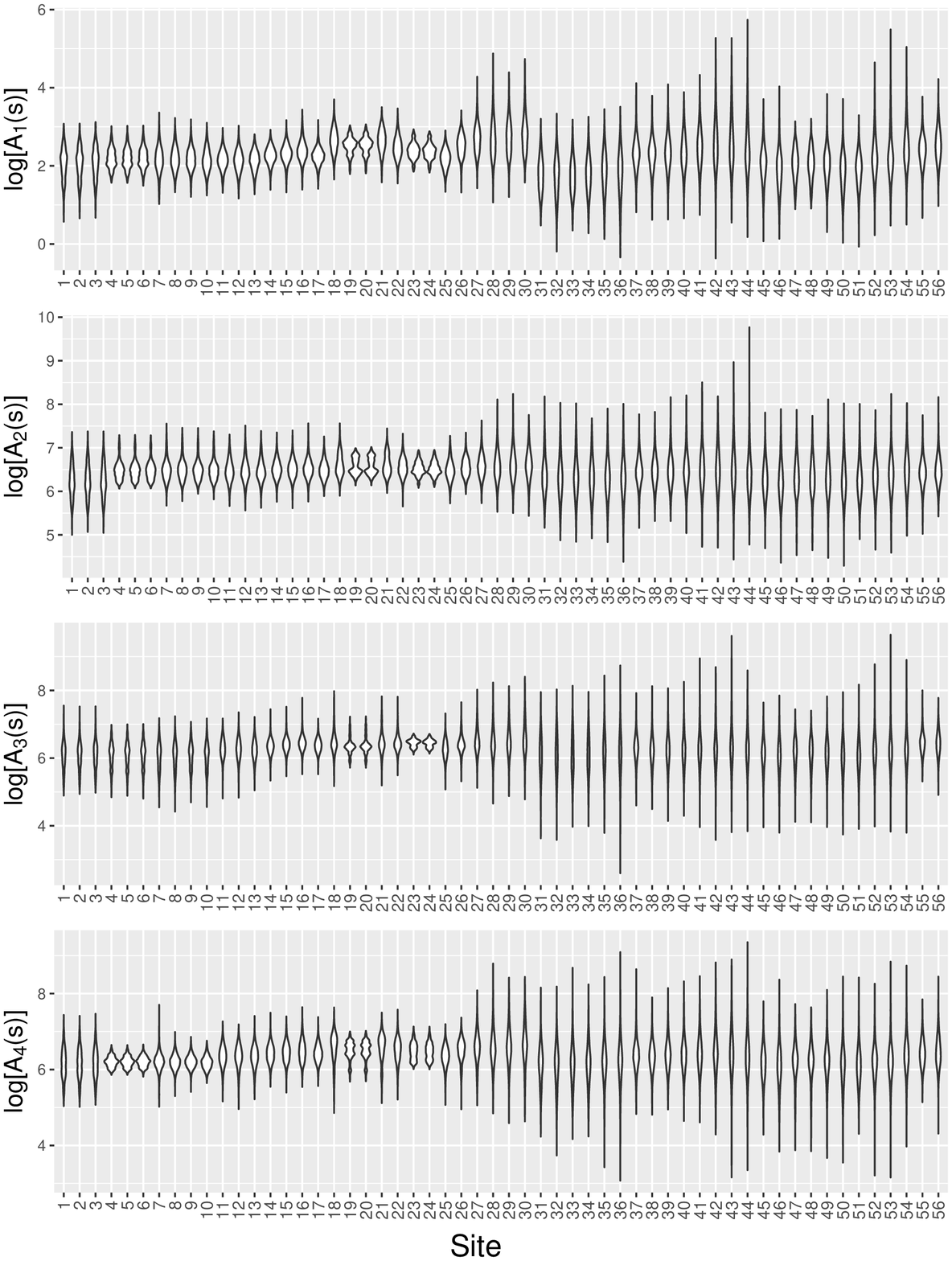}
\end{center}
\vspace{-4mm}

\caption{Posterior distributions for site-specific log-transformed pre-exponentiation factors.}\label{fig:A_bonus}
\end{figure}

\begin{figure}[H]
\begin{center}
\includegraphics[width=\textwidth]{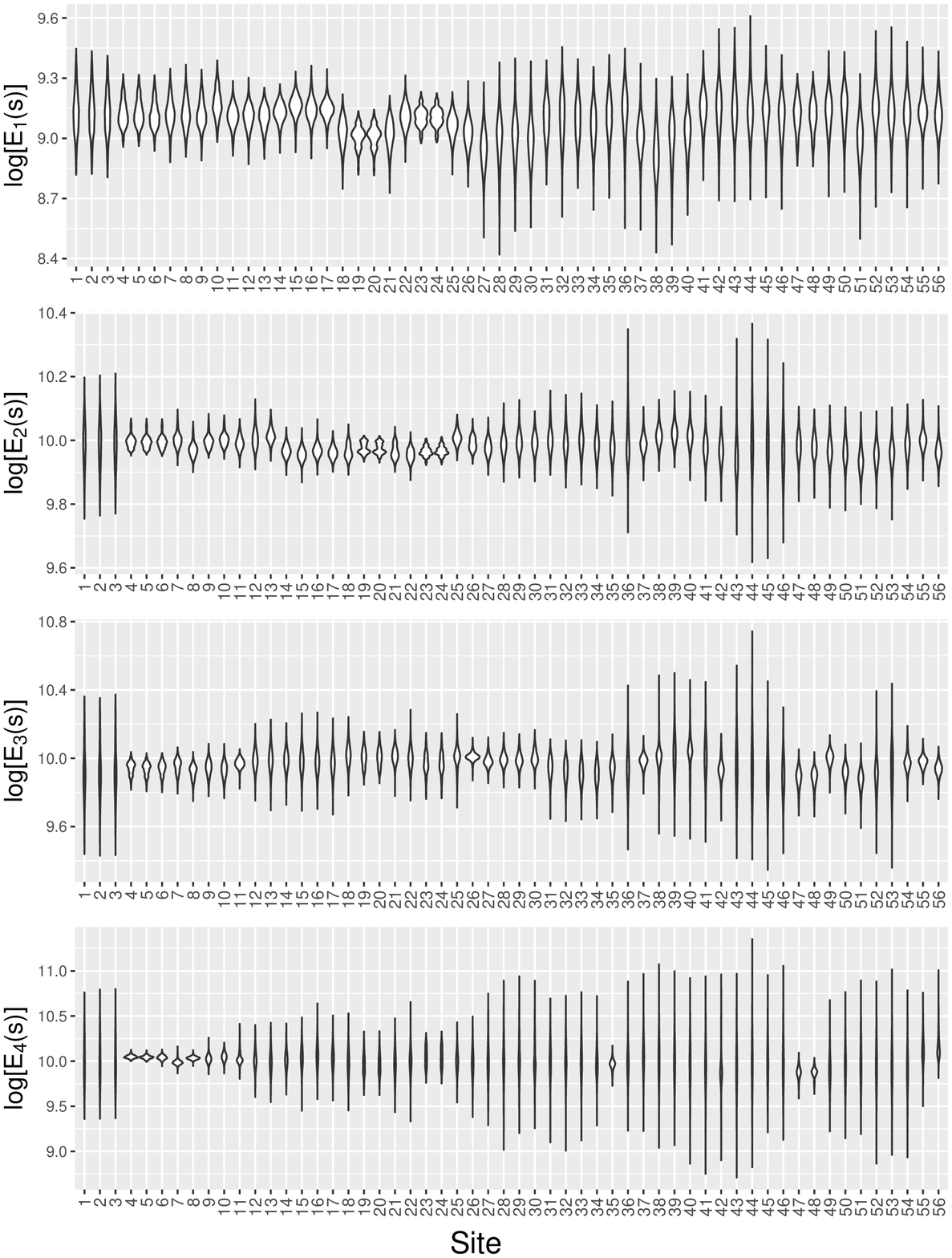}
\end{center}
\vspace{-4mm}

\caption{Posterior distributions for site-specific log-transformed energies of activation.}\label{fig:E_bonus}
\end{figure}

We give violin plots for site-specific posterior distribution in Figures \ref{fig:dens_bonus}, \ref{fig:A_bonus}, and \ref{fig:E_bonus}. From these plots, we see three clear and unsurprising patterns: (1) Arrhenius parameters that govern densities at shallower depths have lower posterior variance compared to parameters for deeper densification stages. (2) Cores with more observations have better resolved posterior distributions. (3) Cores distant from other cores have higher variance.  

\begin{figure}[H]
\begin{center}
\includegraphics[width=0.24\textwidth]{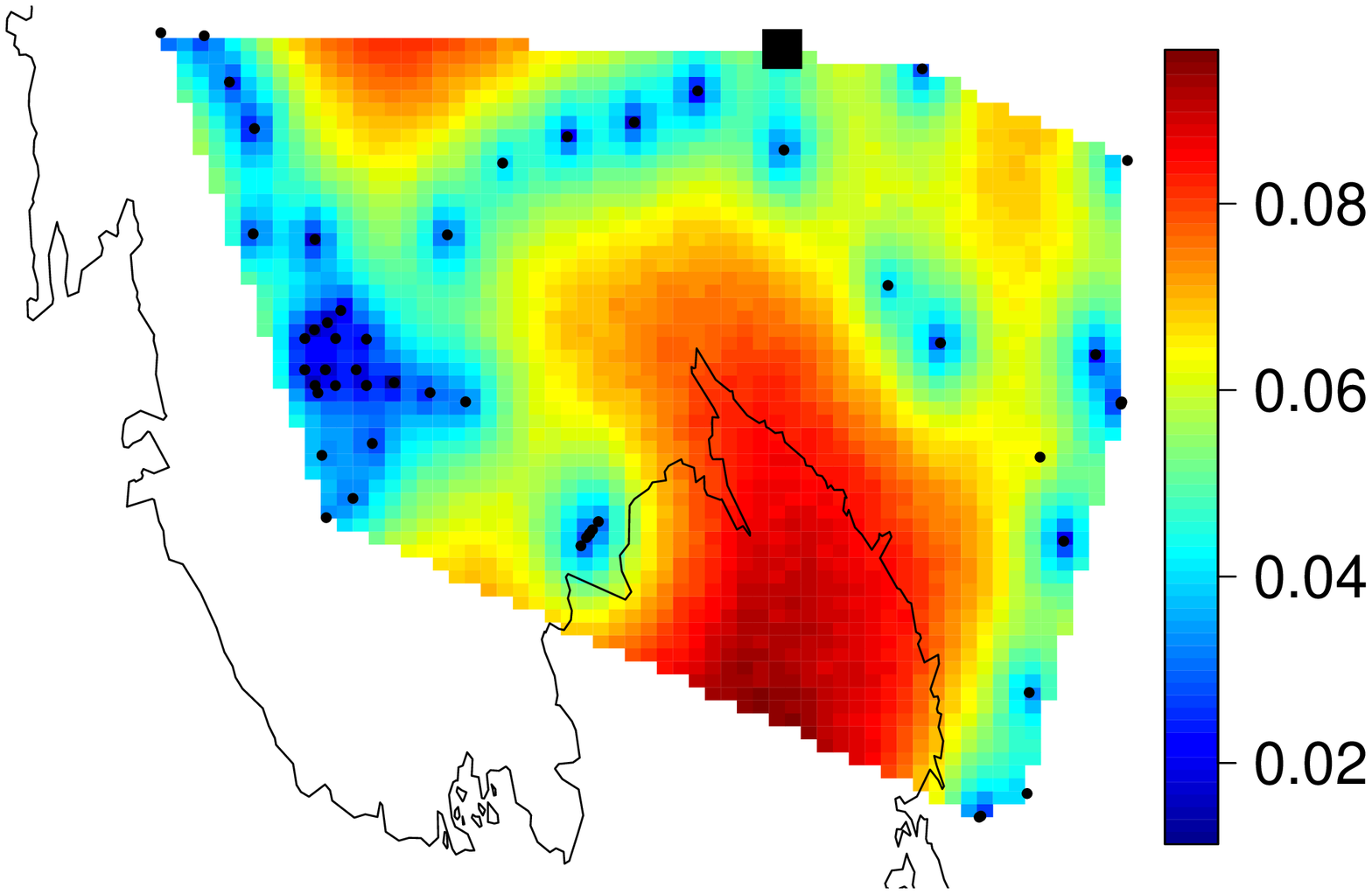}
\includegraphics[width=0.24\textwidth]{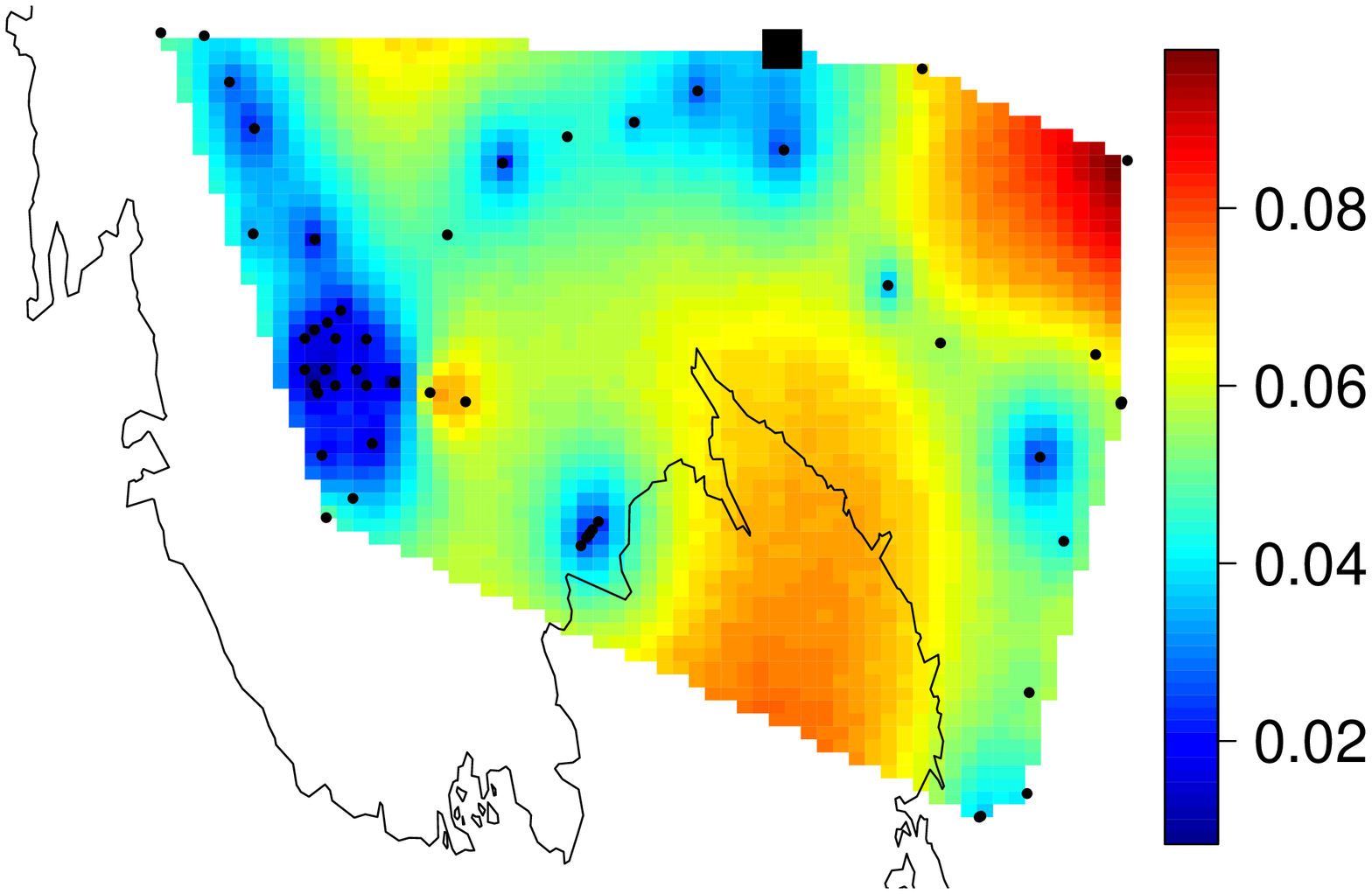}
\includegraphics[width=0.24\textwidth]{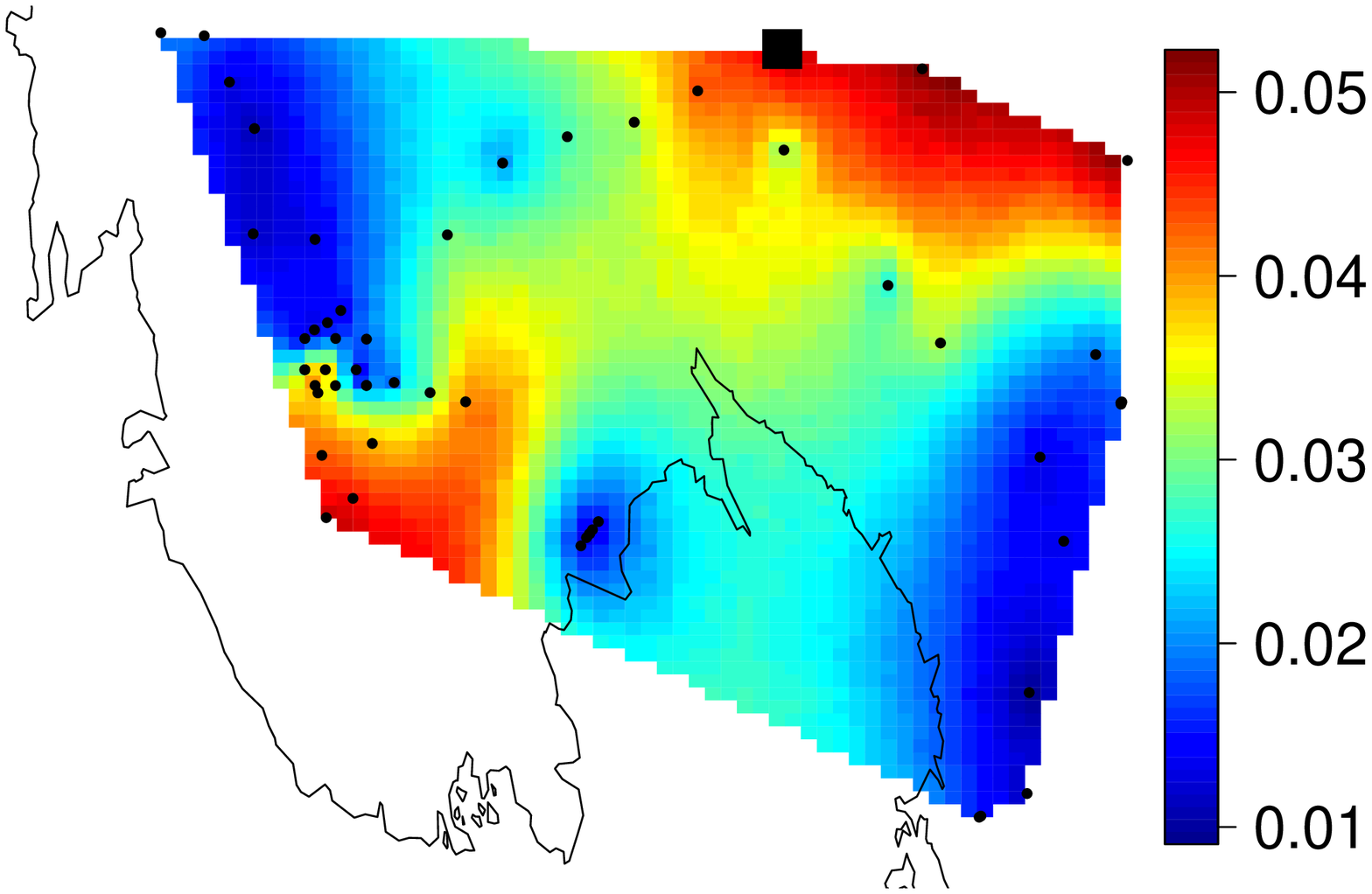}
\includegraphics[width=0.24\textwidth]{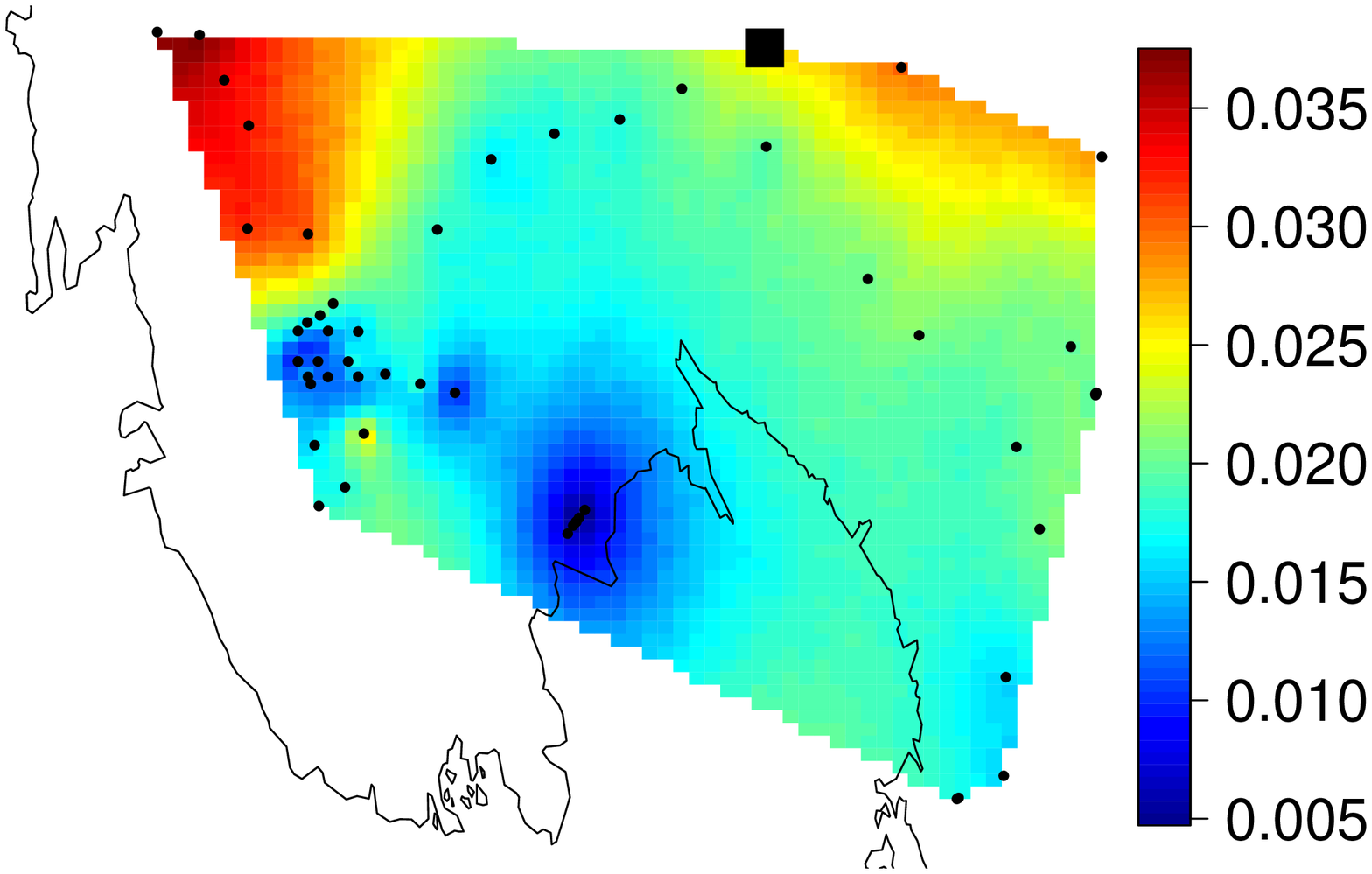} \\
\includegraphics[width=0.24\textwidth]{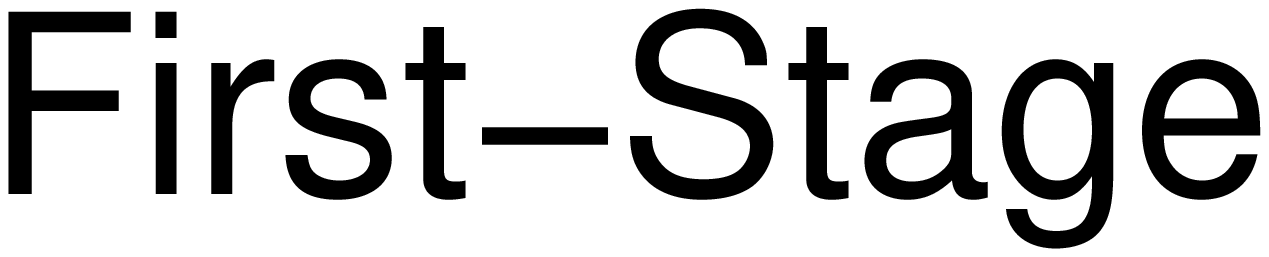}
\includegraphics[width=0.24\textwidth]{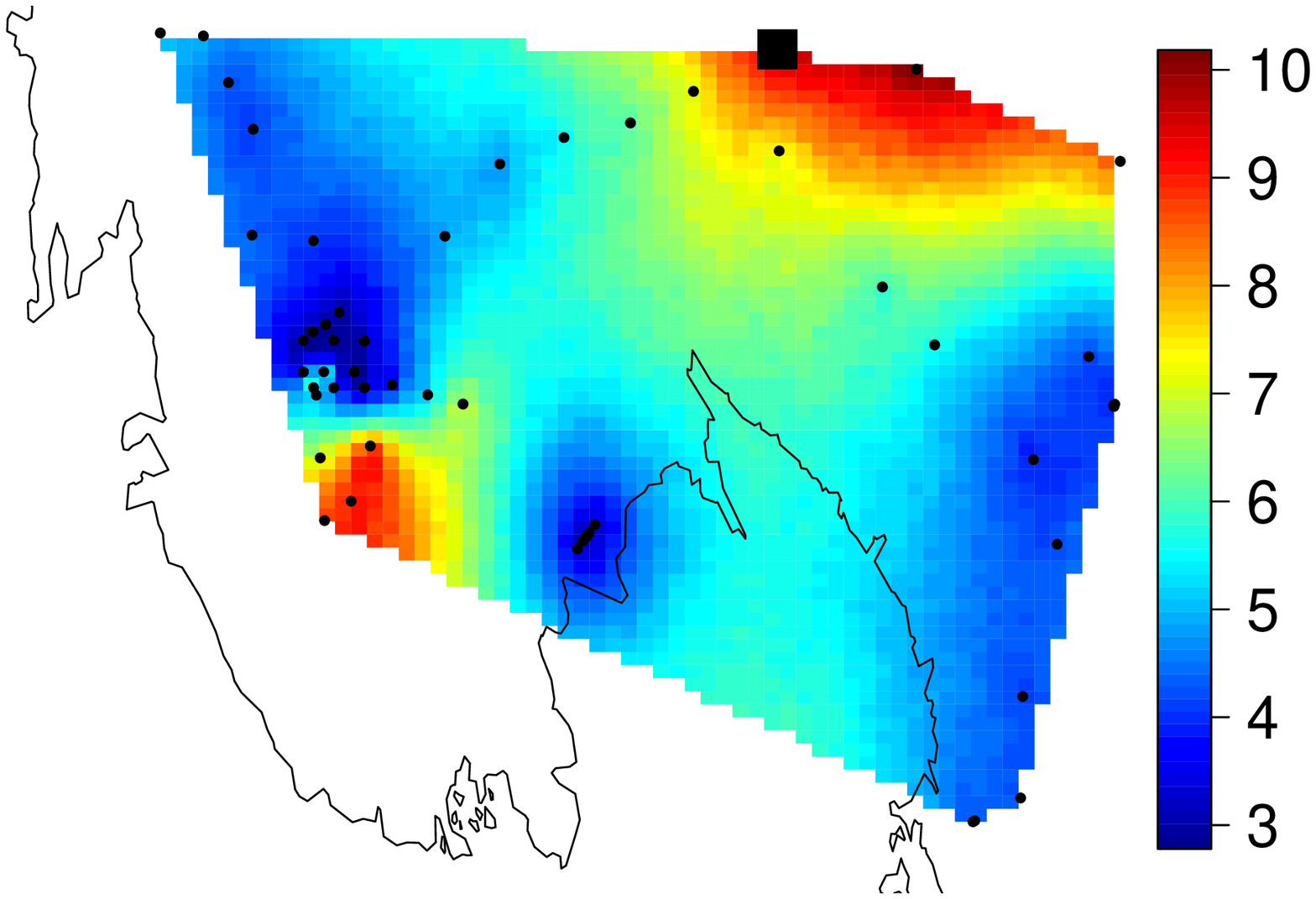}
\includegraphics[width=0.24\textwidth]{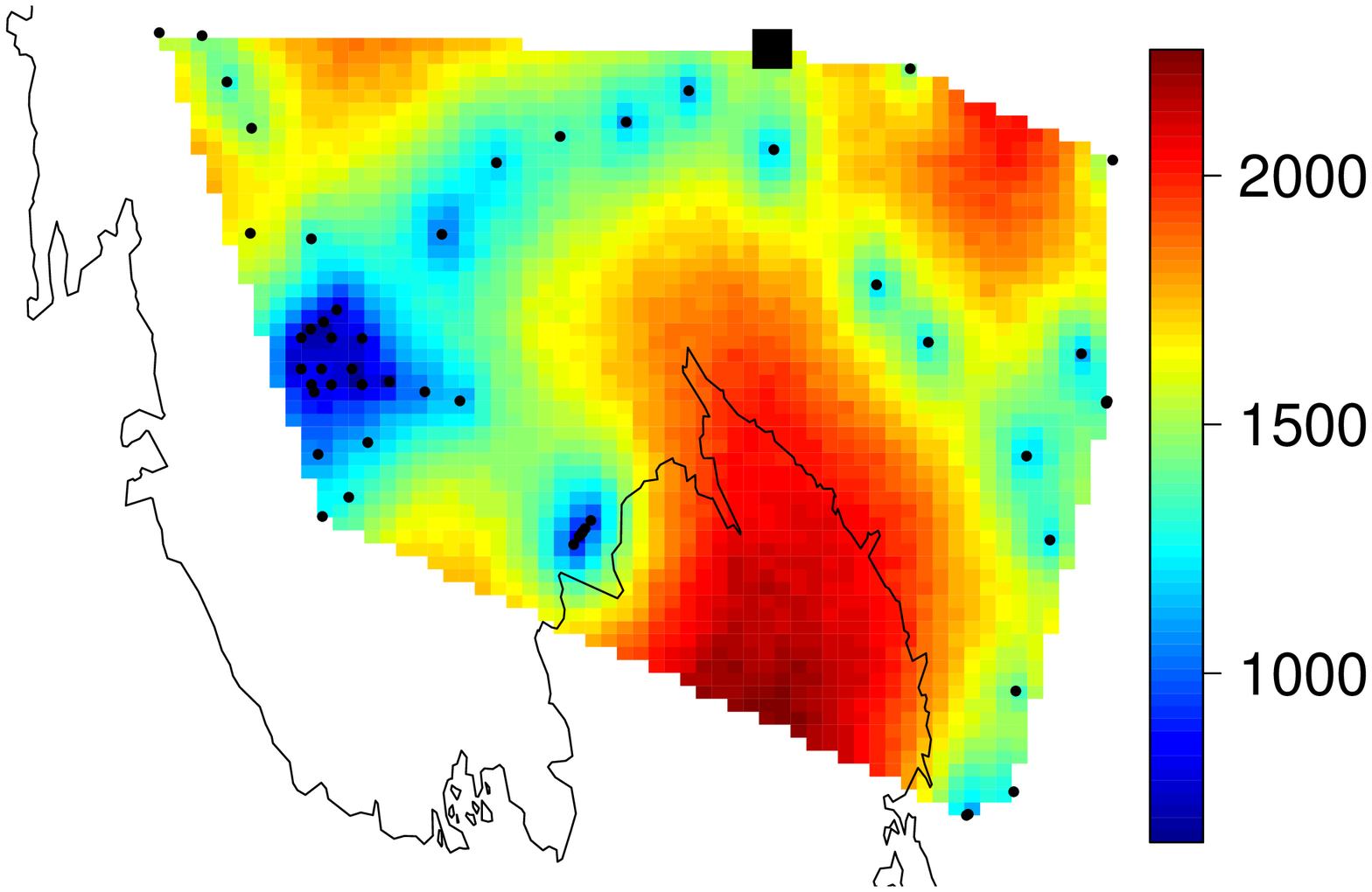}
\includegraphics[width=0.24\textwidth]{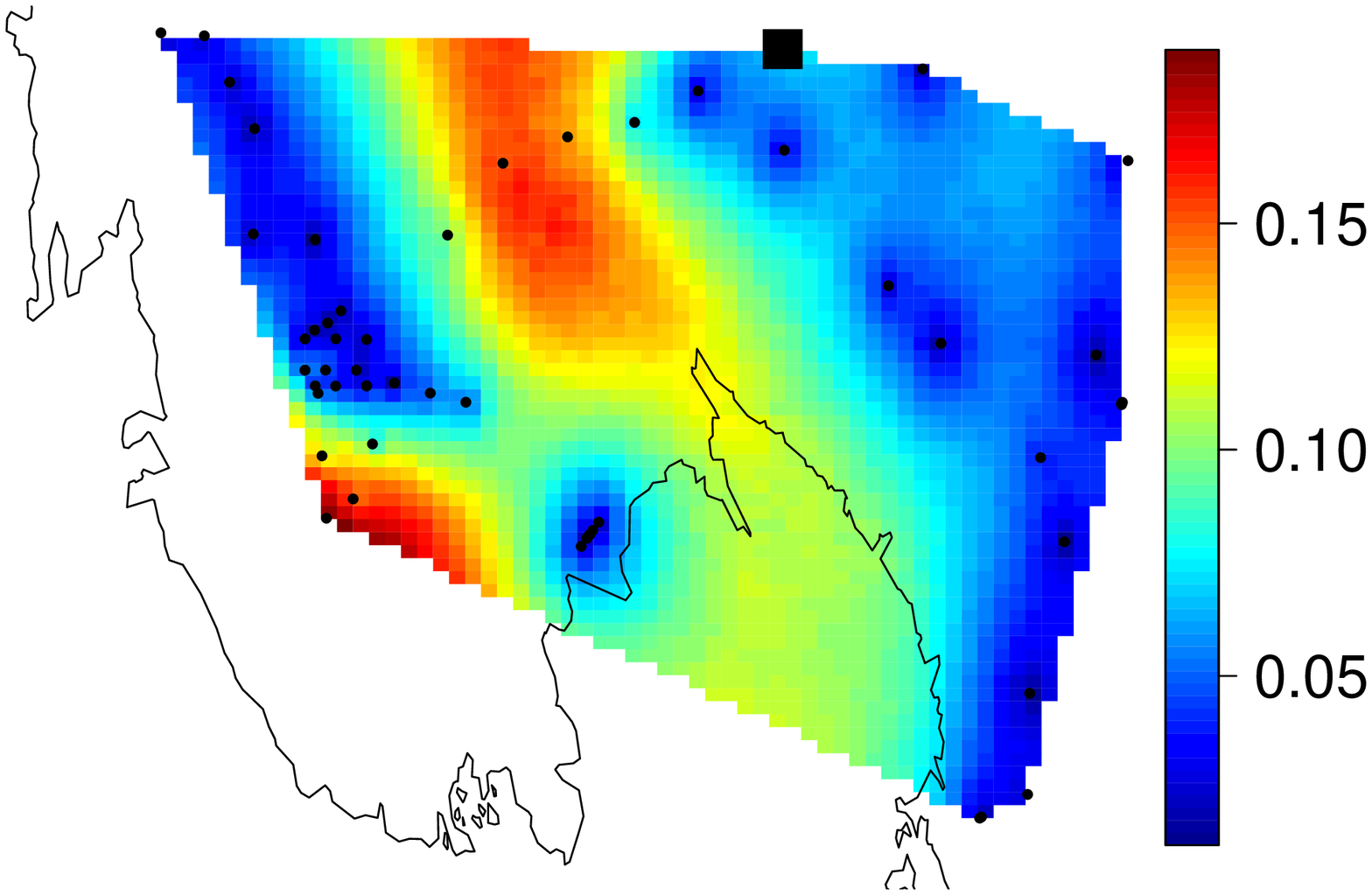}\\
\includegraphics[width=0.24\textwidth]{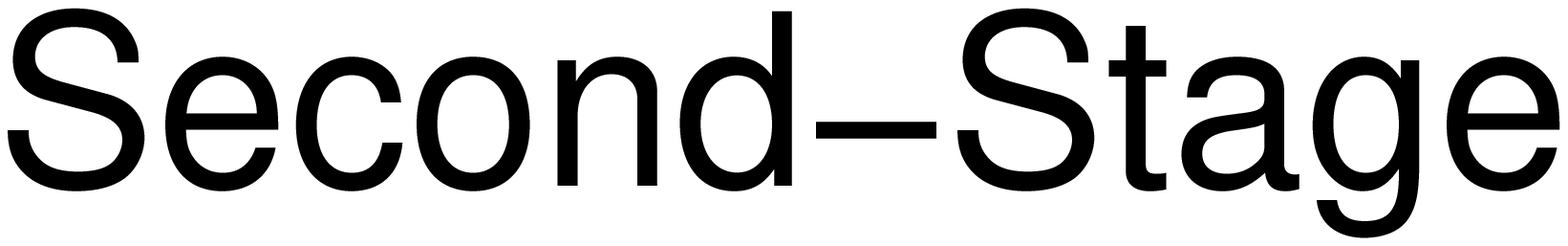}
\includegraphics[width=0.24\textwidth]{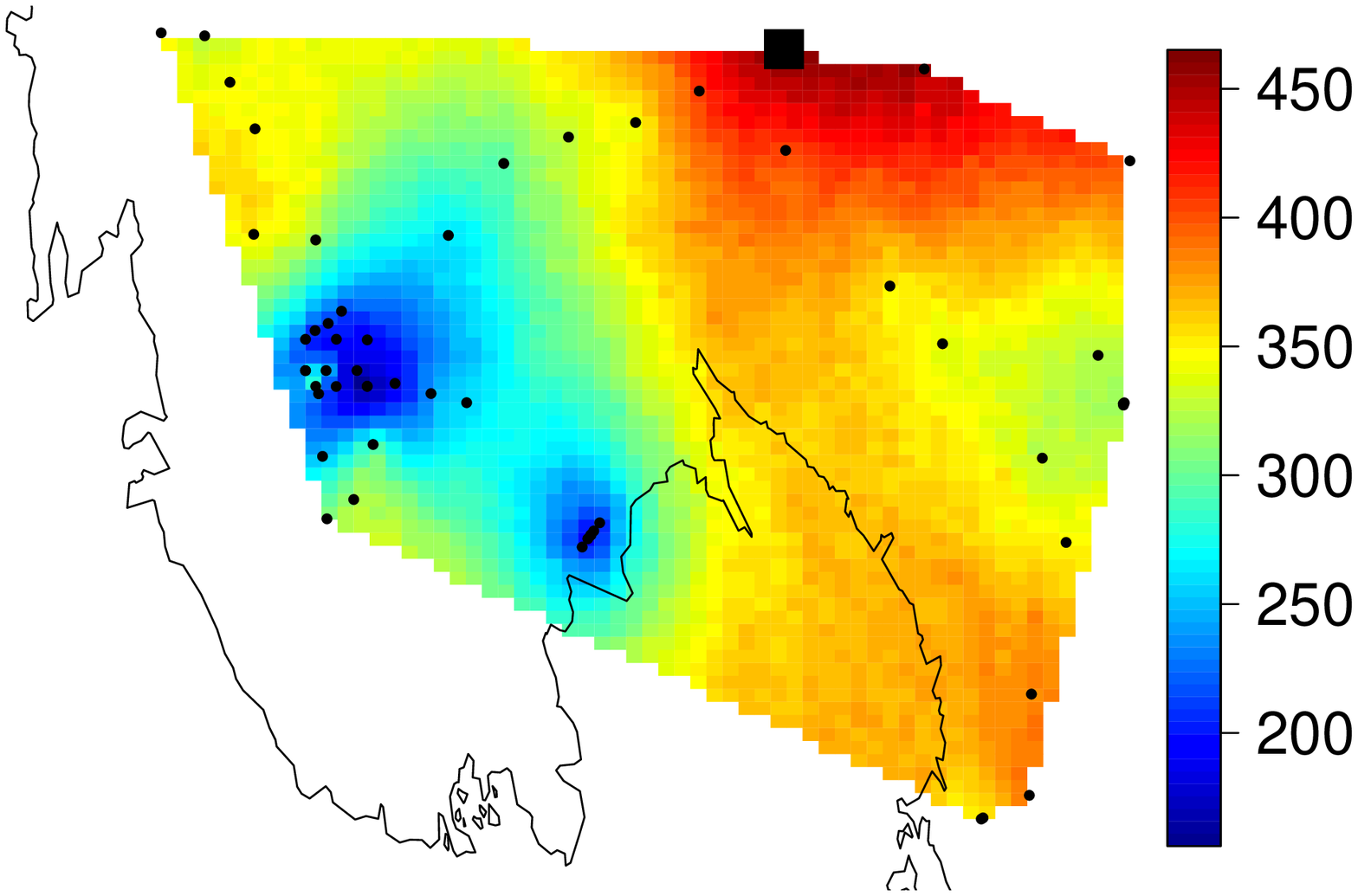}
\includegraphics[width=0.24\textwidth]{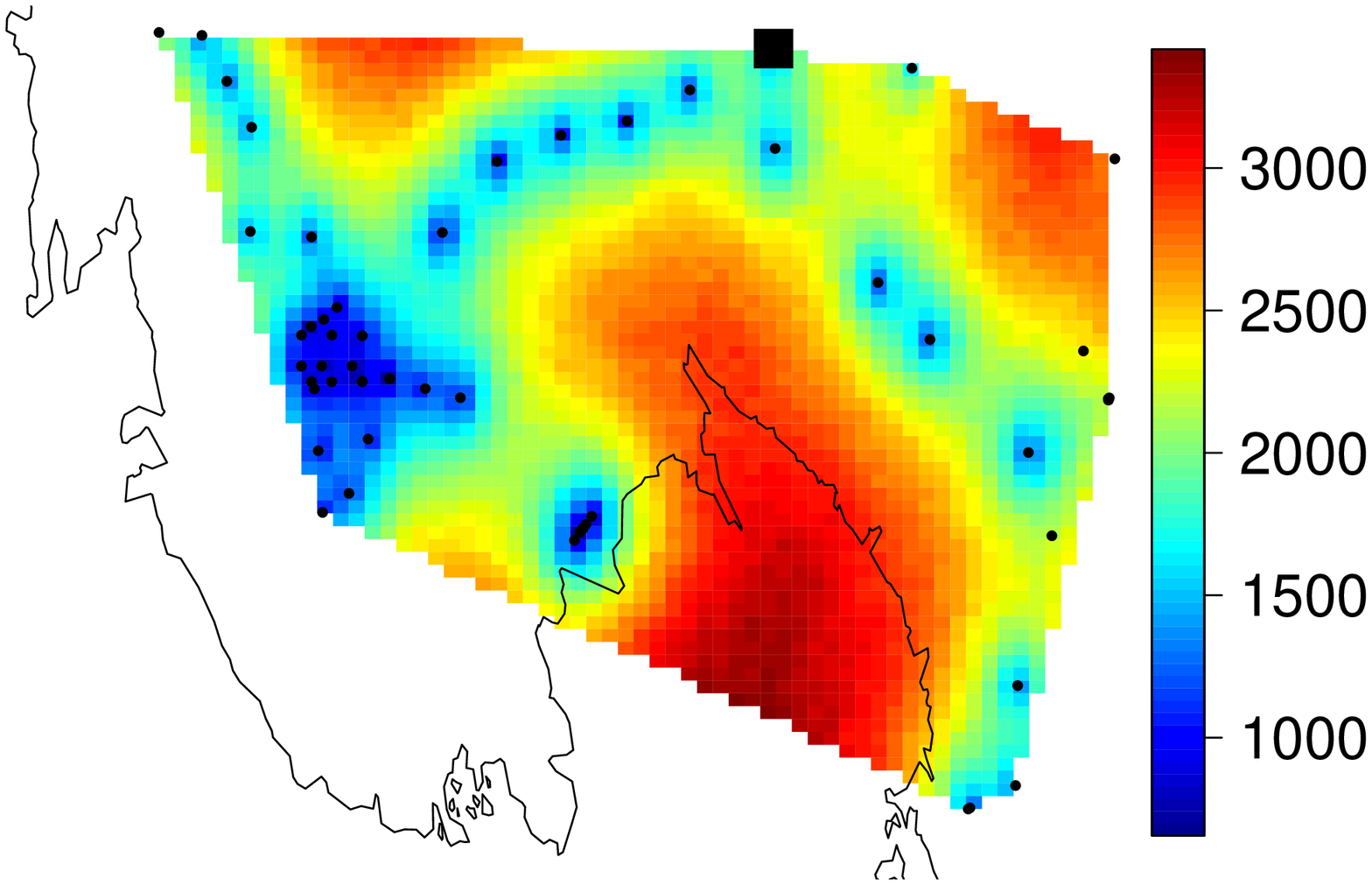}
\includegraphics[width=0.24\textwidth]{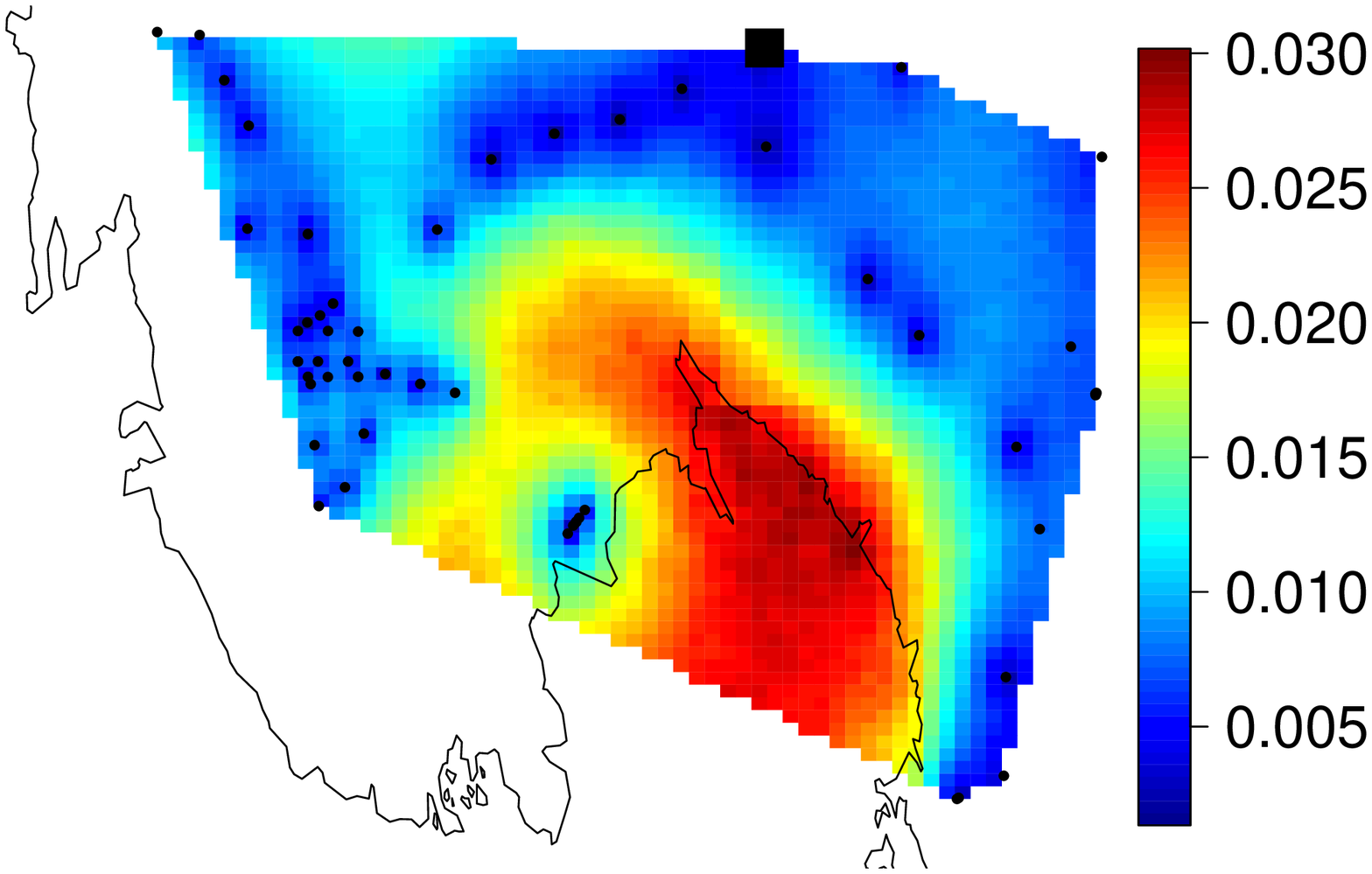}\\
\includegraphics[width=0.24\textwidth]{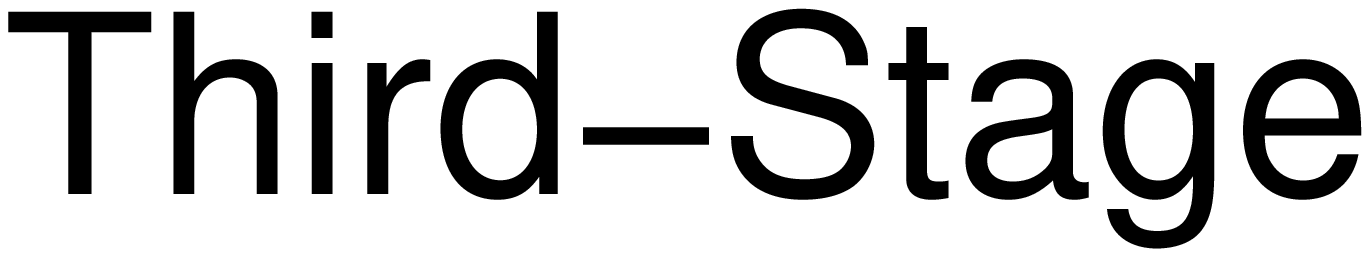}
\includegraphics[width=0.24\textwidth]{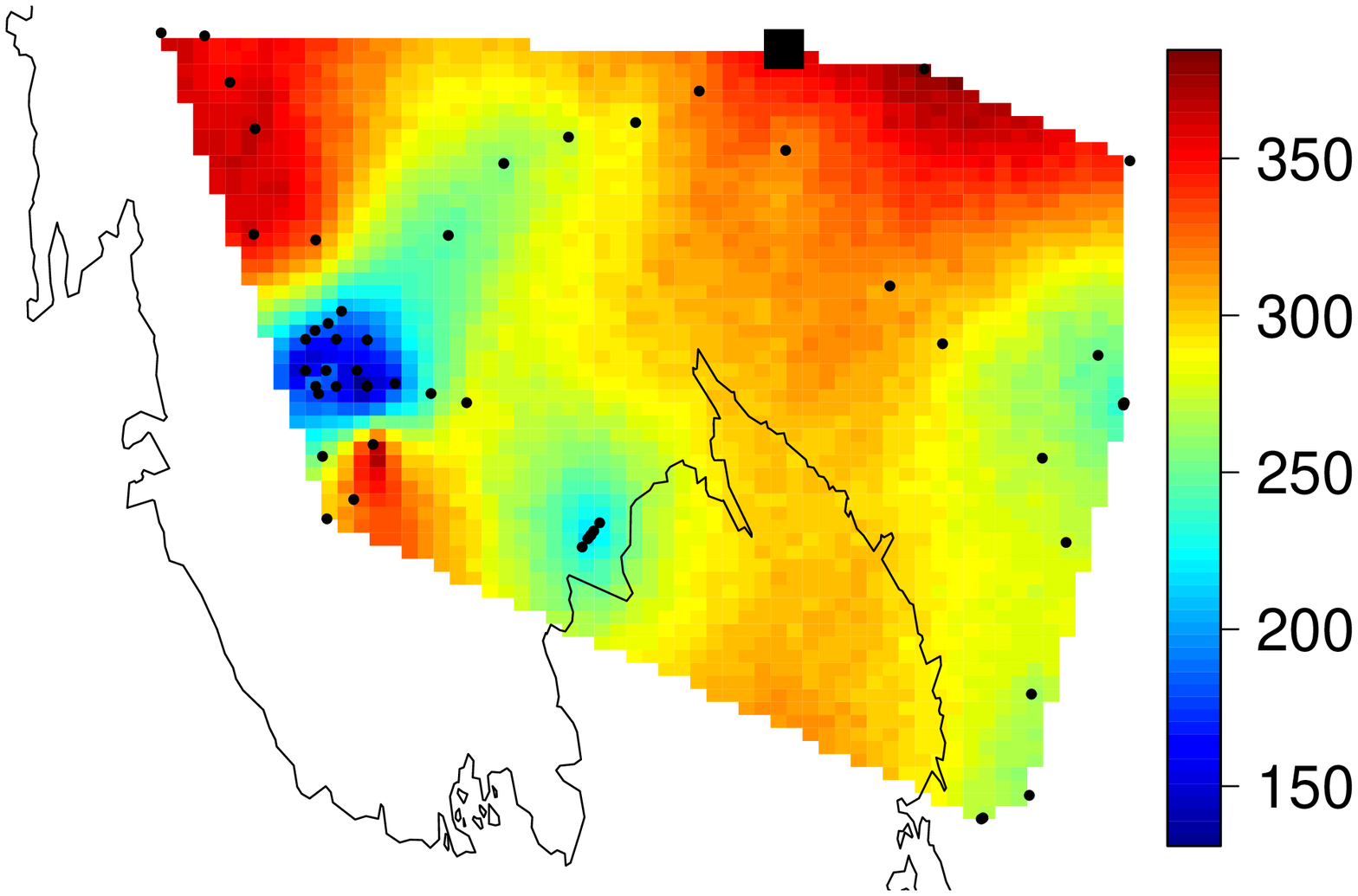}
\includegraphics[width=0.24\textwidth]{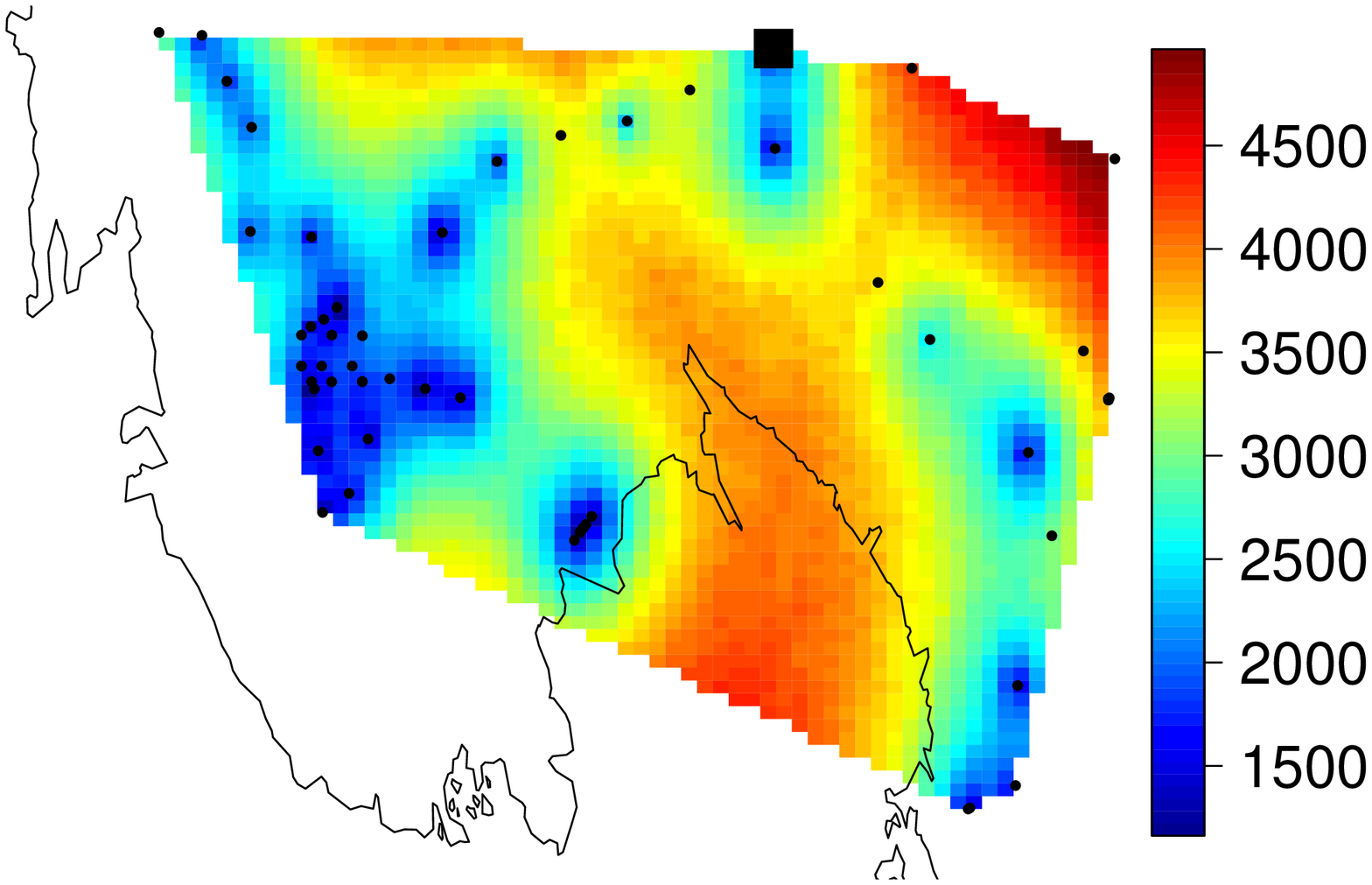}
\includegraphics[width=0.24\textwidth]{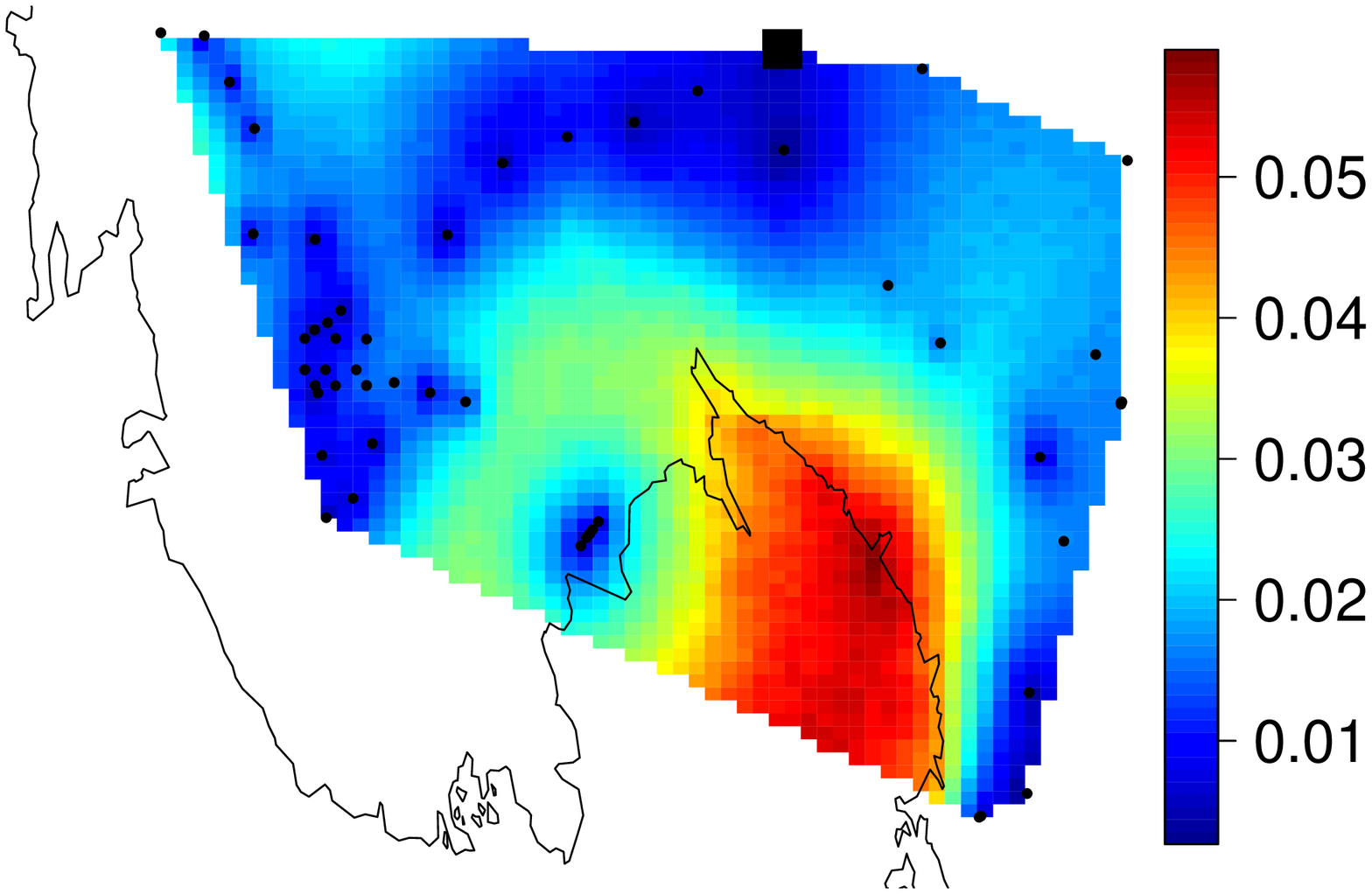}\\
\includegraphics[width=0.24\textwidth]{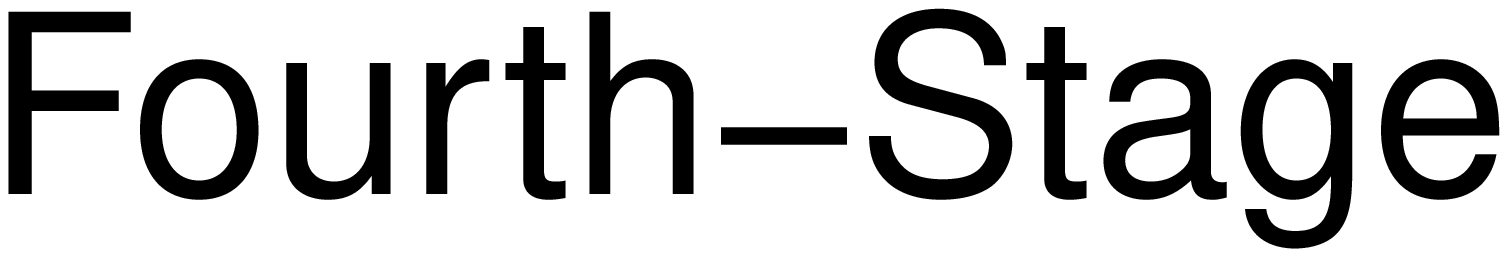}
\includegraphics[width=0.24\textwidth]{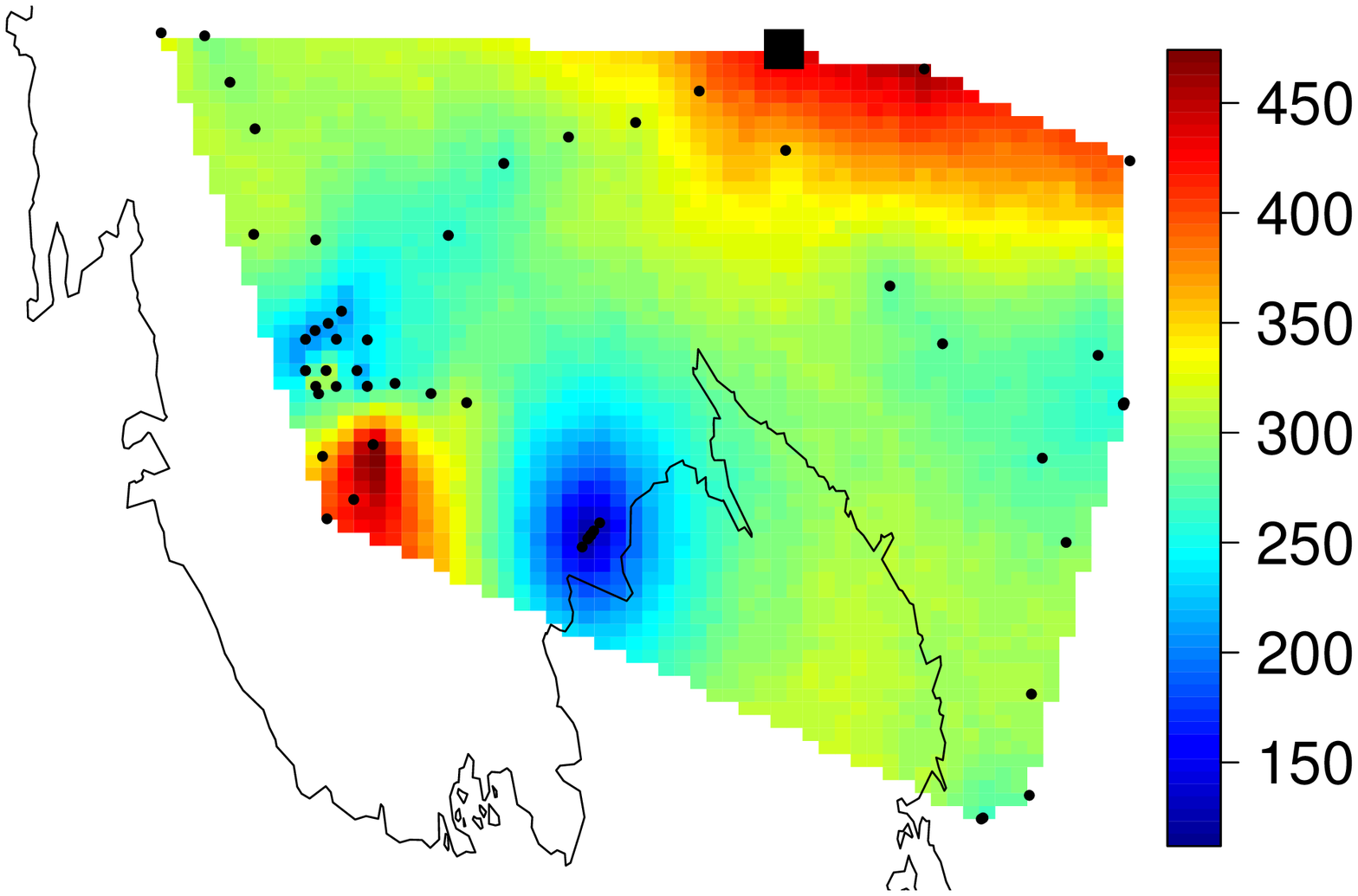}
\includegraphics[width=0.24\textwidth]{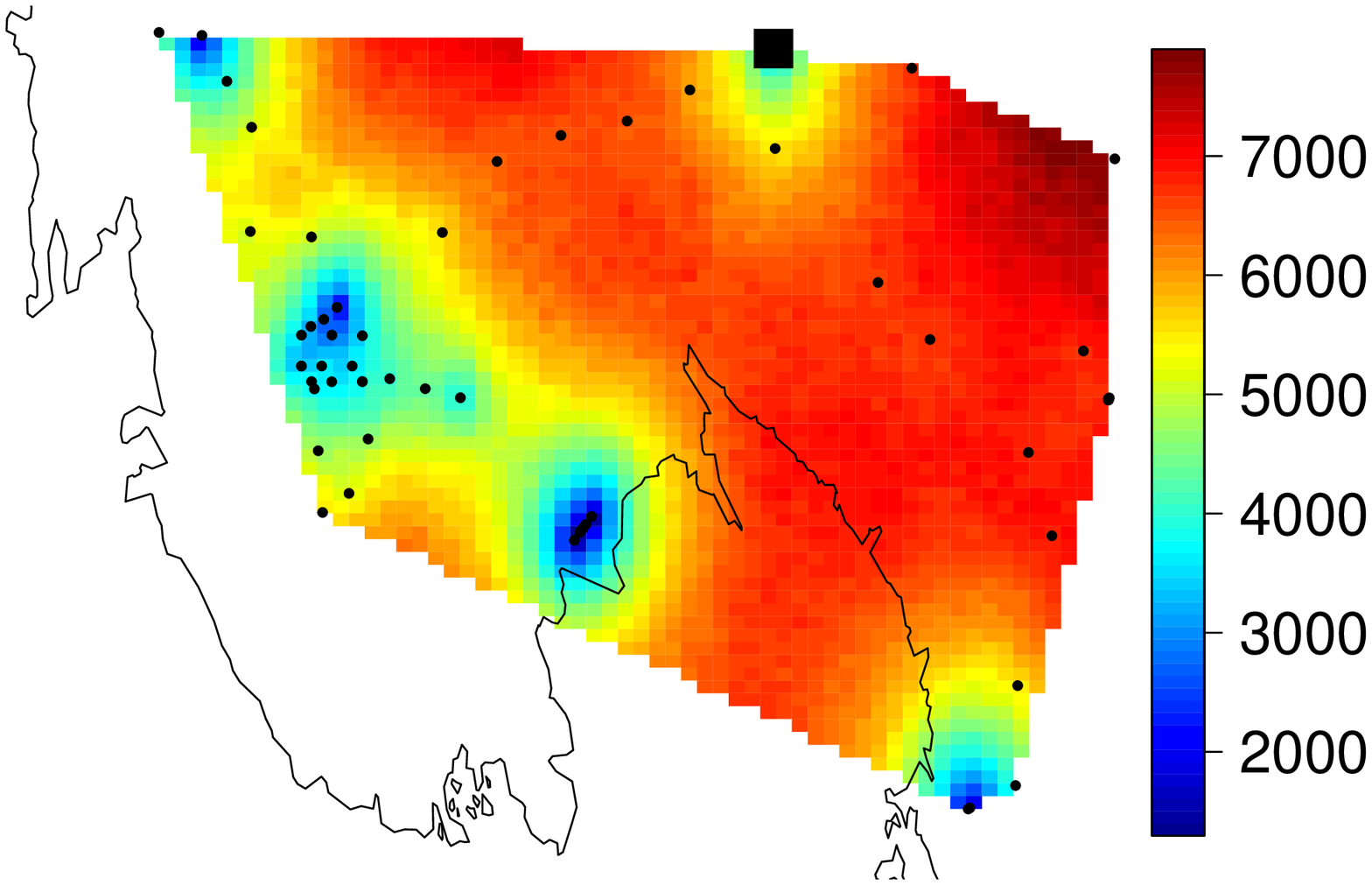}
\includegraphics[width=0.24\textwidth]{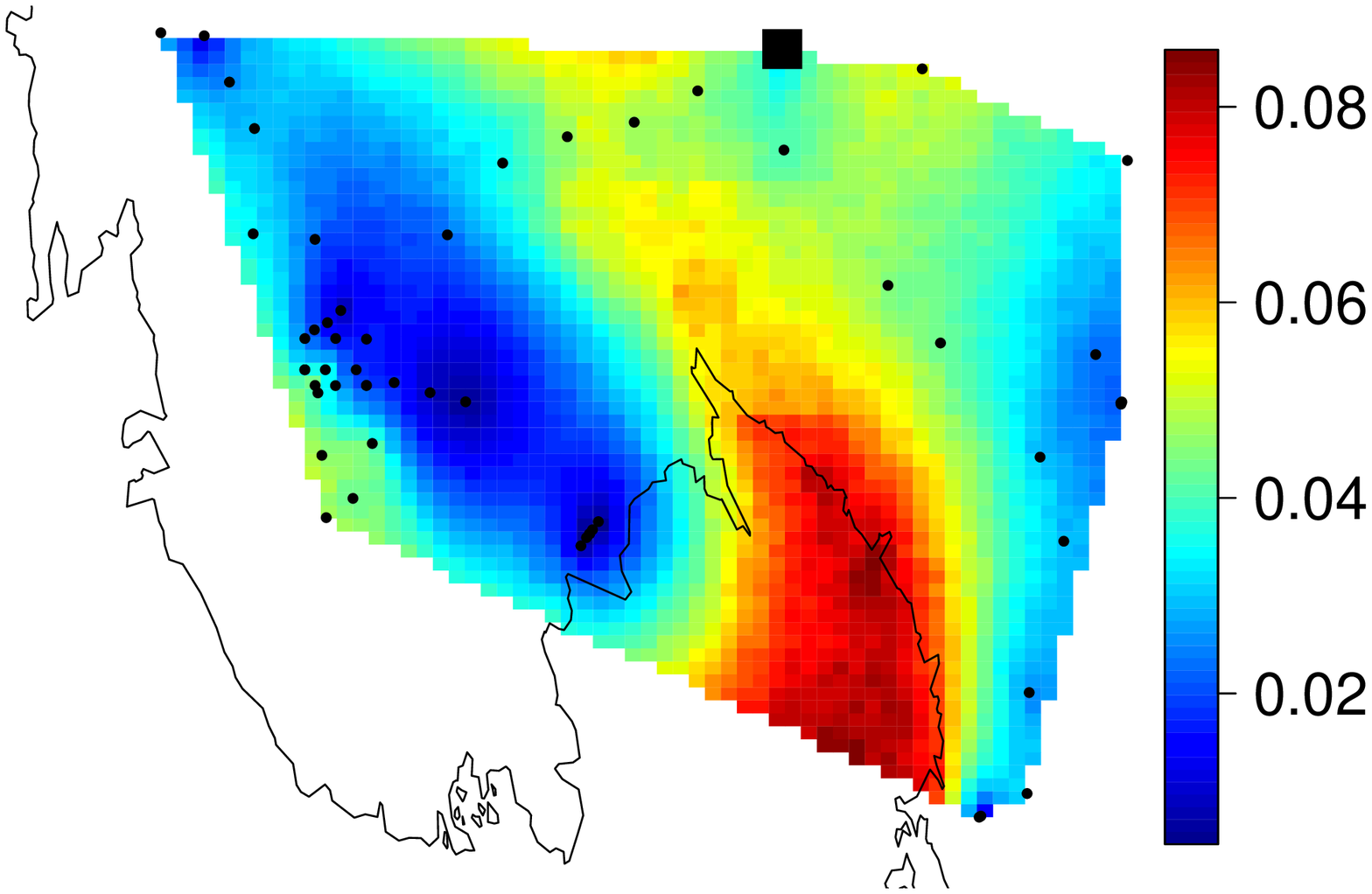}
\end{center}
\vspace{-4mm}

\caption{Interquartile range for (Row 1) surface density and the three critical densities, all in g/cm$^3$, from left to right. Interquartile range for (Rows 2-5) $A_l(\bs)$, $E_l(\bs)$, and $k_l(\bs) = A_l(\bs) \exp\left(-\frac{E_l(\bs)}{RT(\bs)} \right)$, for stages $l = 1,2,3,4$.}\label{fig:spat_unc}
\end{figure}

In Figure \ref{fig:spat_unc}, we plot the interquartile range of each spatially-varying parameter over the convex hull of our dataset. We use the interquartile range because untransformed parameters have extreme values. In general, uncertainty is highest near the boundaries of our interpolation region and far from drilled cores.

\vspace{-10mm}

